\documentclass[11pt,a4paper]{article}
\pdfoutput=1
\usepackage{myjheppub}
\usepackage{latexsym,amsfonts,amsmath,amssymb,textgreek,upgreek}
\usepackage{enumitem}
\usepackage{amsthm}
\usepackage{mathtools}
\usepackage{bbm}
\usepackage{graphicx}
\usepackage{color}
\usepackage{amsfonts}
\usepackage{caption}
\usepackage{subcaption}
\usepackage[normalem]{ulem}
\usepackage{comment}
\usepackage{url}
\usepackage{slashed}
\usepackage{bm}
\usepackage{hhline}
\usepackage{physics}
\usepackage{booktabs}
\usepackage{siunitx}
\usepackage{tikz}
\usepackage{tabu}
\usepackage{dsfont}

\usepackage{footmisc}

\usepackage{tikz}
\usetikzlibrary{decorations.pathreplacing, decorations.markings,calc,shapes.misc,decorations.pathmorphing,patterns.meta, math}
\usetikzlibrary{arrows}

\newcommand{\modd}{\text{ mod }2}

\newcommand\cc{s}

\renewcommand{\Im}{{\rm Im}}
\renewcommand{\Re}{{\rm Re}}
\renewcommand{\Tr}{\mbox{Tr}}

\newcommand{\SL}{\mathrm{SL}}

\def\Z{\mathbb{Z}}

\def\R{\mathrm{R}}
\def\NS{\mathrm{NS}}

\newcommand{\ii}{\text{i}}

\newcommand\be{\begin{equation}}
\newcommand\ee{\end{equation}}
\newcommand\bea{\begin{eqnarray}}
\newcommand\eea{\end{eqnarray}}

\renewcommand{\dd}{\mathrm{d}}

\renewcommand{\d}{\mathrm{d}}

\renewcommand{\a}{\alpha}
\renewcommand{\b}{\beta}

\renewcommand{\Re}{\text{Re}}
\renewcommand{\Im}{\text{Im}}

\newcommand{\p}{\partial}

 \renewcommand{\i}{\text{i}}

\renewcommand{\Im}{\mbox{Im}}
\renewcommand{\Re}{\mbox{Re}}

\newcommand{\beq}{\begin{equation}}
\newcommand{\eeq}{\end{equation}}
\newcommand{\ea}{\end{eqnarray}}
\newcommand{\barr}{\begin{array}}
\newcommand{\earr}{\end{array}}

\renewcommand{\a}{\mathfrak{a}}
\renewcommand{\b}{\mathfrak{b}}
\def\r{\mathsf{R}}
\def\T{\mathsf{T}}
\def\C{\mathsf{C}}

\def\H{\mathcal{H}}
\def\KB{{\rm KB}}
\def\K{\mathbb{K}}
\def\spin{\mathrm{spin}}
\def\pinp{{\mathrm{pin}^+}}
\def\pinm{{\mathrm{pin}^-}}
\def\PD{\mathrm{PD}}
\def\ABK{\mathrm{ABK}}
\def\Hom{\mathrm{Hom}}

\title{3D Gravity and Chaos in CFTs with Fermions 
\vspace{-0.8cm}}

\author{Jan Boruch${}^1$, Elisa Tabor${}^2$ and Gustavo J. Turiaci${}^3$}

\affiliation{
${}^1$Leinweber Institute for Theoretical Physics and Department of Physics, University of California, Berkeley, California 94720, USA
\\
${}^2$Leinweber Institute for Theoretical Physics, Stanford University, Stanford, CA 94305, USA
\\
${}^3$Physics Department, University of Washington, Seattle, WA, USA
}

\abstract{
Pure 3d gravity in AdS is believed to admit a holographic description in terms of 2d CFT. We introduce a theory of fermionic 3d gravity where we sum over geometries equipped with spin structure, and propose it is holographically described by fermionic 2d CFT data. We evaluate the leading contributions to the gravity path integral with one and two torus boundaries, extracting both the spectrum and its spectral statistics from the torus wormhole. Strikingly, the theory has fermionic black hole microstates, even in the absence of bulk fermionic matter. We then incorporate subtle bulk topological field theories, classified by appropriate cobordism groups, and evaluate the one and two-boundary torus partition functions. The spectral statistics we derive from gravity are shown, in all cases, to be consistent with the pattern of anomalies expected from classifications of fermionic 2d CFT. We also define a version of RMT$_2$, a random-matrix framework compatible with the symmetries of 2d CFTs, which naturally accommodates fermionic spectra and reproduces our gravitational results across all cases we analyze. 
}

\begin{document}

\maketitle 

\section{Introduction}

Random matrix theory provides a simple measure of chaos in quantum systems: a Hamiltonian is said to be chaotic if the statistics of its energy levels are distributed as if it were a random matrix. These spectral statistics share universal characteristics depending on the symmetries of the Hamiltonian \cite{Haake:2010fgh}.

One example of a chaotic quantum system is given by black holes, and as such the statistics of their microstates are expected to be well-approximated by random matrix theory \cite{Cotler:2016fpe,Saad:2018bqo}. In two dimensions, this has been shown explicitly through an exact duality between 2d quantum gravity and random matrix models \cite{Saad:2019lba}. A natural next step is to investigate what aspects of this duality generalize to pure gravity in three dimensions, expected to be dual to a 2d CFT \cite{Maldacena:1998bw, Maloney:2007ud, Witten:2007kt}. In this context, progress was made on understanding correlations between microstates in theories with only bosonic degrees of freedom \cite{Cotler:2020ugk,Belin:2020hea, Maxfield:2020ale, Chandra:2022bqq, DiUbaldo:2023qli, Belin:2023efa}, but in order to probe the random matrix statistics of black holes in explicit top-down holographic constructions \cite{Maldacena:1997re, Aharony:1999ti}, we would need to consider theories that also admit fermionic degrees of freedom, and eventually supersymmetry. Relatedly, from the boundary side, if one is interested in developing refinements of chaos that incorporate the symmetries of CFTs then the possibility of incorporating fermions is important.

In this paper, we take the first step in this direction. Just as pure bosonic 3d gravity captures the spectral statistics of 2d bosonic CFTs, we propose a theory of ``fermionic'' 3d gravity capturing statistics of CFTs with fermions. This restricts the definition of the gravitational path integral such that it only sums over geometries admitting spin structures, and generalizes a similar construction in 2d gravity by Stanford and Witten \cite{Stanford:2019vob}. The resulting theory can be intuitively interpreted as  gravity coupled to a gapped fermionic phase of matter, in which the remaining degrees of freedom are purely topological and described by an invertible topological quantum field theory \cite{Witten:2015aba}. The classification of such theories is given by equivalence classes of gapped symmetry-protected topological (SPT) phases of matter, which characterize anomalies of a theory in one dimension lower \cite{Atiyah:1968mp, Witten:1985bt, Witten:1985mj, Kapustin:2014tfa, Kapustin:2014dxa, Witten:2015aba, Freed:2016rqq}. We consider the effect of these topological field theories on the gravitational path integral and on spectral correlations, extending analogous results in lower dimensions \cite{Stanford:2019vob, Turiaci:2023jfa}.

More concretely, the diagnostic of random matrix statistics that we consider is the spectral form factor, an analytic continuation of two copies of the partition function that captures both the short- and long-range repulsion of energy eigenvalues characteristic of random matrix theory (RMT) \cite{Haake:2010fgh}. The duality between 2d quantum gravity and matrix integrals has led to gravitational computations of the spectral form factor \cite{Saad:2019lba}, which in this context probes spectral correlations between black hole microstates and consists of an early-time decaying slope followed by a universal ramp/plateau structure at late times \cite{Cotler:2016fpe, Garcia-Garcia:2016mno, Gharibyan:2018jrp}.

We focus on the transition from slope to linear ramp, which at the level of the gravitational path integral corresponds to a dominance of off-shell two-boundary connected contributions over factorized ones \cite{Saad:2018bqo}. We compute the analogue of these two-boundary wormholes in 3d fermionic gravity, where the two-boundary torus wormhole now captures correlations between energy levels of a putative dual 2d CFT with fermions. In this case, the path integral is constrained by modular invariance \cite{Maloney:2007ud, Witten:2007kt, Benjamin:2020zbs} and the wormhole admits a richer family of Euclidean geometries than its 2d analogue.

For pure bosonic 3d gravity, this off-shell wormhole has been computed either through a gravitational path integral \cite{Cotler:2020ugk} or through a modular completion of near-extremal data \cite{Cotler:2020hgz, DiUbaldo:2023qli}, see also \cite{deBoer:2025rct} for a different approach. The latter approach is part of a larger framework, referred to as RMT$_2$, which reorganizes chaotic microscopic data in a manner consistent with modular invariance \cite{Benjamin:2021ygh, DiUbaldo:2023qli, Haehl:2023xys, Haehl:2023tkr, Benjamin:2020zbs, Boruch:2025ilr, Perlmutter:2025ngj}. In this paper, we extend the framework of RMT$_2$ to account for fermionic fields, and we use this framework to compute spectral correlations between putative fermionic 2d CFT partition functions that are consistent with both RMT data and modular invariance on a torus with a given spin structure. On the one hand this leads to a prediction for gravitational two-boundary torus wormholes in the presence of fermions, and on the other hand this provides a benchmark for the signatures of quantum chaos in fermionic 2d CFTs.

We analytically continue these fermionic Euclidean wormholes, and at late times, we find agreement with random matrix statistics of either unitary, orthogonal, or symplectic symmetry classes, corresponding to the three Dyson ensembles \cite{Dyson:1962es}. The choice of Dyson ensemble depends on the algebra of the random matrix that approximates the Hamiltonian and the global symmetries of the dual CFT. In this paper, we consider putative 2d CFTs with fermion number symmetry $(-1)^{\sf F}$, angular momentum $J$, and $\r\T$ symmetry\footnote{In this paper, we will refer to $\r\T$ as the symmetry implied by the CPT theorem. References on the CPT theorem in quantum field theory include \cite{Luders:1954zz, Pauli:1955xn, Bell:1955djs, Jost:1957, Haag:1992hx, Streater:1989vi}. Some recent relevant discussions can be found in \cite{Freed:2016rqq, Hason:2020yqf, Witten:2023snr, Harlow:2023hjb, Witten:2025ayw, Seiberg:2025zqx}. We follow the convention of \cite{Witten:2023snr} in calling $\C\T$ symmetry $\T$. \label{foot:CPT footnote}}, and we later incorporate (1) time-reversal symmetry $\T$ such that $\T^2=1$, (2) time-reversal symmetry $\T$ such that $\T^2=(-1)^{\sf F}$, and (3) a unitary $\mathbb{Z}_2$ symmetry which can be interpreted as left-moving fermion number $(-1)^{\sf F_L}$.

The classical algebra of the Hamiltonian and the global symmetries can be anomalous at the quantum level, and the classification of these 2d anomalies precisely parallels the classification of 3d topological field theories \cite{Witten:2015aba, Seiberg:2025zqx}. The correct Dyson symmetry class depends not only on the classical global symmetry algebra, but also on its anomalies. In other words, the presence of a topological quantum field theory can lead to different RMT statistics than one might naively expect from the classical symmetries analysis \cite{Stanford:2019vob, Kapec:2019ecr, Turiaci:2023jfa}. In addition to the RMT analysis, we discuss the effect of gauging these (anomalous) global symmetries on the bulk two-boundary wormholes included in gravity.

For the cases we consider, the groups classifying anomalies are identical for free and interacting fermionic QFTs \cite{Kapustin:2014dxa}. Hence in the absence of a solvable holographic non-supersymmetric CFT dual to work with, we compare our predictions for the late-time behavior of two-boundary wormholes with expectations from 2d free fermion partition functions \cite{Kitaev:2009mg, Delmastro:2021xox, Seiberg:2023cdc, Seiberg:2025zqx}. We would like to emphasize that we use the free fermions just to infer the anomalies in the symmetry algebra, but obviously a free CFT would not display the spectral statistics we find from gravity.

\smallskip 

The layout of the paper is as follows. In Section \ref{sec:comments}, we comment on the relation between gravity and spin structures in a general setting and define what is meant by fermionic 3d gravity. In Section \ref{sec:Path_integral_fermionic_3d_gravity}, we compute the solid torus density of states and two-boundary torus wormholes for each choice of spin structure on the torus. In Section \ref{sec:rmt2}, we uplift the framework of RMT$_2$ to 2d CFTs with fermions, and we recover the two-boundary torus wormholes from modular uplifts of RMT data. In Section \ref{sec:discretesymm}, we discuss discrete spacetime symmetries in the gravitational path integral and consider the possible topological field theories on both $\spin$ and $\pinp$ structures. In Section \ref{sec:FL}, we introduce an internal $\Z_2$ symmetry given by left-moving fermion parity $(-1)^{\sf F_L}$ and discuss the corresponding topological field theories at the level of the solid torus and the two-boundary torus wormhole. Finally, we summarize our results and conclude with future directions in Section \ref{sec:discussion}.

\section{Comments on gravity and spin structures} \label{sec:comments}

Einstein gravity is usually characterized by presenting the action
\beq
I = - \frac{1}{16 \pi G_N} \int \d^D x \, \sqrt{g} ( R - 2\Lambda),
\eeq
together with the idea that we should sum over all geometries and topologies. But this information is not quite enough to fully specify the theory, we also need to specify which class of manifolds we are really summing over. In this section we describe how this subtle point that is often ignored can give rise to interesting effects, in particular regarding spin structures.

In its simplest realization, we want to sum over Riemannian manifolds $X$ equipped with a metric $g_{\mu\nu}$. Such a manifold $X$ can be covered with coordinate charts $U_i$ with $i$ an index labeling the set of charts $i\in I$ with the following properties. If the manifold admits a metric, one can select an orthonormal basis of tangent vectors $e^{(i)}_a$ with $a=1,\ldots, D$ for each chart $U_i$. For each intersection $U_i \cap U_j$  there exists transition matrices $M_{ij}$ which takes values in the group $\text{SO}(D)$ and map $e^{(j)}_a$ to $e^{(i)}_a$. They are defined up to orthogonal rotations of the frame vectors within each chart, namely $M_{ij} \simeq M_i M_{ij} M_j^{-1}$ where $M_i\in \text{SO}(D)$ and act on $e_a^{(i)}$ for all $i\in I$. On triple intersections $U_i \cap U_j \cap U_k$ the transition functions should satisfy the cocycle condition 
$
M_{ij} M_{jk} M_{ki}=1.
$
This data, together with the metric, specifies a given orientable Riemannian manifold $X$.

\smallskip

We refer to such a theory as ``bosonic Einstein gravity''. The choice of transition matrices being in $\text{SO}(D)$ is arbitrary and we could have chosen $\text{O}(D)$ or $\text{Spin}(D)$. As explained recently in \cite{Harlow:2023hjb}, the first choice can be interpreted as gauging parity transformations leading to what we refer to as ``unorientable bosonic Einstein gravity'' \cite{Stanford:2019vob,Harlow:2023hjb,Witten:2025ayw}. The second choice restricts the space of manifolds to those that admit a spin structure and will be our main focus. Usually this choice is imposed on us by the presence of fermionic matter fields, but we would like to emphasize that this should be seen as an ingredient in the precise definition of the gravity theory, and not on the matter content. This is similar to the situation for gauge theories, a precise definition requires making some choices regarding global properties of the group. 

\smallskip

The fermionic version of Einstein gravity is defined in the path integral formulation as summing over geometries where the transition matrices admit a lift from $\text{SO}(D)$ to its double cover $\text{Spin}(D)$ (or Pin groups if one wants to gauge parity). Let us explain this briefly. For a given $M_{ij} \in \text{SO}(D)$, its lift to $\widehat{M}_{ij} \in \text{Spin}(D)$ is defined only up to a sign. A spin structure is specified by a choice  of such signs $\widehat{M}_{ij} = \epsilon_{ij} M_{ij}$ with $\epsilon_{ij}\in \{ -1, 1\}$. Moreover, the overall sign of the spinors in each individual chart is also arbitrary. If we rescale spinors in chart $U_i$ by a sign $\epsilon_i \in \{-1,1\}$ then one can define an equivalence class of transition matrices lifts $\epsilon_{ij} \to \epsilon_i \epsilon_{ij} \epsilon_j$. The cocycle condition on triple overlaps $U_i \cap U_j \cap U_k$ further imposes the constraint 
$
\epsilon_{ij}\epsilon_{jk}\epsilon_{ki}=1,
$
such that the $\text{Spin}(D)$ uplift of the transition matrices satisfies the cocycle condition. Each spin structure can be mapped to an element of the cohomology group $H^1(X,\mathbb{Z}_2)$.\footnote{For a two-dimensional Riemann surface $\Sigma$, this definition is equivalent to the familiar holomorphic description of spin structures as square roots of the canonical bundle, $S^{\otimes 2}\cong K_\Sigma$. In particular, the Arf invariant agrees with the parity of the number of holomorphic sections of $S$, which is independent of the complex structure \cite{Atiyah1971}.}

\smallskip

One can show that a manifold admits a spin structure if and only if its second Stiefel-Whitney class $w_2(X) \in H^2(X,\mathbb{Z}_2)$ vanishes. For example, in $D\geq4$ not all manifolds admit a spin structure, with a famous example being the complex projective plane $\mathbb{CP}^2$. The existence of a spin structure restricts the set of geometries that appear in the gravitational path integral in a non-trivial way. It also includes a sum over $|H^1(X,\mathbb{Z}_2)|$ spin structures whenever they exist. We will label the set of spin structures of a spacetime by $\mathfrak{s}$.

\smallskip

It might be useful to have a more physical way to think about such a theory. For example, what can be a concrete way to implement this prescription in the path integral? We can couple the metric to a dynamical Dirac fermion of mass $M$ and then send the mass of the fermion to infinity $M\to\infty$. 
\beq \label{eq:couple_to_M}
I = - \frac{1}{16 \pi G_N} \int \d^D x \, \sqrt{g} (R-2\Lambda) + \int \d^D x \sqrt{g} \, \{ \overline{\Psi}  \slashed{D} \Psi - M \overline{\Psi} \Psi \},
\eeq
That way, the fermion would be integrated out, but the information that the sum over geometries only includes those with smooth spin structures remains.\footnote{This procedure could lead to leftover non-trivial dependence on spacetime geometry after the fermion is integrated out. We will study the effect of such terms in section \ref{sec:discretesymm}.} The discussion above intended to define the same theory without the need introducing a fermionic matter field.

\smallskip
\begin{figure}
    \centering
    \includegraphics[width=0.4\linewidth]{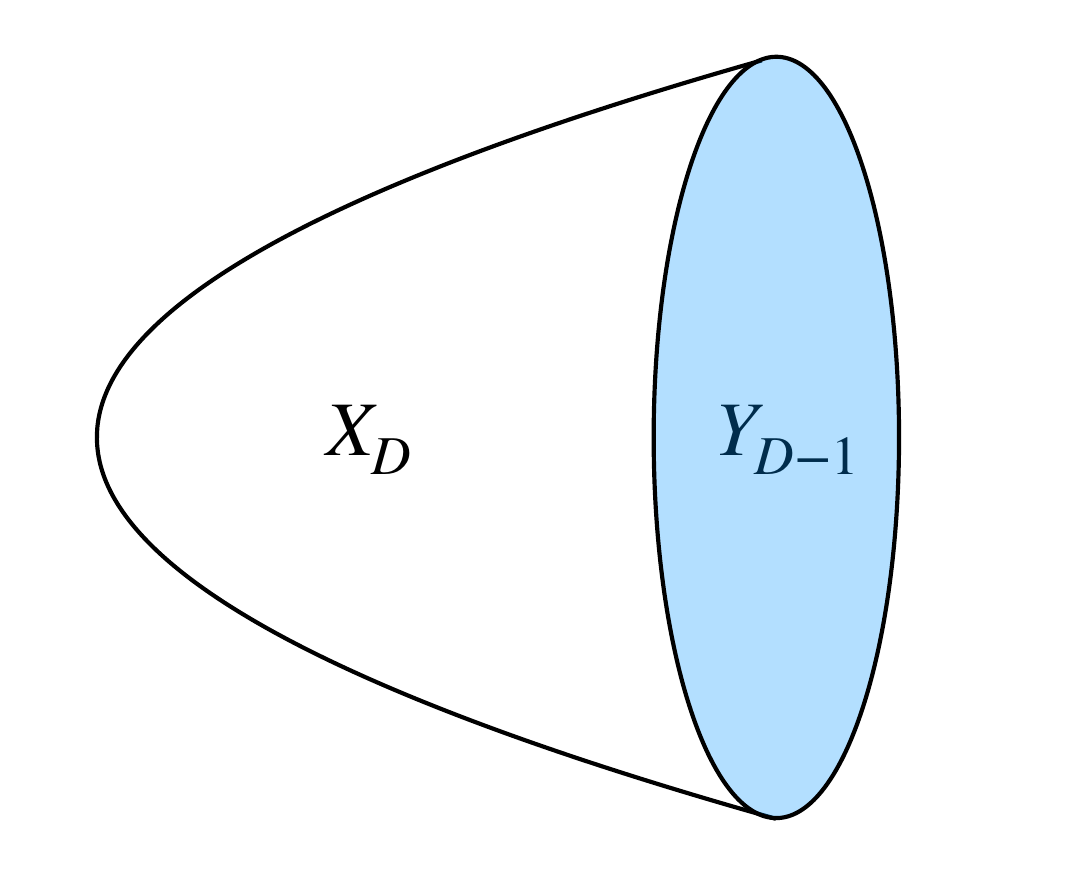}
    \caption{The $D-1$ boundary $Y_{D-1}$ is labeled by its topology, metric, and spin structure $\mathfrak{s}$. In the gravitational path integral we integrate over all manifolds $X_D$ with a bulk spin structure compatible with that of $Y_{D-1}$.}
    \label{fig:XY}
\end{figure}

The restriction over spin manifolds has another consequence, which will be relevant for us. In holography, the gravitational path integral can be thought of as an object that takes as input a $D-1$ dimensional space $Y_{D-1}$, performs an integral over all $D$ dimensional $X_D$ such that $\partial X_D = Y_{D-1}$, and produces a number. In the simplest orientable version of Einstein gravity, the boundary conditions on $Y_{D-1}$ are specified solely by its topology and Riemannian metric which we call $\gamma_{\mu\nu}$ (and other sources if matter fields are present). The path integral maps\footnote{We would like to compare the RHS with the partition function of the boundary. This can be more subtle if the boundary theory has anomalies since the phase of the partition function can be ambiguous.}
$$
\{ Y_{D-1},\gamma_{\mu\nu} \} \to \mathbb{C}.
$$
When we consider fermionic gravity one needs to enlarge the set describing the boundary conditions on $Y_{D-1}$ to include spin structures. Now the path integral provides a map
$$
\{ Y_{D-1}, \gamma_{\mu\nu}, \mathfrak{s}_Y \} \to \mathbb{C}.
$$
In particular, the path integral is only defined for boundaries that themselves admit spin structures. As we will see later in 3d, even though in $D=2,3$ all manifolds have $w_2=0$, the interplay between the sum over the bulk spin structures and the constraint on the boundary spin structure will lead to interesting effects. 

\smallskip 

This Euclidean discussion can be extended to the Lorentzian case. To construct bosonic gravity we replace the structure group $\text{SO}(D)$ by $\text{SO}(1,D-1)$, the proper Lorentz group. Although $\text{SO}(D)$ has a single connected component, $\text{SO}(1,D-1)$ has two connected components, one including the identity and another including ${\sf RT}$, the symmetry implied by the ``CPT theorem''. Therefore in gravity one should always think of ${\sf RT}$ as being gauged \cite{Harlow:2023hjb}. The Lorentzian continuation of the gravity theory including unoriented spaces has all four components of $\text{O}(1,D-1)$, meaning that ${\sf RT}$ and parity are gauged. One can also consider unoriented surfaces with spin structure, see section \ref{sec:discretesymm}.

\subsubsection*{Example: $S^1 \times S^{D-2}$ boundaries}

As an illustration of the importance of refining boundary conditions in the presence of a spin structure, consider the case $Y_{D-1}= S^1 \times S^{D-2}$ with a round sphere metric $\d s^2 = \d t^2 + \d \Omega_{D-2}^2$. The comments below apply to asymptotically AdS or flat gravity but the AdS case might be conceptually cleaner. 

\smallskip

$Y_{D-1}= S^1 \times S^{D-2}$  has two spin structures for $D>2$, specified in terms of transition functions in the spin group without any reference to a matter field. Nevertheless, it is conceptually convenient to introduce a spectator fermion and distinguish them by whether the fermion is periodic or antiperiodic along $S^1$, which we interpret to be the thermal circle. We refer to them as $\mathfrak{s}=\R$ (periodic) or $\mathfrak{s}=\NS$ (antiperiodic) structures respectively. 

\smallskip

$X_D$ with $\partial X_D = Y_{D-1}$ could be a static spherically-symmetric black hole. It contributes to the partition function in an ensemble with inverse temperature $\beta$ and angular velocity $\Omega=0$. Consider also a rotating black hole with the same temperature but $\Omega = 2 \pi \i /\beta$, and ignore spin structures for the moment. The corresponding boundary metric is
$$
\d s^2|_{Y} = \d t^2 + \d \theta^2 + \sin^2 \theta (\d \varphi- 2\pi \d t/\beta)^2 + \cos^2 \theta \d \Omega_{D-4}^2,
$$
where $\partial_\varphi$ generates the angular momentum conjugate to the angular velocity. This metric can be put in the original form without angular velocity by a large diffeomorphism
$$
(t,\varphi) \to (t, \varphi'=\varphi - 2\pi t/\beta).
$$
For generic values of the angular velocity this is incompatible with the identifications $t \sim t+ \beta$ and $\varphi \sim \varphi + 2\pi$. Therefore both the static black hole and the rotating one with $\Omega= 2\pi \i/\beta$ contribute to the same partition function if spin structures are ignored. In fact, for any angular velocity, the resulting gravitational path integral should be periodic in $\Omega \to \Omega + 2 \pi \i \mathbb{Z}/\beta$ implying that all states, and in particular all black holes, are bosonic. 

\smallskip

In fermionic gravity, the large diffeomorphism above is not allowed. It exchanges the boundary spin structure, and the two geometries with $\Omega=0$ and $\Omega=2\pi \i/\beta$ correspond to genuinely different observables. The former measures the total number of states and the latter the number of bosonic minus fermionic states. Implementing this large diffeomorphism twice brings us back to the same boundary spin structure and therefore the path integral is now periodic in $\Omega \to \Omega + 4 \pi \i \mathbb{Z}/\beta$. Therefore a theory of gravity where we sum over manifolds with transition functions in the spin group has fermionic black hole microstates \cite{Chen:2023mbc}. Importantly, this conclusion is true regardless of whether the matter spectrum has fermions in it or not. This is a striking conclusion since we usually think of black holes as being made of collapsing matter.

\subsubsection*{Example: 2d gravity}

In 2d gravity, Stanford and Witten constructed a fermionic version of JT gravity in this fashion \cite{Stanford:2019vob}. Since AdS JT gravity is related to $\text{PSL}(2,\mathbb{R})$ $BF$-theory one can also achieve this by replacing the group by its double-cover $\text{SL}(2,\mathbb{R})$. The most general boundary has $n$ disconnected circles each with either NS or R boundary conditions. 

\smallskip

Evaluating the gravitational path integral of the JT fermionic theory is straightforward. When the number of circles with R structure is odd, one can show that there is no 2d space filling these circles so the answer identically vanishes.\footnote{This elementary fact can be explained in a way that connects with section \ref{sec:discretesymm}. The bordism group of a set of circles is $\mathbb{Z}_2$ and they are distinguished by the 1d mod 2 index counting fermionic zero-modes mod 2 \cite{Kirby:1991}. When the number of R boundaries is even then the number of zero-modes is even and the circles can be filled. When the number of R boundaries is odd the number of zero-modes is odd and it cannot be filled. \label{footnote: bordism}} When the number of R boundaries is even or zero, one simply obtains the bosonic JT amplitude computed in \cite{Saad:2019lba} multiplied by the number of spin structures compatible with the boundary one
$
\sum_{\text{spin}} 1 = 2^{2g+n-1},
$
where $g$ is the genus of the surface. This is easy to compute by identifying the number of spin structures with the number of elements of the group $H^1(X,\mathbb{Z}_2)$.

\smallskip

The innocent factor of $2^{2g+n-1}$ has an interesting implication to the RMT dual. The path integral of bosonic JT gravity is dual to a double-scaled random self-adjoint Hamiltonian $H$. Instead, the RMT dual to fermionic JT gravity is given by a random Hamiltonian compatible with  fermion parity $\mathbb{Z}_2$. The Hilbert space decomposes in bosonic and fermionic components $\mathcal{H} = \mathcal{H}_b \oplus \mathcal{H}_f$ 
\beq\label{minf}
H=\left(\begin{array} {c|c}  H_b & 0\cr \hline 0 & H_f\end{array}\right),~~~~~~(-1)^{\sf F}=\left(\begin{array} {c|c}  1 & 0\cr \hline 0 & -1\end{array}\right).
\eeq
Both $H_b$ and $H_f$ are identically distributed, but otherwise statistically independent, random matrices in the GUE ensemble. The fact that they are identically distributed is the manifestation that the gravity path integral with an odd number of R boundaries vanishes identically. The states in the component $\mathcal{H}_f$ correspond to fermionic black hole states in $\text{AdS}_2$. This provides a very simple manifestation of the features described earlier in this section. In the rest of this paper we will more complicated versions of this construction that appear in 3d gravity and 2d CFTs with fermions.

\section{Path integral of fermionic 3d gravity} \label{sec:Path_integral_fermionic_3d_gravity}

In this section we evaluate the gravitational path integral of fermionic 3d gravity on the solid torus as well as the two-boundary torus wormhole. As we will see, the effect of incorporating spin structure will have more interesting interplay with gravity than its 2d counterpart.

\subsection{Setup and boundary conditions}

Pure fermionic 3d gravity includes spaces where the transition functions can be lifted to $\text{Spin}(3)$. There is an interesting feature in this case which doesn't happen in higher dimensions. If we work in Lorentzian signature $\pi_1(\text{SO}(1,2))=\mathbb{Z}$ and therefore the double cover $\text{Spin}(1,2)$ is not simply connected. This allows for the possibility of defining anyonic versions of 3d gravity. We will comment on this later but for most of the paper the focus will be in the fermionic setup.

\smallskip

We begin by enumerating the possible 2d boundaries according to their spin structure, restricting for simplicity to conformal boundaries which are one or several 2-tori. Although all 3d manifolds admit a spin structure, there is an interesting interplay between the choice of spin structure and the modular group of large diffeomorphisms. 

\smallskip

Let us focus on any such conformal boundary and introduce coordinates $(t,x)$ and metric
\beq
\d s^2 = \d t^2 + \d x^2 ,~~~~(t,x) \sim (t,x+2\pi) \sim (t+\beta, x + \theta).
\eeq
The parameter $\theta$ is proportional to the angular velocity $\theta= \i \beta \Omega$. The moduli parameters are
$$
\tau = \frac{\theta + \i \beta}{2\pi} ,~~~~\overline{\tau}=\frac{\theta - \i \beta}{2\pi} .
$$
We can define the boundary spin structure in terms of the signs that appear in the transition functions lifted to the spin group, but it is easier to introduce a spectator fermion $\Psi(t,x)$ whose periodicity conditions characterize the spin structure 
\bea
\Psi ( t+\beta, x+\theta) &=& e^{\pi \i \mu}\, \Psi( t, x), \\
\Psi ( t, x+ 2\pi ) &=& e^{\pi \i \nu}\, \Psi ( t , x),
\ea
and therefore $\mathfrak{s}\in\{ (\mu,\nu), \text{with $\mu,\nu\in  \mathbb{Z}/2\mathbb{Z}$}\}$. Antiperiodic boundary conditions (NS) correspond to $\mu$ or $\nu$ taking the value $1$. Periodic boundary conditions (R) correspond to the value $0$.\footnote{The choice of $\nu$ corresponds, in the canonical quantization of a putative dual 2d CFT, to two different Hilbert spaces $\mathcal{H}_\NS$ ($\nu=1$) and $\mathcal{H}_\R$ ($\nu=0$). Instead the choice of $\mu$ specifies whether the partition function has a $(-1)^{\sf F}$ insertion ($\mu=0$) or not ($\mu=1$).}

\smallskip

\begin{figure}
    \centering   \includegraphics[width=0.5\linewidth]{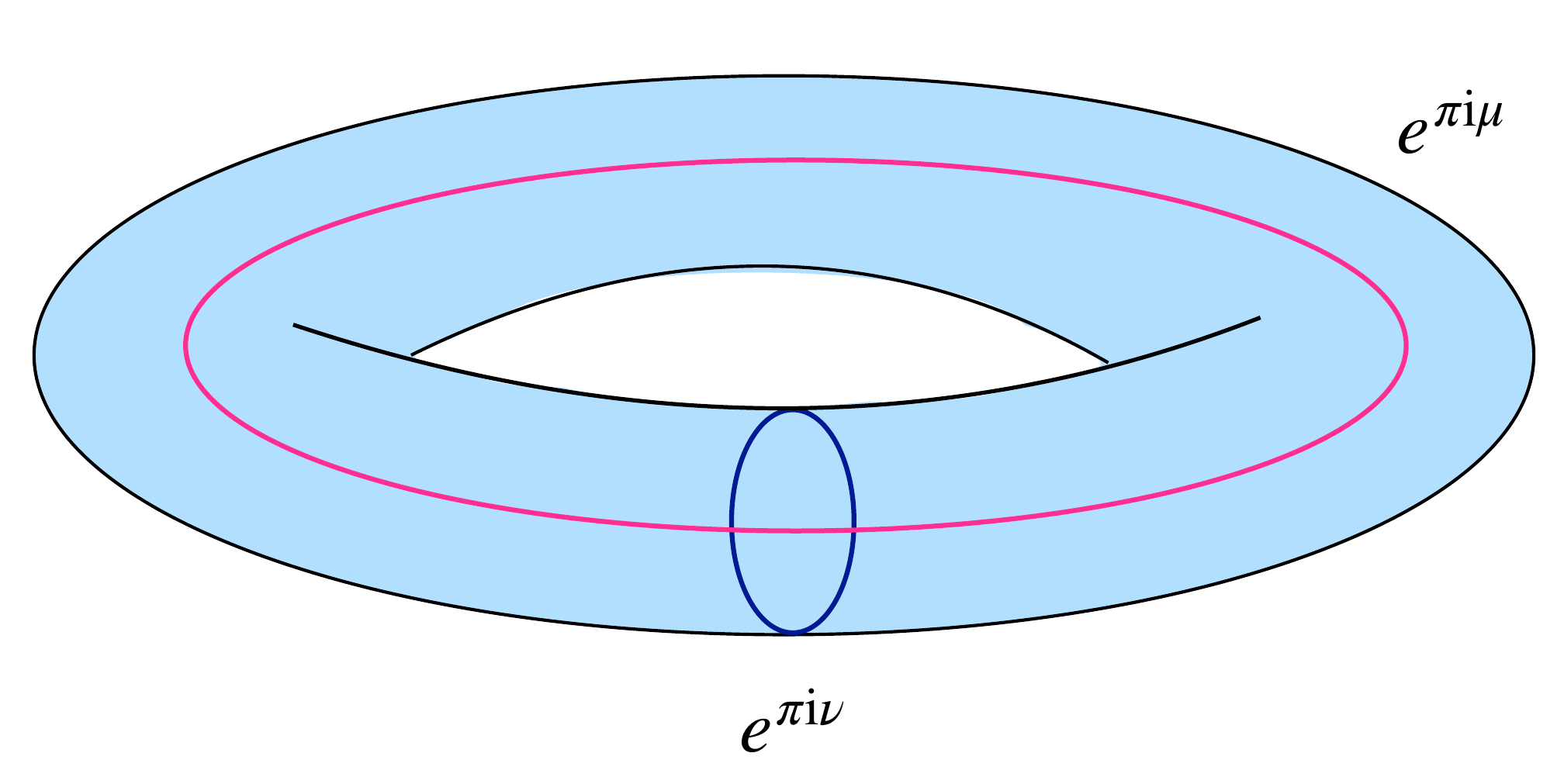}
    \caption{Conventions for the torus spin structures. R and NS Hilbert spaces correspond to $\nu=0$ and $\nu=1$ respectively. $\mu=0$ corresponds to a $(-1)^{\sf F}$ insertion.}
    \label{fig:placeholder}
\end{figure}

There is another description of $\mathfrak{s}$ that is very natural from the 2d CFT point of view. When $\nu=0$ ($\nu=1$) we refer to it as a R (NS) boundary. When $\mu=0$ ($\mu=1$) we label it by $+$ ($-$) indicating fermions are periodic/antiperiodic around the thermal circle. In this notation $\mathfrak{s}\in\{ \R+, \R-, \NS+, \NS-\}$. It will be useful to keep both descriptions. The interpretation in the putative boundary 2d CFT of the gravity path integral with a single such torus boundary is
\bea
Z_{\R+}(\tau,\bar{\tau}) &=& {\rm Tr}_{\mathcal{H}_\R} \left[ ( -1)^{\sf F} \, q^{L_0 - \frac{c}{24}} \bar{q}^{\bar{L}_0 - \frac{c}{24}} \right],
\label{eq:R+_partition_function}
\\
Z_{\R-}(\tau,\bar{\tau}) &=& {\rm Tr}_{\mathcal{H}_\R} \left[  q^{L_0 - \frac{c}{24}} \bar{q}^{\bar{L}_0 - \frac{c}{24}} \right],
\label{eq:R-_partition_function}
\\
Z_{\NS+}(\tau,\bar{\tau}) &=& {\rm Tr}_{\mathcal{H}_\NS} \left[ ( -1)^{\sf F} \, q^{L_0 - \frac{c}{24}} \bar{q}^{\bar{L}_0 - \frac{c}{24}} \right],
\label{eq:NS+_partition_function}
\\
Z_{\NS-}(\tau,\bar{\tau}) &=& {\rm Tr}_{\mathcal{H}_\NS} \left[  q^{L_0 - \frac{c}{24}} \bar{q}^{\bar{L}_0 - \frac{c}{24}} \right].
\label{eq:NS-_partition_function}
\ea
The expectation in pure gravity is that the RHS corresponds to (perhaps an ensemble of) 2d CFTs whose statistics we want to characterize. 

\smallskip

The gravitational path integral with torus boundary conditions should be invariant under the subgroup $\Gamma$ of $\SL(2,\Z)$ leaving the spin structure unchanged. We shall denote the subgroup corresponding to each partition function by the same labels $\Gamma_{\R/\NS\, \pm} \subset \SL(2,\Z)$. To determine the subgroups in a language appropriate to gravity without relying on a putative dual CFT we follow \cite{Maloney:2007ud}. The spin structure after a modular transformation is
\beq
\left(\begin{array}{c}
     \mu  \\
     \nu
\end{array} \right) \to \left(\begin{array}{cc}
     a & b \\
     \cc & d
\end{array} \right) \left(\begin{array}{c}
     \mu  \\
     \nu
\end{array} \right).
\eeq
where $a,b,s,d$ are always integers such that $ad-bs=1$. We can analyze the relevant subgroup by finding the matrices that preserve the column vector mod $2\mathbb{Z}$. The answer for each boundary condition is:

\begin{itemize}
\item $(\mu,\nu)=(0,0)$ corresponding to $\R+$. Any transformation preserves the vanishing of both $\mu$ and $\nu$ and therefore the full modular group preserves the spin structure
\beq
\Gamma_{\R + } = \text{PSL}(2,\Z). 
\label{eq:R+_modular_group}
\eeq

\item $(\mu,\nu)=(1,0)$ corresponding to $\R-$. In this case $\mu \to a \mu$ and $\nu \to \cc \mu$. Preserving the spin structure imposes $\cc = 0\modd$ and $a = 1\modd$, and since $ad-b\cc =1$ it is enough to impose $\cc =0 \modd$. To summarize the modular groups are
\beq
\Gamma_{\R-} = \Gamma_0(2) \equiv \{ g \in \text{PSL}(2,\Z)\, | \, s = 0 \modd\}.
\label{eq:R-_modular_group}
\eeq

\item $(\mu,\nu)=(0,1)$ corresponding to $\NS+$. This structure is preserved by $b=0\modd$ and $d = 1 \modd$ so that
\beq
\Gamma_{\NS+} = \Gamma^0(2) \equiv \{ g\in \text{PSL}(2,\Z) \, | \, b=0 \modd\}
\label{eq:NS+_modular_group}
\eeq

\item $(\mu,\nu)=(1,1)$ corresponding to $\NS-$. We require $a+b = \cc + d = 1 \modd$. The modular subgroups are 
\beq\Gamma_{\NS-} = \Gamma_ \theta \equiv \{ g\in\text{PSL}(2,\Z) \, | \, a+b = \cc + d = 1 \modd\}.
\label{eq:NS-_modular_group}
\eeq

\end{itemize}

The three even spin structures are not independent of each other and are related in a simple way
\beq\label{eq:relbtwpff}
Z_{\NS-}(\tau) = Z_{\NS +} (\tau + 1),~~~Z_{\R-}(\tau)=Z_{\NS+}(-1/\tau). 
\eeq
The left relation is non-trivial since the modular transformation $T:\tau \to \tau+1$ is not an element of $\Gamma^0(2)$. This shows that in the NS Hilbert space spin-statistics relation holds $(-1)^{\sf F}= e^{2\pi \i J}$. This is not true in the Ramond Hilbert space. All states of the putative dual CFT, bosonic or fermionic, have integer momentum. In a theory of free-fermions this is due to the presence of fermion zero-modes in the Ramond sector, but the argument above applies to any theory.

\subsection{Path integral on the solid torus} \label{subsec: solid torus}

In this section we compute the gravitational path integral with a single torus boundary. We also restrict ourselves to the 3d solid torus including cases such as thermal AdS or the BTZ black hole. We know that pure 3d gravity requires off-shell geometries \cite{Maxfield:2020ale} but this is a reasonable quantity to evaluate first. In particular we will see how the quantization of the theory leads to the presence of fermionic black holes regardless of the matter content.   

\smallskip

Start with thermal AdS, a choice of solid torus such that it is the boundary spatial circle that contracts. The path integral is given by \cite{Maloney:2007ud,Giombi:2008vd, Keller:2014xba}
\beq
Z_{\text{TAdS}}(\tau,\bar{\tau}) = \frac{q^{-\frac{c}{24}} }{\prod_{n=2}^\infty (1-q^n) }\,\frac{\bar{q}^{-\frac{c}{24}}}{\prod_{n=2}^\infty(1-\bar{q}^n)}.\label{eq:vacchar}
\eeq
The numerator corresponds to the classical action and the denominator to its perturbative quantum corrections arising from the graviton. All other solid tori can be obtained by a modular transformation
$
\gamma = \left(\begin{smallmatrix}
    a & b \\
    s & d
\end{smallmatrix}\right) \in \text{PSL}(2,\mathbb{Z})$.
The contribution of each solution is given by \eqref{eq:vacchar} with the replacement $\tau \to \gamma \cdot \tau = \frac{a\tau+ b}{ s \tau + d}$ \cite{Dijkgraaf:2000fq,Manschot:2007ha}. Not all modular transformations generate genuinely different geometries. The solid torus is specified by a choice of a unique boundary cycle that contracts inside the bulk, the spatial one in the case of thermal AdS. On the other hand there is no unique choice for the non-contractible cycle and the bulk solution is invariant under $\Gamma_\infty$ generated by $\gamma \to T \gamma$ or $(a,b) \to (a,b) + \mathbb{Z} (s,d)$ which attaches the contractible cycle some integer number of times. In bosonic gravity the sum is over $s,d$ and since integer $a,b$ should exist such that $ad-bs=1$, $s$ and $d$ should be coprime.

The sum over the modular group in fermionic gravity is further restricted by the spin structure. Moreover, thermal AdS does not always allow for a smooth bulk spin structure and one might have to find another seed. The answers in each sector are listed below:

\begin{figure}
    \centering
\includegraphics[width=0.6\linewidth]{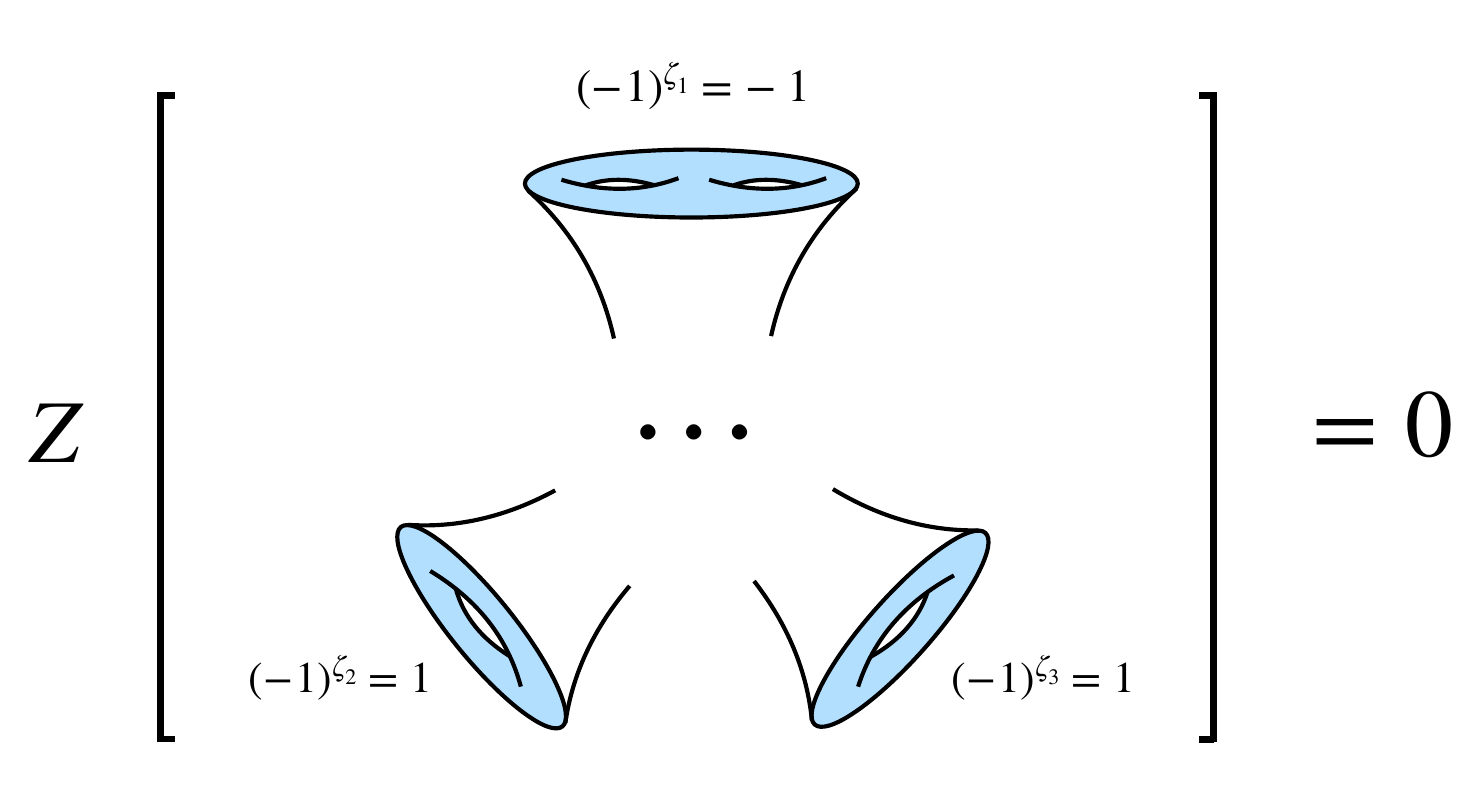}
    \caption{Example of the fact that the gravitational path integral with boundary 2d spin surfaces $Y=Y_1 \cup Y_2 \cup Y_3 \cup \ldots$ vanishes if $\sum_i \zeta_i = 1~\text{mod}~2$. When $\zeta \neq 0 $ no 3d spin surface exists with such a boundary. This applies to a union of tori but also more general surfaces.}
    \label{fig:zero}
\end{figure}
\subsection*{Ramond sector with $(-1)^{\sf F}$}

In this case, there is no 3d geometry that can allow the boundary spin structure to smoothly extend to the interior and the path integral vanishes
\beq \label{ZRP}
Z_{\R+} = 0.
\eeq
One can argue that no asymptotically AdS$_3$ spin geometry exists, either on-shell or off-shell, with an odd spin structure boundary. Intuitively if all cycles have periodic fermions then none of them can smoothly contract. To see this in a way that connects with section \ref{sec:discretesymm} we evaluate the boundary 2d mod 2 index $\zeta$ that distinguishes the trivial and non-trivial component of the 3d spin bordism group. For an R+ structure on the torus, there is a single chiral zero-mode and therefore $\zeta = 1~\text{mod}~2$, implying it is in the non-trivial component of the bordism group and is not the boundary of any 3d spin geometry. In fact, take a union of 2d boundaries of any kind $Y = Y^1 \cup Y^2 \cup \ldots \cup Y^n$ such that $n_+$ have $\zeta_i =0 ~\text{mod}~2$ and $n_-=n-n_+$ have $\zeta=1~\text{mod}~2$. Then the mod 2 index is
$$
\zeta = n_-~\text{mod}~2
$$
and we can immediately deduce the following 
$$
Z(Y) =0 ,~~~~\text{where}~~~~Y=\underbrace{Y^1_{\zeta=1} \cup \ldots \cup Y^{n_-}_{\zeta=1}}_{\text{odd number}} \,\cup\, Y^{n_-+1}_{\zeta=0}\cup  \ldots .
$$
since no $X_3$ exists such that $\partial X_3 = Y$ when $\zeta_Y = 1~ \text{mod}~2$. In particular there is no wormhole connecting an odd number of R+ boundaries.

\smallskip

\subsection*{NS sector with $(-1)^{\sf F}$} 

Thermal AdS is compatible with this spin structure. Since there are no propagating fermionic degrees of freedom (as opposed to e.g. $\mathcal{N}=1$ supergravity) the quantum corrections are precisely the same as in bosonic 3d gravity and the only modification is the restriction to modular transformations that preserve the spin structure
\beq\label{ZNSP}
Z_{\NS + }(\tau,\bar{\tau}) = \sum_{\gamma \in \mathbb{Z} \backslash \Gamma^0(2)} Z_{\text{TAdS}}(\gamma\tau,\gamma\bar{\tau}),~~~Z_{\text{TAdS}}(\tau,\bar{\tau})=\frac{q^{-\frac{c}{24}} }{\prod_{n=2}^\infty (1-q^n) }\,\frac{\bar{q}^{-\frac{c}{24}}}{\prod_{n=2}^\infty(1-\bar{q}^n)}.
\eeq
The restriction $\gamma \in \mathbb{Z} \backslash \Gamma^0(2)$ identifies $\gamma \sim T^{2n} \gamma$ with $n\in\mathbb{Z}$ which preserves the spin structure. In the sum over coprime $s$ and $d$ we restrict to only odd $d$. Moreover to complete the modular transformation we should choose a solution of $a d- b s =1$ with odd $a$ and even $b$. It is worth emphasizing that the BTZ black hole is not part of this sector since $S=\left(\begin{smallmatrix}
    0 & -1\\
    1 & 0
\end{smallmatrix}\right)
$
is not an element of $\Gamma^0(2)$. 

\smallskip

\subsection*{NS sector without $(-1)^{\sf F}$}

Thermal AdS also contributes in this sector leading to a similar answer but the modular subgroup is modified
\beq\label{ZNSM}
Z_{\NS - }(\tau,\bar{\tau}) = \sum_{\gamma \in \mathbb{Z} \backslash \Gamma_\theta} Z_{\text{TAdS}}(\gamma \tau,\gamma \bar{\tau}).
\eeq
Elements in $\mathbb{Z}\backslash \Gamma_\theta$ are identified by $\gamma \sim T^{2n} \gamma$ preserving spin structure. The modular sum can be reduced to a sum over $s$ and $d$ coprime with $s+d$ odd. Moreover we should choose $a$ and $b$ such that $a+b$ is odd but otherwise the summand does not depend on the choice. One can show that
\beq
T^{-1} \, \Gamma_\theta \, T = \Gamma^0(2),~~~~T=\begin{pmatrix}
    1 & 1 \\
    0 &1
\end{pmatrix} 
\eeq
meaning that $\gamma \in \Gamma^0(2)$ if and only if $T^{-1}\,\gamma \, T \in \Gamma_\theta$. This implies that $Z_{\NS+}(\tau,\bar{\tau}) = Z_{\NS-}(T\tau,T\bar{\tau})$ in accordance with the spin-statistic connection in the NS Hilbert space. To show this one rewrites the argument of the RHS of \eqref{ZNSM}, applying $T$ on $\tau$, as $\gamma T \tau \to T T^{-1} \gamma T \tau$. The sum over $\gamma \in \Gamma_\theta$ is replaced by a sum over $\gamma'=T^{-1} \gamma T\in \Gamma^0(2) $. The remaining factor of $T$ multiplying $\gamma'$ on the left leaves the partition function of thermal AdS invariant and has no effect.\footnote{This is not true for $\mathcal{N}=1$ supergravity. The thermal AdS partition function (given by the super-Virasoro vacuum character) is not invariant under $\tau \to T \tau$ due to the presence of fermionic modes. The effect of acting with $T$ is precisely to exchange the perturbative path integral in the $\NS+ \leftrightarrow \NS-$ sectors. }

\smallskip

\subsection*{R sector without $(-1)^{\sf F}$}
The spin structure rules out thermal AdS and it cannot be used as a seed. The BTZ black hole does admit a smooth bulk spin structure and its path integral is
$$
Z_{\text{BTZ}}(\tau, \bar{\tau}) = Z_{\text{TAdS}}(S \cdot \tau, S \cdot \bar{\tau}),~~~~\text{with}~~S = \begin{pmatrix} 0&-1\\1&0 \end{pmatrix}.
$$
The full path integral is
\bea\label{ZRM}
Z_{\R - }(\tau,\bar{\tau}) &=& \sum_{\gamma \in \mathbb{Z} \backslash \Gamma_0(2)} Z_{\text{BTZ}}(\gamma \cdot \tau, \gamma \cdot \bar{\tau}),\nonumber\\
&=&\sum_{\gamma \in \mathbb{Z} \backslash \Gamma_0(2)} Z_{\text{TAdS}}(S\gamma\tau,S\gamma\bar{\tau}).
\ea
The sum over modular images is parameterized by $\gamma$ and the factor of $S$ acting on $\tau$ is imposing that the seed is BTZ and not thermal AdS. The modding by $\mathbb{Z}$ identifies $S\gamma \sim T^{2n} S \gamma$. The sum in this case reduces to $s$ and $d$ coprime with even $s$ and this condition looks slightly different when written in terms of the components of $S \, \gamma$. Notice that 
\beq
S \, \Gamma_0(2)\,  S^{-1} = \Gamma^0(2).
\eeq
This can be used to show that $Z_{\NS+}(\tau,\bar{\tau}) = Z_{\R-}(-1/\tau, -1/\bar{\tau})$.

\smallskip

\subsection*{Leading 2d CFT spectrum}

The path integral with a conformal torus as a boundary can be interpreted as the partition function of a putative 2d CFT as explained earlier. We can extract the contribution from the solid torus, following \cite{Maloney:2007ud, MaxfieldUnp,Benjamin:2020mfz}, to the (average) density of states in the R and NS Hilbert spaces. The Hilbert spaces decompose as
\beq
\mathcal{H}_{\NS} = \bigoplus_{j\in \mathbb{Z}/2} \mathcal{H}_{\NS, j},~~~~\mathcal{H}_\R = \bigoplus_{j\in\mathbb{Z}} \mathcal{H}_{\R,j}^{\text{bos}} \oplus \mathcal{H}_{\R,j}^{\text{fer}},
\eeq
and the spectrum for each component of each Hilbert space is defined through 
\bea
Z_{\R-}(q,\bar{q}) &=&  \sum_{j\in  \mathbb{Z}} \,\int_{-\frac{c}{12}}^{\infty}\, \d E\,(\rho^{\text{bos}}_{\R}(E, j) +\rho^{\text{fer}}_{\R}(E, j) ) \, \chi_{E,j}(q,\bar{q})  ,
\label{eq:R-_partition_function_spectrum}
\\
Z_{\NS-}(q,\bar{q}) &=& \sum_{j\in \frac{1}{2} \mathbb{Z}} \, \int_{-\frac{c}{12}}^{\infty} \,\d E\,\rho_{\NS} (E, j) \, \chi_{E,j}(q,\bar{q}) .
\label{eq:NS-_partition_function_spectrum}
\ea
The characters on the RHS include the sum of a primary of energy $E$ and spin $j$ and its descendants. It is given by $\chi_{E,j}(q,\bar{q}) =\chi_h(q) \chi_{\bar{h}}(\bar{q})$ where $E=h+\bar{h}-c/12$, $j= h-\bar{h}$, and 
\beq
\chi_{h} = \frac{q^{h-\frac{c-1}{24}}}{\eta(\tau)},~~~\chi_{\text{vac}}=\frac{(1-q)q^{-\frac{c-1}{24}}}{\eta(\tau)}.
\eeq
If either $h$ or $\bar{h}$ vanish that component is replaced by $\chi_{\text{vac}}$. We consider theories where the only degenerate state is the vacuum with $E=-c/12$ and $j=0$. For unitary theories $\rho_{\R}(E,j)$ and $\rho_{\NS}(E,j)$ should be non-negative for all angular momenta. The angular momenta in the R Hilbert space is integral and unrelated to fermion parity. Since we argued that the partition function with an odd number of $\R+$ boundaries vanish we deduce that
$$
\langle \rho^{\text{bos}}_{\R}(E, j)\rangle   =\langle \rho^{\text{fer}}_{\R}(E, j)\rangle  = \langle \rho_{\R}(E, j)\rangle  ,
$$
and all energies/spin come with a two-fold degeneracy of bosons and fermions. Let us emphasize that this is only true on average for the ensemble, we will discuss this further when considering wormholes.

\smallskip 

\begin{figure}[t!]
    \centering
    \includegraphics[width=\linewidth]{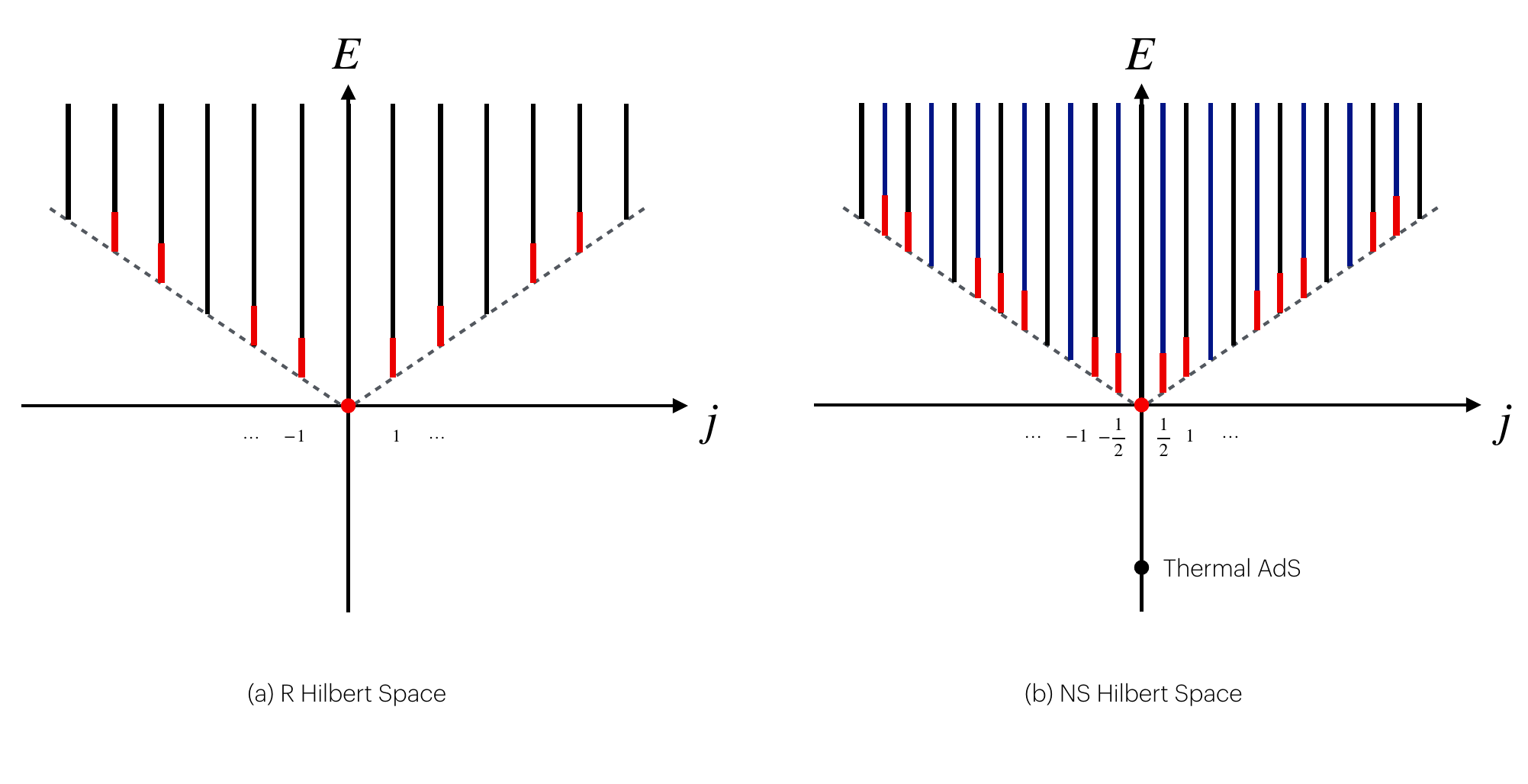}
    \caption{(a) Spectrum of pure fermionic gravity in $\mathcal{H}_\R$ according to the gravitational path integral on a solid torus. The vertical black lines denote black hole states and appear for $E\geq |j|$ and $j\in \mathbb{Z}$. Each comes in two identical copies of bosonic and fermionic states. (b) Same result for $\mathcal{H}_{\NS}$. Bosonic black holes with $E\geq |j|$ and $j\in \mathbb{Z}$ are denoted by the black lines and fermionic $j\in \mathbb{Z}+1/2$ by the blue lines. In (a) and (b) negativities are denoted by the red segments, but all states have a $1/\sqrt{E-|j|}$ edge.}
    \label{fig:spectrum}
\end{figure}

Here we quote the final results and leave a derivation for Appendix \ref{app:spectrum}. A graphical summary of the spectra and its properties in both Hilbert spaces can be found in Figure \ref{fig:spectrum}. The density of states, implied by the gravitational path integral on the solid torus, can be written as
\bea
   \langle \rho_{a}(E,j)\rangle &=&  \sum_{s=1}^\infty \frac{1}{s\sqrt{E^2 - j^2}}\Big\{ S_{a}(j, 0;s) \prod_{\pm}\cosh\Big(\frac{\pi}{s} \sqrt{\frac{(c-1)(E\pm j)}{3}}\Big) \nonumber\\
    && - S_{a}(j, 1; s)\cosh\Big(\frac{\pi}{s} \sqrt{\frac{(c-25)(E+|j|)}{3}}\Big) \cosh\Big(\frac{\pi}{s}\sqrt{\frac{(c-1)(E-|j|)}{3}}\Big) \nonumber\\
    &&- S_{a}(j, -1; s) \cosh\Big(\frac{\pi}{s} \sqrt{\frac{(c-25)(E-|j|)}{3}}\Big) \cosh\Big(\frac{\pi}{s} \sqrt{\frac{(c - 1)(E+|j|)}{3}}\Big) \nonumber\\
    && + S_{a}(j, 0; s) \prod_{\pm}\cosh\Big(\frac{\pi}{s} \sqrt{\frac{(c - 25)(E\pm j)}{3}}\Big) \Big\},\label{eq:rhoFG}
\ea
where $a=\NS$ or $\R$. This expression only holds for $E\geq |j|$ and vanishes otherwise. Other than this continuum the spectrum also has a delta-function at $j=0$ and $E=-c/12$ in the NS Hilbert space, corresponding to the vacuum state. This state is not present in the R Hilbert space. 

\smallskip

The Kloosterman sum appearing in the density of states of the Ramond Hilbert space is given by 
\beq
S_{\R}(j,j';s) = \begin{cases}
    \sum_{\substack{0\leq d<s\\ \gcd(s,d)=1}} \exp\left(2\pi\i \frac{dj+a_{\R}(d,s)j'}{s}\right) ~~~~\text{for $s$ odd}\\
    0 ~~~\text{for $s$ even}
\end{cases}
\eeq
where $a_{\R}(d,s)= (d^{-1})_s$ if $(d^{-1})_s$ is even or $(d^{-1})_s+s$ otherwise. The Kloosterman sum appearing in the NS Hilbert space is
\bea
S_{\NS}(j,j';s) &=& (-1)^{2j+2j'}\sum_{\substack{0\leq d<2s\\ \gcd(s,d)=1\\ d\text{ odd}}} \exp\left(2\pi\i \frac{dj+(d^{-1})_{2s}j'}{s}\right),\\
&=& \sum_{\substack{0\leq d<2s\\ \gcd(s,d)=1\\ d+s\text{ odd}}} \exp\left(2\pi\i \frac{dj+a_{\NS}(d,s)j'}{s}\right)
\ea
where
\begin{align}
    a_{\NS}(d,s) &= \left\{\begin{matrix}
        (d^{-1})_{2s} & \text{ if $s$ is even} \\
        (d^{-1})_{s} &\text{ if $s$ is odd and $(d^{-1})_{s}$ is even} \\
        (d^{-1})_{s} + s &\text{ else}
    \end{matrix}\right.
\end{align}
It is a non-trivial check to verify that the first and second line defining $S_{\NS}$ are equal and it is related to the fact that the path integral with $\NS\pm$ are related by a $T$ transform. The sign $(-1)^{2j'}$ in the first line is irrelevant when $j'$ is an integer, which is the case here.\footnote{This extra factor is important to keep in mind when considering $\mathcal{N}=1$ supergravity, which gets contributions from $j'=\pm 1/2$.} 

\smallskip

Although the density of states in each $\mathcal{H}_\R$ and $\mathcal{H}_\NS$ is different than its bosonic counterpart, the most salient features of the spectrum are very similar:

\smallskip 

For fixed $j$ expression \eqref{eq:rhoFG} naively behaves as $1/\sqrt{\tau}$ for small twist $\tau=E-|j|$. Nevertheless, the most dominant contribution in the large $c$ limit, with $s=1$, satisfies
    $$
 2 S_a ( j,0;1) = S_a(j,-1;1)+S_a(j,1;1)
    $$
    This implies that the sum over hyperbolic cosines vanish linearly with twist and the density of states vanishes as $\sqrt{\tau}$, typical of near-extremal black holes \cite{Ghosh:2019rcj}. 

\smallskip

For a given $j$, contributions with $s>1$ are suppressed by a power of $1/s$ in the exponent but they diverge as $1/\sqrt{\tau}$ as we approach $\tau=0$. The term with smallest $s\neq 1$ therefore becomes dominant at twists that are exponentially small in the central charge \cite{Benjamin:2019stq}. The coefficient in front of $1/\sqrt{\tau}$, for $j>0$, is controlled by the Kloosterman sums 
    $$
 S_a(j,0;s) - S_a(j,j';s)
    $$
    with $j'=1$ or, if this combination vanishes, with $j'= - 1$. The leading contribution comes from the lowest value of $s$ with non-zero $S_a(j,0;s) - S_a(j,1;s)$ or $S_a(j,0;s) - S_a(j,-1;s)$. In the Ramond answer this comes from $s=3$ and the coefficient is negative for $j=1,2~\text{mod}~3$. In the NS answer, for integer $j$, the contribution comes from $s=2$ and is negative for $j=1~\text{mod}~2$. For half-integer spin in the NS Hilbert space, the leading contribution is $s=3$ and the negativity arises at odd $2j = 1, 2~\text{mod}~3$. 

\smallskip

Our perspective on this issue is the following. Pure gravity should be defined not only on geometries that are connected to on-shell configurations but also on off-shell geometries \cite{Cotler:2020ugk,Maxfield:2020ale}. This is believed to explain the fact that the spectrum is continuous possibly as an ensemble over 2d CFT although the precise nature of this ensemble is largely unknown. The near-threshold negativity is not as much of an issue as the fact that the density of states diverges as $1/\sqrt{\tau}$. This is inconsistent with the spectral edge of a chaotic quantum system with the relevant symmetries. More importantly there is no reason to believe that for exponentially small twists, the leading contribution from the solid torus is dominant. In fact, it was proposed in \cite{Maxfield:2020ale} that off-shell geometries with the topology of Seifert manifolds can be dominant at exponentially small twists and that they moreover restore the $\sqrt{\tau}$ edge.\footnote{A different proposal by \cite{Benjamin:2020mfz} involves adding conical deficits to the gravity path integral. One can repeat this for fermionic gravity. It is non-trivial that a new positive term in the $\NS-$ boundary does not spoil positivity in the $\R-$ boundary or vice versa. We thank Jonah Librande for verifying that the negativities can be simultaneously removed by a special choice of spectrum of conical deficits.}

\smallskip

Finally, it is interesting to gauge fermion parity. The torus partition function would be the sum over all sectors
\beq
Z_{\text{gauged $(-1)^{\sf F}$}} = \frac{1}{2}(Z_{\R+}+Z_{\R-}+Z_{\NS+}+Z_{\NS-})=\frac{1}{2}(Z_{\R-}+Z_{\NS+}+Z_{\NS-}),
\eeq
where we used on the RHS that $Z_{\R+}=0$. This gauging automatically projects out fermionic black holes in the NS sector since, as we saw, they contribute with a relative $(-1)^{2j}$ between $Z_{\NS+}$ and $Z_{\NS-}$. It keeps fermionic black holes in the R sector since they have integer spin.  One can prove the identity between the Kloosterman sums 
\beq
2S_{\R}(j,j';s) + S_{\NS}(j,j';s) + (-1)^{2j+2j'}S_{\NS}(j,j';s) = 4S(j,j';s), ~~~~j,j'\in\mathbb{Z}
\eeq
where $S(j,j';s)=\sum_{\substack{0\leq d<s\\ \gcd(s,d)=1}} \exp(2\pi\i \frac{dj+(d^{-1})_{s}j'}{s})$ is the Kloosterman sum appearing in the Maloney-Witten partition function. The factor of $2$ in $S_{\R}$ arises since the partition function gets contributions both from fermionic and bosonic states in $\mathcal{H}_\R$. This immediately implies the following identity
\beq
Z_{\text{gauged $(-1)^{\sf F}$}} (\tau,\bar{\tau}) = \sum_{\gamma \in \Gamma_\infty \setminus\text{PSL}(2,\mathbb{Z})} | \chi_{\text{vac}}(\gamma \tau)|^2.
\eeq
Gauging fermion parity results in the original Maloney-Witten bosonic pure 3d gravity theory. There is a second way to gauge fermion parity with the inclusion of the mod 2 index in the sum over spin structures, analogous to the two types of GSO projections in type-0 string theory. Since all geometries with $\R+$ or an odd number of $\R+$ boundaries vanish, the two gaugings are equivalent.

\subsection{Two-boundary torus wormholes}
\label{subsec:two-boundary-torus-wormholes}

Evaluating path integrals around fully off-shell 3d geometries is largely an open problem in this field. Instead, we consider the simplest off-shell geometry computed so-far, the torus wormhole of Cotler and Jensen \cite{Cotler:2020ugk}. We extend their results to fermionic gravity and in section \ref{sec:discretesymm} we incorporate bulk 3d invertible TQFTs that encode fermionic 2d CFT anomalies.

\smallskip

The Cotler-Jensen wormhole has the topology of $T^2 \times I$ with two 2-torus conformal boundaries. The path integral in this geometry can be separated into two contributions. The first includes the path integral over perturbative quantum gravity (which is a subtle task given the lack of saddle point). The second includes a sum over large diffeomorphisms, the modular group, that transforms the bulk geometry but preserves the conformal boundaries. The information about the spin structure will be relevant when carrying out the second step since the torus are decorated by their spin structure. 

\smallskip

We begin quoting the result for perturbative quantum gravity around a geometry joining two conformal tori with moduli $\tau_1$ and $\tau_2$. At this level, the calculation of \cite{Cotler:2020ugk} is unchanged. The result is\footnote{From now on we omit the dependence on $\bar{\tau}$ in the argument to avoid cluttering, but none of the partition functions appearing in this paper are holomorphic.}
\beq
Z_{0,\text{CJ}}(\tau_1,\tau_2) = \, Z_0(\tau_1) Z_0(\tau_2)\,  \frac{\text{Im}(\tau_1)\text{Im}(\tau_2)}{2\pi^2|\tau_1 + \tau_2|^2},
\eeq
where $Z_0(\tau)= 1/ \sqrt{\text{Im}(\tau)} |\eta(\tau)|^2$ is the modular invariant partition function of a non-compact boson. For the case of bosonic gravity the final answer is a sum of this building block over boundary modular transformations. Since the answer is invariant under $(\tau_1,\tau_2) \to \left(\left(\begin{smallmatrix}
    a & b \\
    s & d
\end{smallmatrix}\right)\cdot \tau_1 , \left(\begin{smallmatrix}
    a & -b \\
    -s & d
\end{smallmatrix}\right) \cdot \tau_2\right)$, for $\left(\begin{smallmatrix}
    a & b \\
    s & d
\end{smallmatrix}\right)$ in $\text{PSL}(2,\mathbb{Z})$, the remaining sum can be written as 
\beq
Z_{\text{CJ}}(\tau_1,\tau_2) = 2 \,  Z_0(\tau_1) Z_0(\tau_2) \sum_{\gamma \in\text{PSL}(2,\mathbb{Z})}\frac{\text{Im}(\tau_1)\text{Im}(\gamma \tau_2)}{2\pi^2|\tau_1 + \gamma \tau_2|^2}
\label{eq:CJwormhole_bosonic}
\eeq
The overall factor of two was identified in \cite{Yan:2023rjh} and we will explain its origin in section \ref{sec:discretesymm}, among other things. 

\smallskip
\begin{figure}
    \centering
    \includegraphics[width=0.5\linewidth]{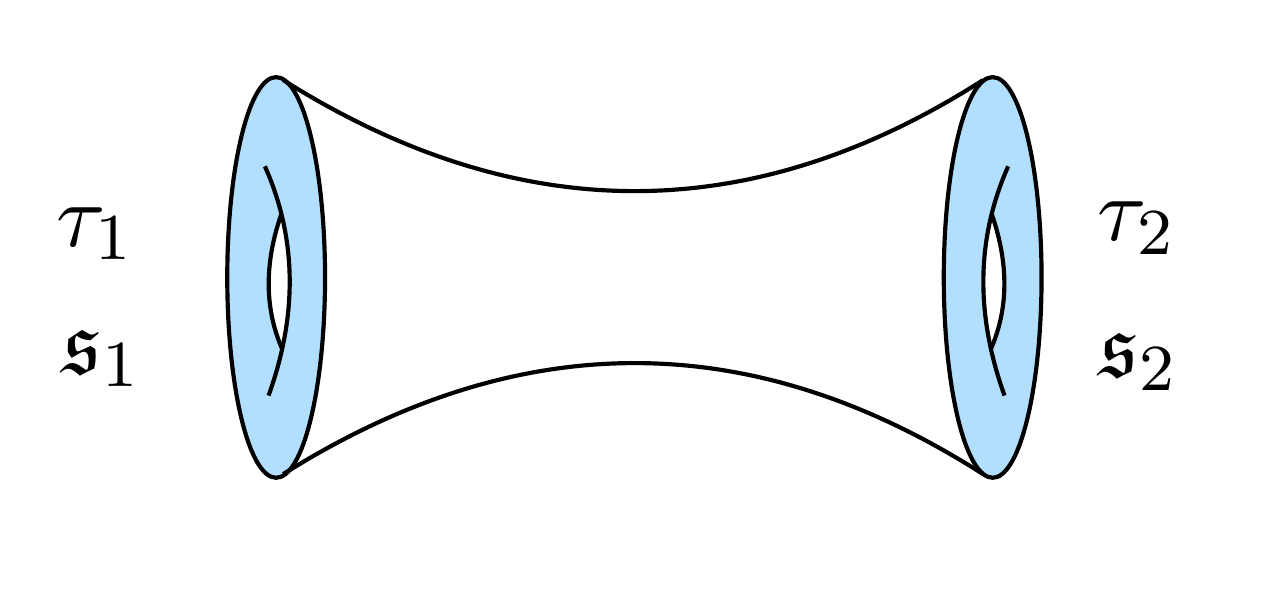}
    \caption{In this section we compute the two-boundary torus wormhole with boundary moduli $(\tau_1,\bar{\tau}_1)$ and $(\tau_2,\bar{\tau}_2)$ as well as spin structure $\mathfrak{s}_1$ and $\mathfrak{s}_2$.}
    \label{fig:CJS}
\end{figure}

The result for fermionic gravity is slightly different. The gravitational path integral now again depends on the boundary spin structure. We first consider the case where both boundaries have the same spin structure $\mathfrak{s}_1=\mathfrak{s}_2=\mathfrak{s}$\footnote{The form of the wormholes studied in this section was independently derived in the upcoming work \cite{Jensen:wip}.}
\beq
Z_{\mathfrak{s}\mathfrak{s}}(\tau_1,\tau_2) = 4 \, Z_0(\tau_1) Z_0(\tau_2) \sum_{\gamma \in G_{\mathfrak{s}\mathfrak{s}}}\frac{\text{Im}(\tau_1)\text{Im}(\gamma \tau_2)}{2\pi^2|\tau_1 + \gamma \tau_2|^2}, \label{eq:NSMNSM}
\eeq
where $Z_{\mathfrak{s}\mathfrak{s}}(\tau_1,\tau_2)$ is the torus wormhole with both boundaries of spin structure $\mathfrak{s}$ which could be $\NS\pm$ or $\R\pm$ (we consider the case $Z_{\mathfrak{s}_1\mathfrak{s}_2}$ with $\mathfrak{s}_1\neq \mathfrak{s}_2$ later in this section). The group of modular images $G_{\mathfrak{s}\mathfrak{s}}$ relevant for each boundary is
\beq
G_{\R+\R+} =\text{PSL}(2,\mathbb{Z}),~~~G_{\R-\R-}=\Gamma_0(2),~~~G_{\NS+\NS+}=\Gamma^0(2),~~~G_{\NS-\NS-} = \Gamma_\theta.
\eeq
The extra factor of $2$ arises from a sum over bulk spin structures and will be explained in section \ref{sec:discretesymm}. The two cycles on the boundary torus are non-contractible in the $T^2 \times I$ geometry and there is no need to correct with an $S$ transformation in the $\R-$ boundary, which was necessary in the case of a solid torus. The mod 2 index of $Y=T^2 \cup T^2$ both with $\R+$ structure is $\zeta=0~\text{mod}~2$ and therefore there is no issue in finding a bulk filling. Similarly any path integral with an even number of $\R+$ boundaries is in principle not vanishing.   

\smallskip

These results are consistent with the modular transformations of the even spin structures, namely $\NS+$, $\NS-$ and $\R-$. These obviously need to hold in any gravity calculation but is good to check. Take first the relation between the two NS boundaries
\bea
Z_{\NS-\NS-}(T \cdot \tau_1,T\cdot \tau_2) &=& Z_0(\tau_1) Z_0(\tau_2) \sum_{\gamma \in\Gamma_\theta}\frac{2}{\pi^2}\frac{\text{Im}(T\cdot\tau_1)\text{Im}(\gamma T\cdot\tau_2)}{|T\cdot\tau_1 + \gamma T\cdot\tau_2|^2}  \nonumber\\
&=&  Z_0(\tau_1) Z_0(\tau_2) \sum_{\gamma \in\Gamma_\theta}\frac{2}{\pi^2}\frac{\text{Im}(\tau_1)\text{Im}(T^{-1}\gamma T\cdot\tau_2)}{|\tau_1 + T^{-1} \gamma T\cdot\tau_2|^2}\nonumber\\
&=& Z_{\NS+\NS+}(\tau_1,\tau_2),
\ea
since we can easily check that $T$ conjugates elements of $\Gamma_\theta$ into $\Gamma^0(2)$ and vice-versa. A very similar derivation relates the $\NS+$ partition function and $\R-$ since $S$ conjugates elements of $\Gamma^0(2)$ into $\Gamma_0(2)$ and vice-versa.

\smallskip

\subsubsection*{NS Hilbert space}

The physical information encoded in these path integrals is the two-point correlation between energy levels of the 2d CFT. For example from the NS sector we can extract 
$$
Z_{\NS-\NS-}(\tau_1,\tau_2) = \sum_{j_1,j_2 \in \frac{1}{2} \cdot \mathbb{Z}} \int \d E_1 \d E_2 \, \langle \rho_{\NS}(E_1,j_1) \rho_{\NS}(E_2,j_2)\rangle \chi_{E_1,j_1}(\tau_1) \chi_{E_2,j_2}(\tau_2)
$$
from a Laplace transform. Since the quantum statistics of the state is distinguished by $j~\text{mod}~\mathbb{Z}$ the two sectors $\NS\pm$ encode the same information once we fix the spin.  The only difference with the calculation in the sector $\NS+$ is the presence of some factors of $(-1)^{2j_1+2j_2}$ on the RHS of the expansion above.

\smallskip

It is instructive to rewrite \eqref{eq:NSMNSM} in the following way. Decompose the sum over $\Gamma_\theta$ into $\Gamma_\infty$ and $\Gamma_\infty \setminus \Gamma_\theta$, such that 
\beq
\sum_{\gamma\in \Gamma_\theta} \frac{\text{Im}(\tau_1)\text{Im}(\gamma \tau_2)}{|\tau_1 + \gamma \tau_2|^2} = \sum_{\gamma\in \Z \setminus \Gamma_\theta} \sum_{n \in \mathbb{Z}} \frac{\text{Im}(\tau_1)\text{Im}(\gamma \tau_2+2n)}{|\tau_1 + \gamma \tau_2+2n|^2},
\eeq
where $\Z \setminus \Gamma_\theta$ has identification $\gamma \sim T^{2n} \gamma$. The sum over $n$ is a sum over integer shifts of the relative angular velocity between the two boundaries. Therefore its Poisson resummation should lead to an expansion in spin. Indeed, for the $\gamma = \text{id}$ contribution
$$
 \sum_{n \in \mathbb{Z}} \frac{\text{Im}(\tau_1)\text{Im}(\tau_2+2n)}{|\tau_1 + \tau_2+2n|^2} = \frac{\pi}{2}\sum_{j \in \frac{1}{2} \mathbb{Z}} \frac{y_1 y_2}{y_1 + y_2} e^{-2\pi |j| (y_1+y_2)} e^{2\pi \i j (x_1 + x_2)} 
$$
where $\tau_1 = x_1 + \i y_1$ and $\tau_2 = x_2 + \i y_2$. One can identify $x_1 = \i \beta_1\Omega_1/2\pi$ and $x_2 = - \i \beta_2 \Omega_2 / 2\pi$, while $y_1 = \beta_1/2\pi$ and $y_2 =\beta_2/ 2\pi$. The contribution to the full answer is
\beq
|\eta(\tau_1)|^2|\eta(\tau_2)|^2\, Z_{\NS-\NS-}(\tau_1,\tau_2) = \sum_{j \in \frac{1}{2} \mathbb{Z}}   \frac{1}{\pi} \frac{\sqrt{\beta_1 \beta_2}}{\beta_1+\beta_2} e^{-(\beta_1 + \beta_2) |j|} e^{2\pi \i j(x_1 + x_2)} + \ldots
\eeq
If one can ignore contributions from $\gamma \neq T^{2n}$, the torus wormhole reproduces the analogous result in random matrix theory for each fixed-spin sector. \cite{Cotler:2020ugk} shows this is the case in the regime of large $\beta_{1,2}$. Relatedly, the spectrum in each fixed-spin sector are statistically independent in this limit. This property is not true in the full answer. Indeed it is reasonable to expect that states of different spins have to be coupled in order to preserve boundary locality.

If we evaluate the spectral form factor with $\beta_1 = \beta+ \i t$ and $\beta_2 = \beta - \i t$ then we obtain in the large $t$ limit
\beq\label{eq:ZNSMNSMDC}
|\eta(\tau_1)|^2|\eta(\tau_2)|^2\, Z_{\NS-\NS-}(\tau_1,\tau_2) = \sum_{j \in \frac{1}{2} \mathbb{Z}}   \frac{t}{2\pi \beta}\, e^{-2\beta |j|} \, e^{2\pi \i j(x_1 + x_2)} + \ldots
\eeq
This is valid for any $\beta$ as long as $t$ is large. One can verify this by evaluating the summand in \eqref{eq:NSMNSM} and verifying that contributions from $\gamma$ with coefficient $s \neq 0$ decay faster, as $1/t^3$. This is the universal ramp behavior characteristic of level repulsion which can be obtained via the double cone \cite{Saad:2018bqo, Chen:2023hra}. The prefactor is consistent with the GOE/GSE ensemble. We will discuss this in great detail in section \ref{sec:discretesymm}. The same result holds for the $\NS+$ wormhole indicating that in this limit the bosonic and fermionic sectors are statistically independent. This is not true beyond this limit since $Z_{\NS-\NS-}\neq Z_{\NS+\NS+}$ in general.

\subsubsection*{R Hilbert space}

The Ramond sector is special since the statistics of the states are not distinguished by angular momentum. Therefore the $\R+$ and $\R-$ partition functions do encode non-trivial independent information. 
\bea
Z_{\R\pm\R\pm} &=& \sum_{j_1,j_2\in\mathbb{Z}} \int \d E_1 \d E_2 \, \langle (\rho_{\R}^{\text{bos}}(E_1,j_1)\mp \rho_{\R}^{\text{fer}}(E_1,j_1))(\rho_{\R}^{\text{bos}}(E_2,j_2) \mp \rho_{\R}^{\text{fer}}(E_2,j_2) )\rangle \nonumber\\
&&\chi_{E_1,j_1}(\tau_1) \chi_{E_2,j_2}(\tau_2).
\ea
In RMT the bosonic and fermionic microstates would be statistically independent. This is not true any longer for 2d CFTs. Indeed we find
\beq
Z_{\R+\R+}\neq Z_{\R-\R-} ,~~~\Rightarrow~~~\left\langle \rho_{\R}^{\text{bos}}(E_1,j_1) \rho^{\text{fer}}_{\R}(E_2,j_2)\right\rangle \neq 0.
\eeq
Nevertheless in the case of the R sector we argued earlier that the path integral with an odd number of $\R+$ boundaries identically vanishes. This means that, although the bosonic and fermionic sectors are correlated, their joint probability distribution is symmetric between the two sectors. In other words, if we denote by $P(H_{\R, b}, H_{\R, f})$ the probability of having Hamiltonians $H_{\R, b}$ and $H_{\R, f}$ for the bosonic/fermionic states of spin $j$ then
$$
P(H_{\R, b}, H_{\R, f}) = P(H_{\R, f}, H_{\R, b}).
$$
In the case of 2d gravity the probability distribution is not only symmetric but it factorizes. That is not the case in 3d gravity.

\smallskip

Finally it is instructive to take the double-cone limit of the Ramond wormhole. The only modification compared to the NS sector is the relation
$$
 \sum_{n \in \mathbb{Z}} \frac{\text{Im}(\tau_1)\text{Im}(\tau_2+n)}{|\tau_1 + \tau_2+n|^2} =\pi \sum_{j \in  \mathbb{Z}} \frac{y_1 y_2}{y_1 + y_2} e^{-2\pi |j| (y_1+y_2)} e^{2\pi \i j (x_1 + x_2)} 
$$
leading to an extra overall factor of two 
\beq \label{eq:R poisson}
|\eta(\tau_1)|^2|\eta(\tau_2)|^2\, Z_{\R\pm\R\pm}(\tau_1,\tau_2) = \sum_{j \in  \mathbb{Z}}   2 \,\frac{t}{2\pi \beta}\, e^{-2\beta |j|} e^{2\pi \i j(x_1 + x_2)} + \ldots
\eeq
This factor arises because states of all angular momenta come in boson/fermion degenerate pairs. Each sector of fixed spin and statistics is described by a GOE/GSE ensemble, more in section \ref{sec:discretesymm}. Since the bosonic and fermionic sectors are statistically independent in the double-cone limit (but not beyond) the answer is the same for $\R+$ or $\R-$.

\subsubsection*{Mixed Hilbert spaces}

It is also interesting to consider cross-correlations between different Hilbert spaces. This can be easily extracted from wormholes connecting two boundary torus of different spin structure. An immediate observation we can make is that 
$$
Z_{\R+\R-}=Z_{\R+\NS+}=Z_{\R+\NS-}=0
$$
identically. This arises from the fact that the mod 2 index of the boundary is non-trivial. This implies that the joint probability distribution on all sectors NS/R is symmetric under exchange of bosonic/fermionic R-sector Hamiltonians.

\smallskip

The rest of the mixed-boundary wormholes are non-trivial. For example consider one boundary being $\NS+$ and the other $\R-$. A torus wormhole exists as long as the meaning of space vs time cycle is exchanged between the two boundaries, implemented by a relative S transformation
\bea
Z_{\R-\NS+}(\tau_1,\tau_2) &=& Z_0(\tau_1) Z_0(\tau_2) \sum_{\gamma \in\Gamma^0(2)}\frac{2}{\pi^2} \frac{\text{Im}(S\tau_1)\text{Im}(\gamma \tau_2)}{|S\tau_1 + \gamma \tau_2|^2},\nonumber\\
&=&Z_0(\tau_1) Z_0(\tau_2) \sum_{\gamma \in\Gamma_0(2)}\frac{2}{\pi^2} \frac{\text{Im}(\gamma\tau_1)\text{Im}(S^{-1} \tau_2)}{|\gamma\tau_1 + S^{-1}\tau_2|^2}\label{eq:ZRMNSP}
\ea
In the second line we used the fact that $S^{-1}\,\Gamma^0(2)\, S = \Gamma_0(2)$ and redefined $\gamma$. This shows that it does not matter which side of the wormhole the S move is applied to. Similarly one can show that, when $Z_{\mathfrak{s}_1\mathfrak{s}_2}(\tau_1,\tau_2)$ is non-zero it is given by
\beq \label{eq:mixed wormholes}
Z_{\mathfrak{s}_1 \mathfrak{s}_2}(\tau_1,\tau_2) = Z_0(\tau_1) Z_0(\tau_2) \sum_{\gamma \in G_{\mathfrak{s}_1 \mathfrak{s}_2}}\frac{2}{\pi^2} \frac{\text{Im}(\tau_1)\text{Im}(\gamma \tau_2)}{|\tau_1 + \gamma \tau_2|^2}
\eeq
where the sets $G_{\mathfrak{s}_1 \mathfrak{s}_2}$ are given by 
\beq
G_{\R-\NS+} =  S^{-1}\,\,\Gamma^0(2),~~~~G_{\NS-\NS+}=T^{-1}\,\,\Gamma^0(2),~~~~~G_{\NS-\R-} = T^{-1} S^{-1}\,\,\Gamma_0(2).
\eeq
Exchanging the order of the boundary spin structure does not change the answer. Similar to the second line of \eqref{eq:ZRMNSP} one has to use that $T^{-1}\, \Gamma^0(2) T = \Gamma_\theta$  and $T^{-1} S^{-1}\, \Gamma_0(2)\,  ST = \Gamma_\theta$. 

\smallskip

Two of these amplitudes measure correlations between the spectrum in different sectors, namely 
\beq
\left\langle \rho_{\NS} (E_1,j_1) \rho^{\text{bos}}_{\R}(E_2,j_2) \right\rangle~~~\text{and}~~~\left\langle \rho_{\NS} (E_1,j_1) \rho^{\text{fer}}_{\R}(E_2,j_2) \right\rangle.
\eeq
These are non-zero and given by Laplace transforms of the path integrals above. We can verify that in the double-cone limit, the sectors are approximately statistically independent. Indeed one can show that there is no element in $\gamma \in S^{-1}\Gamma^0(2) $ or $\gamma \in T^{-1} S^{-1}\Gamma_\theta $ with vanishing $s=0$ and, as we explained earlier, the contributions to the double-cone limit from $s\neq0$ are suppressed. For example,
$$
\gamma= S^{-1} \cdot \begin{pmatrix}
    a' & b' \\
    s' & d' 
\end{pmatrix}
 = \begin{pmatrix}
    s' & d' \\
    -a' & -b' 
\end{pmatrix},~~~~\gamma= T^{-1} S^{-1}\cdot\begin{pmatrix}
    a' & b' \\
    s' & d' 
\end{pmatrix}
= \begin{pmatrix}
    a'+s' & b'+d' \\
    -a' & -b' 
\end{pmatrix}
$$
In the left equation, for $s=-a'$ to vanish we need $a'=0$, which is incompatible with $a'd'-b's'=1$ and $b'$ being even in $\Gamma^0(2)$. In the right equation, $s=-a'$ cannot vanish since $a'$ is odd in $\Gamma_0(2)$. The double-cone limit of these correlators therefore vanish.

\smallskip

There are elements in the sum over $\gamma \in \Gamma^0(2) T$ with $s=0$. One can show that they lead to a double-cone limit 
\beq
|\eta(\tau_1)|^2|\eta(\tau_2)|^2\, Z_{\NS-\NS+}(\tau_1,\tau_2) = \sum_{j \in \frac{1}{2} \mathbb{Z}}   \frac{t}{2\pi \beta}\,(-1)^{2j}\, e^{-2\beta |j|} \, e^{2\pi \i j(x_1 + x_2)} + \ldots
\eeq
which is compatible with \eqref{eq:ZNSMNSMDC}. In fact since $Z_{\NS-\NS+}(\tau_1,\tau_2) = Z_{\NS-\NS-}(\tau_1,\tau_2+1)$ and given the spin-statistics connection in the NS Hilbert space, there is no new information to be extracted from $Z_{\NS-\NS+}$.

\section{Holographic 2d CFTs with fermions and RMT${}_2$} \label{sec:rmt2}

In this section, we discuss the boundary perspective on the putative fermionic CFTs holographically dual to fermionic pure 3d gravity defined in section \ref{sec:Path_integral_fermionic_3d_gravity}. As in the case of holographic 2d CFTs dual to bosonic pure 3d gravity, there is no known explicit example of such a theory. Nevertheless, due to combined Virasoro and modular symmetries on a torus, it is possible to learn non-trivial information about the spectrum of such CFTs based on minimal assumptions which we now discuss. 

To study holographic CFTs which admit a semiclassical dual gravitational description, we will assume an irrational CFT with large value of the central charge $c$. The spectrum in the $\NS$ sector will consist of bosonic states of integer spin and fermionic states of half-integer spin $j$, and will be assumed to contain an energy gap\footnote{For more generic holographic CFTs it would suffice to assume a sparse spectrum between vacuum and black hole threshold. For our purposes we assume exact energy gap as we are interested in the special case of pure 3d gravity.} between the vacuum state $h = \overline{h} = 0$ and a dense number of states beginning at the black hole threshold $h= \frac{c-1}{24}$, $\overline{h}=\frac{c-1}{24}$. In contrast, the spectrum in the $\R$ sector will consist of bosonic and fermionic states of integer spin\footnote{Recall that, as explained around equation \eqref{eq:relbtwpff}, in the Ramond Hilbert space the usual spin-statistics relation does not hold and correspondingly all states, bosonic and fermionic, have integer quantized angular momentum.}, and will be assumed to be supported only at and above the black hole threshold. 
As will be clear from section \ref{subsec:integrability}, these assumptions suffice to reproduce the coarse-grained densities of states discussed in section \ref{subsec: solid torus}.

As a last part of holographic input, we will assume that in the near-extremal limit, defined as $\beta \sim c$, $c \to \infty$, the spectral statistics of dense spectrum located near the black hole threshold are well-approximated by random matrix statistics. This assumption is motivated by the emergence of JT gravity dynamics, which presents a similar near-threshold behavior, in the near-horizon region of a near-extremal rotating BTZ black hole in AdS$_3$ \cite{Ghosh:2019rcj,Maxfield:2020ale}. As discussed in \cite{Haehl:2023tkr,Haehl:2023xys,Haehl:2023mhf,DiUbaldo:2023qli} for bosonic CFTs, this assumption together with modular symmetry suffices to reproduce the two boundary torus wormhole \eqref{eq:CJwormhole_bosonic} directly from the field theory side. In section \ref{subsec:RMT2_two_boundary_wormhole} we will show that a similar story holds in the case of fermionic CFTs, for any choice of boundary spin structure \eqref{eq:R+_partition_function}-\eqref{eq:NS-_partition_function}, and allows us to reproduce exactly the two-boundary wormholes given in \eqref{eq:NSMNSM}.

\subsection{Torus partition function and square-integrability}
\label{subsec:integrability}

Our starting point is to consider an exact partition function $Z^{\mathfrak{s}}(\tau)$, invariant under modular group $\Gamma_{\mathfrak{s}}$, of a putative CFT dual to pure fermionic 3d gravity subject to assumptions described above. To remove descendant states we will focus on the primary partition functions, defined as\footnote{Similarly to previous sections, for brevity we suppress the explicit $\overline{\tau}$ dependence and denote $Z_{p}^\mathfrak{s}(\tau ,\overline{\tau}) \equiv Z_{p}^\mathfrak{s}(\tau )$. We emphasize that $Z_{p}^\mathfrak{s}(\tau )$ is still a non-holomorphic function.}
\be 
Z_{p}^\mathfrak{s}(\tau) = \sqrt{y} |\eta(\tau)|^2 Z_{\mathfrak{s}}(\tau)
,
\ee
for any choice of spin structure $\mathfrak{s}$. This function is not holomorphic but we omit the $\bar{\tau}$ dependence to avoid cluttering. Here, multiplication by $|\eta(\tau)|^2$ cancels contributions of all of the descendants to the partition function, and the additional factor of $\sqrt{y}$ is included to preserve modular invariance of $Z_{p}^\mathfrak{s}(\tau)$. 
Because we are interested in studying consequences of modular symmetry on spectral correlations, we wish to reorganize the information about the CFT spectrum in a fully modular invariant manner. This can be done conveniently \cite{Benjamin:2021ygh,DiUbaldo:2023qli} by treating $Z_{p}^\mathfrak{s}(\tau)$ as a function on a fundamental domain of $\Gamma_{\mathfrak{s}}$, $\Gamma_{\mathfrak{s}} \backslash \mathbb{H}$, where $\mathbb{H}$ denotes the upper-half plane with coordinates $\tau = x+\i y$, and then decomposing it in terms of harmonic basis of eigenfunctions of hyperbolic laplacian $\Delta_\tau = -y^2(\partial_x^2 + \partial_y^2)$ acting on the space of square-integrable functions $L^2(\Gamma_{\mathfrak{s}} \backslash \mathbb{H})$. 

\begin{figure}
    \centering
    \includegraphics[width=0.8\linewidth]{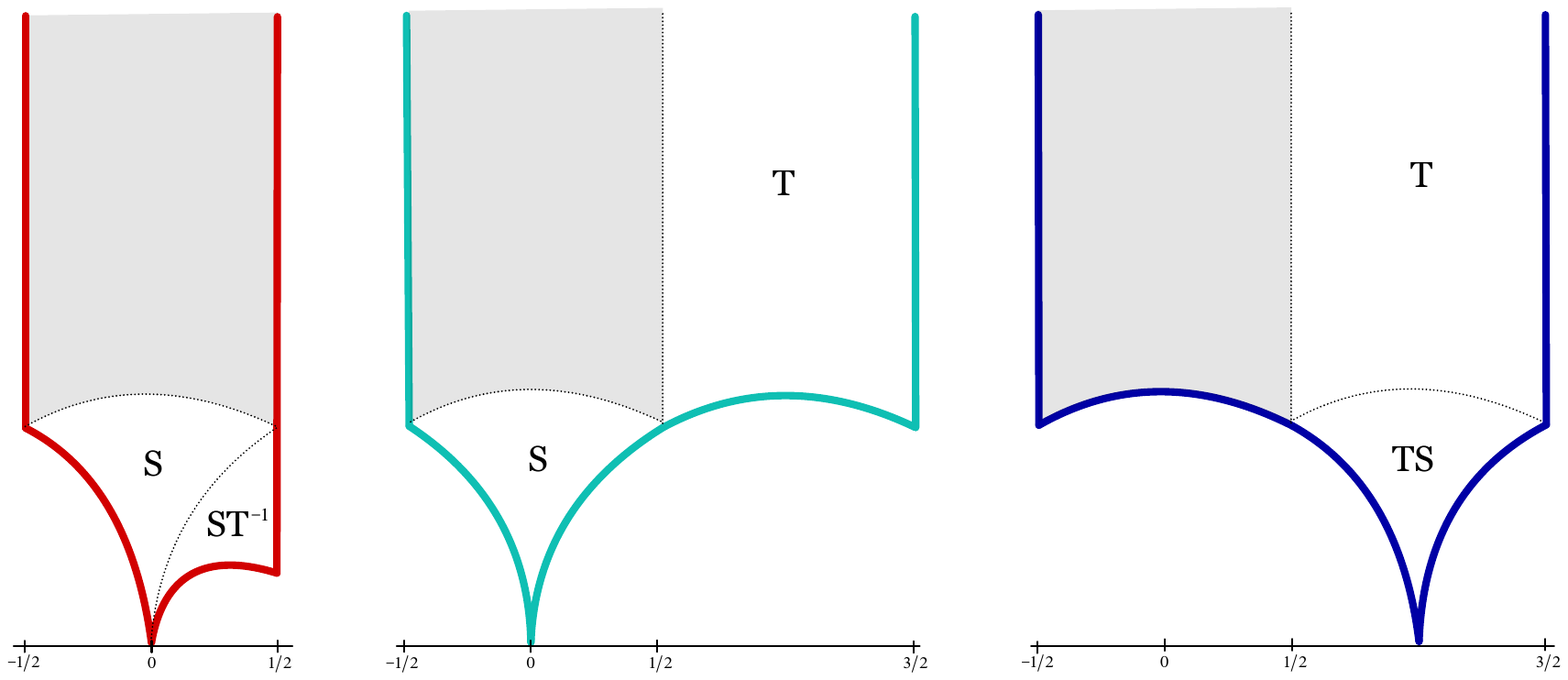}
    \vspace{-3.5cm}
    \caption{Fundamental domains $\Gamma_{\mathfrak{s}}\backslash \mathbb{H}$ considered in this section, contrasted with the fundamental domain $SL(2,\mathbb{Z}) \backslash \mathbb{H}$ (shaded region). \textbf{Left:} Fundamental domain of $\Gamma_0 (2)$ relevant for $Z_{\R -}(\tau)$ with cusps at 0 and $\infty$. \textbf{Middle:} Fundamental domain of $\Gamma^0 (2)$ relevant for $Z_{\NS +}(\tau)$ with cusps at 0 and $\infty$. \textbf{Right:} Fundamental domain of $\Gamma_\theta$ relevant for $Z_{\NS -}(\tau)$ with cusps at 1 and $\infty$.}
    \label{fig:domains}
\end{figure}

As written, however, the function  $Z_{p}^\mathfrak{s}(\tau)$ is not expected to generically be square-integrable. This can be seen clearly already in the bosonic case \cite{Benjamin:2021ygh}, where contributions of states coming from below the black hole threshold -- in pure gravity case, the vacuum -- lead to exponential growth $e^{\frac{c-1}{6}\pi y}$ of the primary partition function at the cusp $y \to \infty$. In that case, to construct a square-integrable part of the partition function, one subtracts then the contribution of the states below the black hole threshold in a modular invariant manner by performing a Poincaré sum. This then allows one to recognize the subtracted ``non-square-integrable" part of the partition function with the gravitational partition function on the solid torus \cite{Maloney:2007ud}, whilst the remaining square-integrable part can be used to analyze chaotic properties of the spectrum \cite{DiUbaldo:2023qli}.  

Following the same logic as in the bosonic case, we will now analyze the square-integrability properties of partition functions $Z_{p}^\mathfrak{s}(\tau)$ on their respective fundamental domains $\Gamma_{\mathfrak{s}} \backslash \mathbb{H}$. Apart from the $\R+$ sector, the fundamental domains contain an additional cusp, in contrast to the bosonic case, where integrability will have to be analyzed separately. Nevertheless, we will find that the identification of the non-square-integrable part of the partition function as the gravitational partition function on the solid torus persists for every choice of spin structure.


\paragraph{(NS$+$, NS$-$, R$-$):}
To begin, we will focus on three spin structures NS$+$, NS$-$, R$-$ connected through $\text{SL} (2,\mathbb{Z})$ transformations as in \eqref{eq:relbtwpff}. These relations will slightly simplify the integrability analysis as different cusps will be easier to analyze in different sectors. 

Starting with the cusp at $\tau=0$ in the NS$+$ sector, to verify square-integrability near the cusp it is useful to perform an $\text{SL}(2,\mathbb{R})$ reparametrization of the fundamental domain such that the zero cusp is located at $y \to \infty$. Because the Petersson measure $\dd x \dd y /y^2$ is invariant under these transformations we can analyze integrability with respect to the same measure. To implement the $\text{SL}(2,\mathbb{R})$ transformations, it is typical to use the so-called scaling matrix $\sigma_0 $ \cite{Iwaniec2002SpectralMO}, defined such that
\be 
\sigma_{0,\NS+} \cdot \infty = 0 , \qquad \sigma_{0,\NS+}^{-1} \Gamma_{0,\NS+} \sigma_{0,\NS+} = \Gamma_{\infty,\NS+}, 
\ee
where $\Gamma_{0,\NS+} \in \Gamma^0 (2)$ denotes the stabilizer of the 0 cusp, and $\Gamma_{\infty,\NS+} \in \Gamma^0 (2)$ denotes the stabilizer of the cusp at $\infty$.\footnote{Note that stabilizer groups at $\infty$ for different spin structures $\mathfrak{s}$ are denoted in this section as $\Gamma_{\infty,\mathfrak{s}}$. In section \ref{sec:Path_integral_fermionic_3d_gravity} these groups appeared collectively as a $\mathbb{Z}$ modding of the modular group.}
In particular, for the NS+ partition function the scaling matrix can be chosen to be exactly the $S$ generator
\be 
\sigma_{0,\NS+} = 
\begin{pmatrix}
0 & -1 \\ 1 & 0 
\end{pmatrix}  
= S 
.
\ee
Note that $S$ is not part of the relevant modular group $\Gamma^0 (2)$ for NS+ but, nevertheless, one can use it to conveniently reparametrize the fundamental domain.

After transformation through a scaling matrix, we are therefore interested in the behavior of $Z^{\NS +}_p(\sigma_0 \tau)$ at $y \to \infty$. Because in this case the scaling matrix is the $S$ generator, we can directly find this behavior by looking instead at the R$-$ partition function at $\infty$  
\be 
Z^{\NS +}_p(\sigma_0 \tau) |_{y \to \infty} = Z_p^{\R -} (\tau) |_{y \to \infty} 
\sim \sqrt{y} \rho_{\text{threshold}}^{\R} 
,
\label{eq:slow_growth_NS+}
\ee
where $\rho_{\text{threshold}}^\R$ denotes the number of degenerate primary states exactly at the black hole threshold $h_p^\R+\overline{h}^\R_p = \frac{c-1}{12}$ in the R Hilbert space. This establishes that the $\NS +$ partition function is of slow growth at the 0 cusp, and therefore close to 0 belongs to the class of functions amenable to harmonic analysis on the fundamental domain \cite{zagier1981rankin}\footnote{See also \cite{Benjamin:2021ygh} for a discussion of harmonic analysis of functions of slow-growth near the cusp.}. 
At the same time, equation \eqref{eq:slow_growth_NS+} also shows that the R$-$ partition function has integrable behavior close to the cusp at $\infty$.

Next, we need to analyze the behaviors of both $\NS+$ and $\R-$ partition functions at their opposite cusps, $\infty$ and 0 respectively. The $\NS+$ partition function blows up exponentially near $\infty$ due to contributions of states below black hole threshold $h_p + \overline{h}_p \leq \frac{c-1}{12}$, 
\be 
Z^{\NS +}_p(\tau)|_{y \to \infty} 
\sim e^{2\pi y \left( \frac{c-1}{12} - h_p - \overline{h}_p \right)}  ,
\ee
and therefore their contribution needs to be subtracted to construct an integrable partition function. 
To remove the exponential behavior it is sufficient to remove the contributions from the states below the black hole threshold
\be 
Z_L^{\NS +} (\tau) = \sqrt{y} |\eta(\tau)|^2 \sum_{h_p , \overline{h}_p \leq \frac{c-1}{12}}  \chi_{h_p} (\tau )\overline{\chi}_{\overline{h}_p}  (\overline{\tau}) 
.
\ee
This, however, would lead to a decomposition that is not modular invariant under $\Gamma^0 (2)$. Modular invariant decomposition can be easily constructed from $Z_L^{\NS +} (\tau)$ by summing over modular images 
\be 
\hat{Z}_L^{\NS+} (\tau) \equiv \sum_{\gamma \in \Gamma_{\infty,\NS +} \backslash \Gamma^{0}(2)} 
Z_L^{\NS +} ( \gamma \tau) 
,
\ee
where we mod out by a stabilizer group of the cusp at $\infty$ under which individual characters themselves are invariant. 
With this, we can now define
\be 
Z^{\NS +}_{\text{spec}}(\tau)
\equiv  
Z^{\NS +}_{p}(\tau) - \hat{Z}_L^{\NS+} (\tau)  ,
\ee
which is modular invariant under $ \Gamma^0 (2)$ and square-integrable (of slow-growth) at the 0 and $\infty$ cusps on the fundamental domain $\Gamma^0 (2) \backslash \mathbb{H}$. As already mentioned at the beginning of the subsection, in the theory of pure gravity we assume that the only primary state below the black hole threshold is the vacuum, and therefore we find 
\be 
\hat{Z}_L^{\NS+} (\tau) =
\sqrt{y} |\eta(\tau)|^2
\sum_{\gamma \in \Gamma_{\infty,\NS +} \backslash \Gamma^{0}(2)} 
\chi_{\text{vac}} (\tau) \overline{\chi}_{\text{vac}}(\overline{\tau}) ,
\ee
which manifestly matches the gravitational partition function on the solid torus in \eqref{ZNSP}.

We now turn to 0 cusp of the $\R -$ partition function.
We first notice that $\R-$ partition function is not integrable at the 0 cusp for the same reason as above: the scaling matrix of $\Gamma_0(2)$ at the 0 cusp is given by 
\be 
\sigma_{0,\R -} = \begin{pmatrix}
0 & -\frac{1}{\sqrt{2}} \\
\sqrt{2} & 0 
\end{pmatrix} 
,
\ee
and correspondingly, the non-integrable behavior near the 0 cusp is captured in terms of $\NS+$ spectrum as (here denoted as $(h_p^{\NS+},\overline{h}_p^{\NS+})$ to keep track of the distinction)
\begin{align}
Z^{\R -}( \sigma_{0,\R -} \tau) |_{y \to \infty} &= Z^{\NS +} (S  \sigma_{0,\R -} \tau) |_{y \to \infty} 
\nonumber
\\
&= Z^{\NS +} (2\tau) |_{y \to \infty} 
\sim e^{4\pi y\left( 
\frac{c-1}{12}-h_p^{\NS+} - \overline{h}_p^{\NS+}
\right)} .
\end{align}
The square-integrable part of $\R -$ can be then obtained by relying on relation to $\NS +$ 
\be 
Z^{\R -}_{\text{spec}} (\tau)  \equiv 
Z^{\NS +}_{\text{spec}} (S \tau) ,
\ee
which can be naturally rewritten as 
\begin{align} 
Z^{\R -}_{\text{spec}} (\tau)  &= 
Z^{\R -}_p (\tau) 
- \sum_{\gamma \in \Gamma_{\infty,\NS +} \backslash \Gamma^{0}(2)} 
Z^{\NS +}_L (\gamma S \tau)
\\
&= 
Z^{\R -}_p (\tau) 
- \sum_{\gamma \in \Gamma_{0,\R -} \backslash \Gamma_0 (2)} 
Z^{\NS +}_L (S \gamma \tau) 
.
\end{align}
In the above, we used the isomorphism
\be
\Gamma_{\infty,\NS +} \backslash \Gamma^{0}(2) \cdot S \simeq 
S \cdot \Gamma_{0,\R -} \backslash \Gamma_0 (2)
,
\ee
where 
\be 
\Gamma_{\infty,\NS +} =  \left\{ \begin{pmatrix}
1 & 2k \\ 0 & 1 
\end{pmatrix} : k \in \mathbb{Z}
\right\} 
,
\qquad 
\Gamma_{0,\R -} =  \left\{ \begin{pmatrix}
1 & 0 \\ 2k & 1 
\end{pmatrix} : k \in \mathbb{Z}
\right\}
.
\ee
As before, we can again recognize the non-square-integrable part subtracted as the gravitational partition function on the solid torus. This follows simply from recognizing that 
\be 
\hat{Z}_L^{\R -} (\tau) \equiv 
\sum_{\gamma \in \Gamma_{0,\R -} \backslash \Gamma_0 (2)} 
Z^{\NS +}_L (S \gamma \tau) 
= \sqrt{y} |\eta(\tau)|^2 
\sum_{\gamma \in \Gamma_{0,\R -} \backslash \Gamma_0 (2)} 
\chi_{\text{BTZ}} (\gamma \tau) 
\overline{\chi}_{\text{BTZ}} (\gamma \overline{\tau}) , 
\ee
which manifestly matches \eqref{ZRM}. Here we introduced a convenient notation for the $S$-transformed vacuum character $\chi_{\text{BTZ}} (\tau) \equiv \chi_{\text{vac}} (S \tau)$.

The discussion for the $\NS -$ partition function follows analogously. We introduce the square-integrable part as
\be 
Z^{\NS -}_{\text{spec}} (\tau) 
\equiv Z^{\NS +}_{\text{spec}} (T\tau) .
\ee
This can be explicitly written in the form
\begin{align}
Z^{\NS -}_{\text{spec}} (\tau) &= 
Z^{\NS -}_p (\tau) 
- \sum_{\gamma \in \Gamma_{\infty,\NS+} \backslash \Gamma^0 (2)} 
Z^{\NS +}_L (\gamma T \tau)  
\\ 
&= 
Z^{\NS -}_p (\tau) 
- \sum_{\gamma \in\Gamma_{\infty,\NS-} \backslash \Gamma_{\theta}} 
Z^{\NS +}_L (T \gamma \tau)  ,
\end{align}
after relying on the isomorphism
\be 
T^{-1} \cdot  \Gamma_{\infty,\NS+} \backslash \Gamma^0 (2) \cdot 
T 
\simeq 
\Gamma_{\infty,\NS-} \backslash \Gamma_{\theta} ,
\qquad
\Gamma_{\infty ,\NS+} = \Gamma_{\infty, \NS-} .
\ee
The subtracted part can now be recognized as \eqref{ZNSM}
\be 
\hat{Z}^{\NS -}_L
\equiv
\sum_{\gamma \in\Gamma_{\infty,\NS-} \backslash \Gamma_{\theta}} 
Z^{\NS +}_L (T \gamma \tau) 
= 
\sqrt{y} |\eta(\tau)|^2
\sum_{\gamma \in\Gamma_{\infty,\NS-} \backslash \Gamma_{\theta}} \chi_{\text{vac}}(\gamma \tau) 
\overline{\chi}_{\text{vac}}(\gamma \overline{\tau}) 
,
\ee
where we used that the vacuum character is invariant under $T$ transformations (note that this still leads to a different partition functions $\NS +$ and $\NS -$ sectors due to different modular groups appearing in the sum). 

\paragraph{(R$+$):} Lastly, we discuss the square-integrable part of R$+$ partition function on $SL(2,\mathbb{Z})\backslash \mathbb{H}$. For holographic theories, we expect there to be no states in the Ramond Hilbert space below the black hole threshold. As such, the partition function is readily integrable at the $\infty$ cusp. On the other hand, the fundamental domain of $SL(2,\mathbb{Z})$ does not contain the cusp at $0$, and therefore we simply have
\be 
Z^{\R +}_{\text{spec}} \equiv Z_p^{\R +}(\tau) . 
\ee
The non-integrable part of the partition function in this case is exactly vanishing,
\be 
\hat{Z}_L^{\R +} (\tau) = 0 ,
\ee
which again matches the gravitational partition function on the solid torus discussed in \eqref{ZRP}.

\subsection{Spectral decomposition for congruence subgroups}

The upshot of the previous subsection is that, for each of the partition functions $Z_{p}^\mathfrak{s}(\tau)$, we have constructed the square-integrable parts of the partition functions $Z_{\text{spec}}^\mathfrak{s}(\tau) \in L^2 (\Gamma_{\mathfrak{s}} \backslash \mathbb{H})$ on their respective fundamental domains $\Gamma_{\mathfrak{s}} \backslash \mathbb{H}$. With this, we are now able to reorganize physical information contained in $Z_{\text{spec}}^\mathfrak{s}(\tau)$ in a modular invariant manner. We now proceed to review some general aspects of the spectral decomposition of square-integrable functions in terms of the harmonic eigenbasis of the hyperbolic Laplacian on each fundamental domain. 
In the next subsection, this will allow us to fully use the power of modular invariance to constrain spectral statistics of holographic CFTs.

The following discussion follows \cite{Iwaniec2002SpectralMO}. 
The space of square-integrable functions $L^2 (\Gamma_{\mathfrak{s}} \backslash \mathbb{H})$ on the fundamental domain of $\Gamma_{\mathfrak{s}}$, equipped with the inner product 
\be 
( f,g )= \int_{\Gamma_{\mathfrak{s}} \backslash \mathbb{H}} \frac{\dd x \dd y}{y^2} f(\tau) \overline{g}(\tau) ,  
\ee
admits a decomposition 
\be 
L^2 (\Gamma_{\mathfrak{s}} \backslash \mathbb{H}) 
= \widetilde{\mathcal{C}} (\Gamma_{\mathfrak{s}} \backslash \mathbb{H}) \oplus \widetilde{\mathcal{E}} (\Gamma_{\mathfrak{s}} \setminus \mathbb{H}) . 
\ee 
Here, $\widetilde{\mathcal{E}} (\Gamma_{\mathfrak{s}} \backslash \mathbb{H})$ denotes the closure of a space spanned by the wavepackets formed out of continuous families of Eisensteins $E_{\mathfrak{a},s}^{\mathfrak{s}} (\tau)$ associated with each cusp (as shown in figure \ref{fig:domains}, for some of the partition functions there will be more than one cusp).
$\widetilde{\mathcal{C}} (\Gamma_{\mathfrak{s}} \backslash \mathbb{H})$ denotes the closure of the space spanned by a discrete family of cusp forms $\phi_{n} (\tau)$. Both the Eisensteins and the cusp forms are eigenfunctions of the hyperbolic Laplacian 
\begin{align}
\Delta = -y^2 (\partial_x^2 + \partial_y^2) ,
\end{align}
satisfying 
\begin{align}
\Delta E_{\mathfrak{a},s}^{\mathfrak{s}} (\tau ) &= s(1-s) E_{\mathfrak{a},s}^{\mathfrak{s}} (\tau ) , 
\qquad s= \frac{1}{2} + \i \omega ,  \qquad \omega \in \mathbb{R} ,
\\ 
\Delta \phi_n^{\mathfrak{s}} (\tau) &= s_n(1-s_n) \phi_n^{\mathfrak{s}} (\tau) , 
\qquad  s_n= \frac{1}{2} + \i r_n , \,\,\,  r_n \in \mathbb{R} , \,\,\, n\in \mathbb{N} .
\end{align}
Denoting as $\Gamma_{\mathfrak{s},\mathfrak{a}} \in \Gamma_{\mathfrak{s}}$ the stabilizer group of the cusp $\mathfrak{a}$,
the Eisenstein series associated to cusp $\mathfrak{a}$ can be defined for $\Re(s)>1$ as 
\be 
E^\mathfrak{s}_{\mathfrak{a}} (\tau ,s) = \sum_{\gamma \in \Gamma_{\mathfrak{s},\mathfrak{a}} \setminus  \Gamma_{\mathfrak{s}}} \Im (\sigma_{\mathfrak{s},\mathfrak{a}}^{-1} \gamma \tau)^s  ,
\ee
where $\sigma_{\mathfrak{s},\mathfrak{a}}$ denotes the scaling matrix of the cusp $\mathfrak{a}$.
The space of cusp forms $\phi^{\mathfrak{s}} (\tau) \in \mathcal{C} (\Gamma_{\mathfrak{s}} \backslash \mathbb{H})$ is defined by imposing that their zero mode at each cusp $\a$ vanishes, 
\be 
\int_{0}^1 \dd x \, \phi^{\mathfrak{s}} (\sigma_{\mathfrak{s},\a} \tau) = 0 
\qquad \Leftrightarrow \qquad ( \phi^{\mathfrak{s}}  , E_{\a,s}^{\mathfrak{s}}  )
= 0 ,
\ee
where we chose parametrization of the fundamental domain such that $x \sim x+1$.
The above also guarantees that any cusp form is orthogonal to space spanned by Eisensteins.
Both cusp forms and Eisensteins admit Fourier-Whittaker decompositions at each cusp $\a$
\begin{align} 
E_{\a}^{\mathfrak{s}} (\sigma_{\mathfrak{s},\b} \tau ,s) &= \delta_{\a \b} y^s + \varphi_{\a \b}^{\mathfrak{s}} (s) y^{1-s} 
+ \sum_{n \neq 0} \varphi_{\a \b}^{\mathfrak{s}} (n,s) 
W_s (n \tau) , 
\label{eq:FourierWhittaker}
\\ 
\phi_n^{\mathfrak{s}} (\sigma_\a \tau) &= 
\sum_{n \neq 0} c_{\a ,n}^{\mathfrak{s}} W_s (n \tau) , 
\end{align}
which also allows to analytically continue Eisensteins to $\Re (s) \geq \frac{1}{2}$. In the above, 
\begin{align}
\varphi_{\a \b}^{\mathfrak{s}} (s) &= 
\pi^{1/2} \frac{\Gamma(s-\frac{1}{2})}{\Gamma(s)} \sum_{c} c^{-2s} \mathcal{S}_{\a \b} (0,0;c) 
,
\\ 
\varphi_{\a \b}^{\mathfrak{s}} (n,s) &= \pi^s \frac{|n|^{s-1}}{\Gamma(s)}
c^{-2s} \mathcal{S}_{\a \b} (0,n;c) 
,
\end{align}
with Kloosterman sums $\mathcal{S}_{\a \b} (m,n;c)$ defined as 
\be 
\mathcal{S}_{\a \b} (m,n;c) = 
\sum_{ B \backslash \sigma_{\mathfrak{s},\a}^{-1} \Gamma_{\mathfrak{s}} \sigma_{\mathfrak{s},\b} / B
} e^{2\pi \i \left(m \frac{d}{c} + n \frac{a}{c} \right)} 
,
\qquad B = \left\{ \begin{pmatrix}
1 & n 
\\ 
0 & 1 
\end{pmatrix} : n \in \mathbb{Z} 
\right\} 
.
\ee
These sums generalize the ones that appear in the previous section when computing the spectrum on the solid torus. For our purposes, the most important properties satisfied by $\varphi_{\a \b}^{\mathfrak{s}}$ will be that it is symmetric, $\varphi_{\a \b}^{\mathfrak{s}} = \varphi_{\b \a}^{\mathfrak{s}}$, and that if we form a so-called scattering matrix $\Phi(s) = \{ \varphi_{\a \b}(s) \}$, it satisfies 
\be 
\Phi (s) \Phi(1-s) = I . \label{eq:scattering_matrix_unitarity}
\ee
This is also known as the unitarity of the scattering matrix (see, for example, Theorem 6.6 of \cite{Iwaniec2002SpectralMO}).
The coefficients $c_{\a ,n}^{\mathfrak{s}}$ denote the Fourier coefficients of the cusp forms (not known analytically), and $W_s(\tau)$ denotes the Whittaker function 
\be 
W_s(\tau) = 2 \sqrt{|y|} K_{s- \frac{1}{2}} (2\pi |y|) e^{2\pi \i x} .
\ee

With this, the spectral decomposition of any square-integrable function $Z_{\text{spec}} (\tau) \in L^2 (\Gamma_{\mathfrak{s}} \backslash \mathbb{H})$ takes the form\footnote{See Theorems 4.7 and 7.3 in \cite{Iwaniec2002SpectralMO}.}
\begin{align} 
Z_{\text{spec}}^{\mathfrak{s}} (\tau) 
= 
&\sum_\a 
\int_{-\infty}^\infty \frac{\dd \omega}{4\pi}  
(
Z_{\text{spec}}^{\mathfrak{s}} , E_{\a,\frac{1}{2}+\i \omega}^{\mathfrak{s}} )
E_{\a,\frac{1}{2}+\i \omega}^{\mathfrak{s}} (\tau ) 
+ \sum_{n\in \mathbb{N}} ( Z_{\text{spec}}^{\mathfrak{s}}, \phi_n^{\mathfrak{s}} ) \phi_n^{\mathfrak{s}}(\tau) 
.
\end{align}
Note that this is a manifestly modular invariant decomposition of the square-integrable part of an exact CFT partition function. The microscopic CFT data is now encoded in the overlaps $( 
Z_{\text{spec}}^{\mathfrak{s}} , E_{\a,s}^{\mathfrak{s}} )$ and $( Z_{\text{spec}}^{\mathfrak{s}}, \phi_n^{\mathfrak{s}} )$. In the following subsection, we will see how this can be used to determine minimal corrections required by modular invariance to a given near-extremal behavior.

\subsection{RMT$_2$ perspective on two-boundary torus wormholes}
\label{subsec:RMT2_two_boundary_wormhole}

We are now ready to discuss torus wormholes from the boundary perspective. The general approach we will follow in this section is that of RMT$_2$ -- a framework \cite{DiUbaldo:2023qli,Boruch:2025ilr} which allows one to construct a modular completion of a given near-extremal random matrix theory data. 
In particular, we will see how fully modular invariant two-boundary wormhole amplitudes can arise as a minimal modular completion of the JT gravity data expected to dominate near-extremality. 
The discussion here generalizes the results of \cite{DiUbaldo:2023qli} to the case of fermionic CFTs for any choice of spin structures.

To make comparison with gravity, we take the view that the gravitational path integral can only be used to extract coarse-grained statistical information about the underlying microscopic CFT spectrum. We will denote the quantities as obtained from the gravitational path integral with brackets $\langle \dots \rangle_{\text{grav}}$. In view of the discussion in subsection \ref{subsec:integrability}, we know that for single boundary partition function the non-square-integrable parts exactly match the solid torus partition functions of \ref{subsec: solid torus},
\be 
\langle Z^\mathfrak{s} (\tau) \rangle_{\text{grav}} = \frac{1}{\sqrt{y}|\eta (\tau)|^2} \hat{Z}_L^{\mathfrak{s}}(\tau) , 
\ee
and we therefore interpret $\hat{Z}_L^{\mathfrak{s}}(\tau)$ as the self-averaging part of the partition function, capturing the average density of states. On the other hand, the above relation also implies that $Z^\mathfrak{s}_{\text{spec}}(\tau)$ satisfies
\be 
\langle Z^\mathfrak{s}_{\text{spec}} (\tau) \rangle_{\text{grav}} = 0 ,
\ee
and is therefore interpreted as capturing the highly oscillatory, non-self-averaging, part of the density of states, responsible for restoring the exact discreteness of the underlying CFT spectrum. 

To study spectral statistics of the chaotic part of the spectrum, we are interested in higher-moments of the oscillatory part of the partition function $Z^\mathfrak{s}_{\text{spec}}(\tau)$. In particular, for comparison with two-boundary wormhole amplitudes we will focus on studying 
\be 
Z_{\text{RMT}_2}^{\mathfrak{s}} (\tau_1, \tau_2)
\equiv 
\langle Z^{\mathfrak{s}}_{\text{spec}}(\tau_1)Z^{\mathfrak{s}}_{\text{spec}}(\tau_2) \rangle_{\text{grav}}. 
\ee 
Our goal now is to determine the coarse-grained moments of overlaps 
\be 
\langle (Z^\mathfrak{s}_{\text{spec}}, E_{\a,s_1}^\mathfrak{s}/\phi^{\mathfrak{s}}_{n_1}) 
(Z^\mathfrak{s}_{\text{spec}}, E_{\a,s_2}^\mathfrak{s}/\phi^{\mathfrak{s}}_{n_2}) 
\rangle_{\text{grav}}, 
\ee 
by imposing the expected near-extremal behavior. In its final form, RMT$_2$ would allow us to determine all of the coarse-grained overlaps, including the cusp form overlaps, by imposing the expected near-extremal behavior of $Z_{\text{RMT}_2}^{\mathfrak{s}|(j_1,j_2)} (y_1, y_2)$ for each spin sector $(j_1,j_2)$. However, this is challenging due to the non-analytic nature of the cusp forms $\phi_n^\mathfrak{s}(\tau)$. As discussed in \cite{Haehl:2023xys}, the Fourier coefficients of cusp forms satisfy the so-called ``arithmetic chaos" and there is no known analytic expression for their values. This makes fixing their overlaps analytically through near-extremal behaviors challenging, and one is forced to make additional assumptions to be able to pursue analytic analysis further. Nevertheless, after obtaining the explicit final answer, it will become clear that the result reproduces the expected near-extremal behavior in every spin sector $(j_1,j_2)$, validating the assumptions made in our analysis. 

Motivated by the discussion of \cite{DiUbaldo:2023qli}, we will make two simplifying assumptions which will allow us to proceed further. First, we will restrict ourselves to the ``diagonal subspace" of Eisensteins and cusp forms: this means we will assume that, at the level of two boundaries, the coefficients in the CFT spectral decomposition in front of distinct eigenfunctions are not correlated with each other\footnote{This assumption can be viewed as the CFT analogue of Berry's diagonal approximation, for more discussion of this point see \cite{DiUbaldo:2023qli}.}. Second, motivated by the bulk expectations we will assume that the two-boundary partition function is invariant under simultaneous modular transformations acting on both boundaries,\footnote{For the following discussion, we switch for convenience to conventions used in \cite{Iwaniec2002SpectralMO} in which relative modular transformations act on the two-boundary partition function as $(\tau_1,\tau_2)\to(\gamma \tau_1 , \gamma \tau_2)$. This differs from conventions used in section \ref{sec:Path_integral_fermionic_3d_gravity}. We will revert back to conventions of section \ref{sec:Path_integral_fermionic_3d_gravity} to compare with gravitational answers at the end of this subsection.}
\be 
Z_{\text{RMT}_2}^{\mathfrak{s}} (\gamma \tau_1, \gamma \tau_2) = Z_{\text{RMT}_2}^{\mathfrak{s}} (\tau_1, \tau_2), 
\qquad \gamma \in \Gamma_{\mathfrak{s}}. 
\ee
This second assumption implies that the two-boundary amplitude depends on $\tau_1$ and $\tau_2$ only through the hyperbolic distance 
\be 
u(\tau_1 , \tau_2) 
= \frac{|\tau_1 - \tau_2|^2}{4\Im \tau_1 \Im \tau_2} .
\ee
Furthermore, as shown in \cite{DiUbaldo:2023qli}, this assumption implies that the functional form of coarse-grained overlaps for the cusp forms is exactly the same as the one for Eisensteins. Under both of the above assumptions, the two-boundary spectral decomposition of $Z_{\text{RMT}_2}^{\mathfrak{s}} (\tau_1, \tau_2)$ takes the form 
\begin{align}
Z_{\text{RMT}_2}^{\mathfrak{s}} ( \tau_1, \tau_2)
&= 
\sum_\a 
\int_{\frac{1}{2}+\i \mathbb{R}} \frac{\dd s}{4\pi \i}  
h^{\mathfrak{s}}(s)
E_{\a,s}^{\mathfrak{s}} (\tau_1) 
E_{\a,1-s}^{\mathfrak{s}} (\tau_2) 
+ \sum_{n \in \mathbb{N}} h^{\mathfrak{s}}(s_n) \phi_n^{\mathfrak{s}}(\tau_1) \phi_n^{\mathfrak{s}}(\tau_2)  
, \label{eq:diagonal_spectral_decomposition}
\end{align}
where we denoted the coarse-grained overlaps as
\be 
h^{\mathfrak{s}}(s) = \langle (Z^\mathfrak{s}_{\text{spec}}, E_{\a,s}^\mathfrak{s}) 
(Z^\mathfrak{s}_{\text{spec}}, E_{\a,1-s}^\mathfrak{s}) 
\rangle_{\text{grav}}
= 
\langle (Z^\mathfrak{s}_{\text{spec}}, \phi_n^\mathfrak{s}) 
(Z^\mathfrak{s}_{\text{spec}}, \phi_n^\mathfrak{s}) 
\rangle_{\text{grav}} . 
\ee

We now proceed to fix $h^{\mathfrak{s}}(s)$ through the expected near-extremal behavior for each spin structure $\mathfrak{s}$. Projecting ourselves onto the scalar sector $(j_1=0,j_2=0)$, the cusp forms drop out and we have 
\be
Z_{\text{RMT}_2}^{\mathfrak{s}|(0,0)}(y_1 , y_2) 
= \sum_{\a}
\int_{\frac{1}{2}+\i \mathbb{R}} \frac{\dd s}{4\pi \i}  
h^{\mathfrak{s}}(s) E_{\a,s}^{(0)}(y_1) E_{\a,1-s}^{(0)}(y_2) ,
\ee
where we denoted the scalar modes of the Eisensteins as $E_{\a,s}^{(0)}(y)$. Explicitly, using the Fourier-Whittaker expansion \eqref{eq:FourierWhittaker} in coordinates associated with the cusp at $\infty$, $\b=\infty$, we have 
\be 
E_{\a,s}^{(0)}(y) = \delta_{\a,\infty} y^s + \varphi_{\a,\infty}^\mathfrak{s} (s) y^{1-s} .
\ee
Plugging this in and using the symmetry of the scattering matrix $\varphi_{\a,\infty}^\mathfrak{s} = \varphi_{\infty,\a}^\mathfrak{s}$, the resulting terms can be combined into 
\begin{align} 
Z_{\text{RMT}_2}^{\mathfrak{s}|(0,0)}(y_1 , y_2) 
&= 
\int_{\frac{1}{2}+\i \mathbb{R}} \frac{\dd s}{4\pi \i}  
h^{\mathfrak{s}}(s) (1+\sum_{\a} \varphi_{\infty,\a}^\mathfrak{s}(s) \varphi_{\a,\infty}^\mathfrak{s}(1-s)) y_1^s y_2^{1-s} 
\\
&+
\int_{\frac{1}{2}+\i \mathbb{R}} \frac{\dd s}{2\pi \i}  
h^{\mathfrak{s}}(s)
\varphi_{\infty,\infty}(1-s) (y_1 y_2)^s  .
\end{align}
From the unitarity of the scattering matrix \eqref{eq:scattering_matrix_unitarity} it follows that 
\be 
\sum_{\a} \varphi_{\infty,\a}^\mathfrak{s}(s) \varphi_{\a,\infty}^\mathfrak{s}(1-s) = 1, 
\ee
and therefore the scalar sector answer can be rewritten as 
\begin{align} 
Z_{\text{RMT}_2}^{(0,0)}(y_1 , y_2) &= 
\int_{\frac{1}{2}+\i \mathbb{R}} \frac{\dd s}{2\pi \i}  
h^{\mathfrak{s}}(s) (r z^{\frac{1}{2}-s} + r^{2s} \varphi_{\infty \infty}(1-s) ) 
\\ 
&= r \sqrt{z} \mathcal{R}(z) + \mathcal{S}(r) , 
\end{align}
where we introduced $r \equiv \sqrt{y_1 y_2}$, $z\equiv y_1/y_2$ and denoted the relevant inverse Mellin transforms as
\be 
\mathcal{R}(z) = \mathcal{M}^{-1} [h^{\mathfrak{s}}(s)(s) ; z]  
, 
\qquad 
\mathcal{S}(r) = 
\mathcal{M}^{-1} [h^{\mathfrak{s}}(s)(s) \varphi_{\infty \infty}(1-s) ; z] 
.
\ee
to explicitly match the notation of \cite{DiUbaldo:2023qli}. The above directly generalizes the computation of \cite{DiUbaldo:2023qli} to the case of fermionic CFTs for any choice of spin structure $\mathfrak{s}$ and in the presence of multiple cusps on the fundamental domain $\Gamma_{\mathfrak{s}}\backslash \mathbb{H}$. 
The key observation now is that in the near-extremal limit, $Z_{\text{RMT}_2}^{(0,0)}(y_1 , y_2)$ is dominated only by the first term
\be 
Z_{\text{RMT}_2}^{\mathfrak{s}| (0,0)} ( y_1, y_2) \simeq 
r \sqrt{z} \mathcal{R}(z) ,
\qquad 
(r \gg 1, z \text{ fixed})
.
\ee
This follows from noticing that to evaluate $\mathcal{S}(r)$ one needs to close the contour to the left, meaning that the final answer obtained through summing over residues will only contain terms behaving as $r^\alpha$ with $\Re (\alpha) < 1$. 
The near-extremal limit only fixes the form of $\mathcal{R}(z)$, whilst the remainder $\mathcal{S}(r)$ collects together corrections required purely by modular invariance.  

Finally, we demand that in the limit $y_1,y_2 \sim c$, $c\to \infty$, $Z_{\text{RMT}_2}^{\mathfrak{s}| (0,0)} ( y_1, y_2)$ takes the form of the JT gravity double trumpet \cite{Saad:2018bqo,Saad:2019lba} 
\be 
Z_{\text{RMT}_2}^{\mathfrak{s}| (0,0)} ( y_1, y_2)|_{y_1,y_2 \sim c\to\infty} \simeq 
C_{\mathfrak{s}}
\frac{y_1 y_2}{y_1+y_2} 
= C_{\mathfrak{s}}\frac{r \sqrt{z}}{1+z}
,
\ee
where $C_{\mathfrak{s}}$ denotes a constant factor that differs for each spin structure.
The overlaps $h^{\mathfrak{s}}(s)$ are now computed through the Mellin transform as 
\be 
h^{\mathfrak{s}}(s) = \mathcal{M}[\mathcal{R}(z) ; s]  =C_{\mathfrak{s}} \frac{\Gamma(s)\Gamma(1-s)}{\pi} .
\label{eq:CJ_overlaps}
\ee
With this, we can now plug in the overlaps back into the full spectral decomposition. The resulting two-boundary wormhole amplitude takes the form \eqref{eq:diagonal_spectral_decomposition} with the above overlaps. To compare with the results of section \ref{subsec:two-boundary-torus-wormholes} we wish to rewrite the final result as a sum over modular images. This can be done conveniently through the so-called Kuznetsov trace formula (see Theorem 7.4 of \cite{Iwaniec2002SpectralMO}), as originally pointed out in \cite{Haehl:2023mhf} for the bosonic case. The formula states that general diagonal amplitudes of the form \eqref{eq:diagonal_spectral_decomposition}, with any overlap function $h^{\mathfrak{s}}(s)$, can be rewritten as a sum over relative modular transformations 
\be 
Z^{\mathfrak{s}}_{\text{RMT}_2} (\tau_1 ,\tau_2) = \sum_{\gamma \in \Gamma_{\mathfrak{s}}} k^{\mathfrak{s}} (u(\tau_1 , \gamma \tau_2)) , 
\label{eq:RMT2_wormhole_answer}
\ee
with the seed $k^{\mathfrak{s}} (u(\tau_1 ,  \tau_2))$ determined from $h^{\mathfrak{s}}(s=\frac{1}{2}+\i t) \equiv \widetilde{h}^{\mathfrak{s}} (t)$ via 
\begin{align}
\widetilde{h}^{\mathfrak{s}} (t) &= 4\pi \int_0^\infty \dd u 
F_{\frac{1}{2}+\ii t}(u) k^{\mathfrak{s}}(u)
, \qquad F_{\frac{1}{2}+\i t}(u) \equiv  
{}_2 F_1 \left(\frac{1}{2}+\i t,\frac{1}{2}-\i t; 1 ; -u \right) ,
\\ 
k^{\mathfrak{s}}(u) &= \frac{1}{4\pi} \int_{-\infty}^\infty \dd t \, t
F_{\frac{1}{2}+\ii t}(u) \tanh (\pi t) \widetilde{h}^{\mathfrak{s}}(t) 
.
\end{align}
Inserting \eqref{eq:CJ_overlaps} into the above relation, we determine the seed as 
\be 
k^{\mathfrak{s}}(u)  = \frac{C_{\mathfrak{s}}}{4\pi^2} 
\frac{1}{1+u}
,
\ee
where, to explicitly match the Cotler-Jensen answer, the distance is taken between $\tau_1 = x_1 + \i y_1$ and $\tau_2 = -x_2 + \i y_2$, to account for the orientation reversal on the $\tau_2$ boundary.  We thus obtain
\be 
k^{\mathfrak{s}}(\tau_1,\tau_2)  = \frac{C_{\mathfrak{s}}}{\pi^2} \frac{y_1 y_2}{(x_1+x_2)^2 + (y_1+y_2)^2} 
.
\ee
With the above seed, \eqref{eq:RMT2_wormhole_answer} manifestly matches the gravitational answers explained in \eqref{eq:NSMNSM}, after choosing $C_{\mathfrak{s}}=2$ and multiplying by factors of $Z_0(\tau_{1/2})$ to reintroduce contributions from descendants. Note that in this section we worked with coordinates with normalized such that $x \sim x+1$. To match more natural normalizations used in previous section, $x_{\NS} \sim x_{\NS} + 2$, one needs to reparametrize the modular coordinates as $\tau = \frac{1}{2} \tau_{\NS}$. 

In the following sections, we proceed to analyze two-boundary wormholes in the presence of additional discrete symmetries. Whilst we will be using mostly gravitational language, we note that for all the cases without gravitational anomalies discussed below the above RMT$_2$ discussion still holds, including some subset of the cases with a unitary $\mathbb{Z}_2$ symmetry where fundamental domains can present three cusps. For the cases with gravitational anomalies the partition functions are not strictly modular invariant and therefore the formalism here has to be correspondingly adapted. This is a very interesting problem which we leave for future work.

\section{Discrete symmetries and bulk topological field theories}\label{sec:discretesymm}

So far, we have mostly discussed the path integral of fermionic 3d gravity on orientable surfaces, with no additional symmetries beyond Lorentz invariance and fermion number symmetry. In this section, we explain how this discussion changes in the presence of anti-unitary spacetime symmetries $\r\T$ and $\T$. We begin by reviewing these anti-unitary $\Z_2$ symmetries and the resulting (non-orientable) contributions to the fermionic path integral.

The CPT theorem tells us that Lorentz-invariant field theories are invariant under $SO(1,D-1)$, including not just the connected group $SO^+(1,D-1)$ but also the anti-unitary transformation $\r\T$ (see footnote \ref{foot:CPT footnote}). As discussed in section \ref{sec:comments}, the Wick rotation of $\r\T$ is a rotation by $\pi$, which is an element of $SO(D)$. When we lift $SO(D)$ to Spin$(D)$, the rotation by $\pi$ squares to $(-1)^{\sf F}$, and its analytic continuation contributes another factor of $(-1)^{\sf F}$ (as recently reviewed in e.g. \cite{Witten:2025ayw}), resulting in the Lorentzian symmetry
\begin{equation}
    (\r\T)^2 = 1 \,.
\end{equation}
$\r\T$ is a global symmetry of the boundary $D$-dimensional field theory that must be gauged in the $D+1$-dimensional bulk dual: the gravitational path integral sums over topologies that include $\r\T$ reflections \cite{Harlow:2023hjb}. Since the analytic continuation of $\r\T$ is simply a rotation by $\pi$, $\r\T$ is an orientation-preserving symmetry and the path integral sums over orientable manifolds. In addition to obeying the appropriate boundary conditions, the manifolds $X$ included in this path integral must (1) admit a spin structure: the second Stiefel-Whitney class $w_2(X)\in H^2(X,\Z_2)$ vanishes, and (2) be orientable: the first Stiefel-Whitney class $w_1(X)\in H^1(X,\Z_2)$ vanishes.\footnote{For example, consider $S^1$, for which $w_1(S^1)=w_2(S^1)=0$. There are two distinct spin structures, corresponding to the two equivalence classes of $H^1(S^1,\Z_2) \cong\Z_2$ and given by periodic versus antiperiodic boundary conditions.}

We now consider field theories with $\r$ and $\T$ symmetry, in addition to $\r\T$. Whereas the group $SO(D)$ has a unique connected double cover Spin$(D)$, the orthogonal group $O(D)$ has two double covers, Pin$^\pm(D)$, with identity component Spin$(D)$. In Euclidean signature, the two double covers are distinguished by the square of the reflection $\r$ of a coordinate axis, where $\r^2=(\pm1)^{\sf F}$ for Pin$^\pm$.\footnote{In particular, going around a crosscap twice amounts to acting with $\r^2$, which means $\pinm$ crosscaps have $\NS$ periodicity while $\pinp$ crosscaps have $\R$ periodicity.} For simplicity we use the same label for the action of a discrete symmetry in spacetime, and the operator in Hilbert space implementing it, as it is clear from the context. In Lorentzian signature, parity analytically continues to $\r^2=(\pm1)^{\sf F}$, while time reversal analytically continues to $\T^2=(\mp1)^{\sf F}$ after Wick rotation.

The existence of pin$\pm$ structures is determined by Stiefel-Whitney classes: a manifold $X$ admits a $\pinp$ structure if and only if $w_2(X)=0$, while it admits a $\pinm$ structure if and only if $(w_1^2+w_2)(X)=0$:\footnote{For example, consider $\mathbb{RP}^n$. The total Stiefel-Whitney class is given by $w(\mathbb{RP}^n)=(1+a)^{n+1}$, where $a$ denotes the generator of $H^1(\mathbb{RP}^n,\Z_2)$ and multiplication denotes a cup product while addition is mod 2 \cite{Milnor:1974}. Since $w(\mathbb{RP}^2)=1+a+a^2$, $\mathbb{RP}^2$ admits a $\pinm$ but no $\pinp$ structure, and since $w(\mathbb{RP}^4)=1+a+a^4$, $\mathbb{RP}^4$ admits a $\pinp$ but no $\pinm$ structure.}
\begin{equation} \label{eq: pin pm}
\begin{split}
    &\pinp: \r^2=1 \text{ and } \T^2=(-1)^{\sf F} \quad\Leftrightarrow\quad w_2(X)=0 \\
    &\pinm: \r^2=(-1)^{\sf F} \text{ and } \T^2=1 \quad\Leftrightarrow\quad (w_1^2+w_2)(X)=0 \,.
\end{split}
\end{equation}
The gravitational path integral now sums over $D+1$-dimensional topologies that include $\r\T$, $\r$, and $\T$ reflections, and includes manifolds that satisfy \eqref{eq: pin pm}. Since reflection of a single coordinate flips the orientation, the path integral now includes not only orientable, but also non-orientable manifolds: the pin$^\pm$ conditions \eqref{eq: pin pm} do not uniquely determine $w_1(X)$ \cite{Witten:2016cio, Harlow:2023hjb}.

For each of these structures (spin and pin$^\pm$), we will classify the different topological terms that can consistently contribute to the gravitational path integral and study their effect on the RMT ensembles in each case.

\subsection{Bordisms and cobordisms in gravity} \label{subsec:bordism}
In this section we review how the bordism group of the $D$-dimensional boundary spin or pin$^\pm$ manifold determines whether it admits a bulk dual \cite{Thom:1954, Atiyah:1961, Anderson:1967, Stong:1968}, and how the cobordism group of the $D+1$-dimensional bulk spin or pin$^\pm$ manifold classifies some  subtle topological terms that can appear in the gravitational path integral \cite{Witten:1985bt, Kapustin:2014tfa, Kapustin:2014dxa, Freed:2016rqq}.

\subsubsection*{Bordism groups of boundary manifolds}
In section \ref{subsec: solid torus} and in footnote \ref{footnote: bordism}, we used the bordism equivalence class of the boundary manifold to determine the existence of a bulk filling. The group $\Omega_D^{\spin}(pt)$ consists of bordism classes of $D$ dimensional $\spin$ manifolds, where two manifolds are bordant if their union is the boundary of a $D+1$ dimensional manifold (see \cite{Freed:bordism} for more formal definitions).

We now specialize to the 2d boundaries relevant for 3d gravity. In this case, the spin bordism group is given by 
\begin{equation}
    \Omega_2^{\spin}(pt) = \Z_2 \,,
\end{equation}
and the topological invariant that distinguishes the two equivalence classes is the Arf invariant, which counts the number of zero modes mod 2 of the chiral Dirac equation for a spinor on a 2-manifold $X$ \cite{Kirby:1991}. This classification applies to surfaces with arbitrary genus, but for simplicity, consider a 2-torus $X=T^2$. The three even spin structures $(\NS\pm,\R-)$ have no zero modes, and hence correspond to the trivial element of $\Omega_2^{\spin}$. Since these spin structures are cobordant to a point, they bound nontrivial 3-manifold fillings, which for example include the on-shell saddles described in section \ref{sec:Path_integral_fermionic_3d_gravity}. On the other hand, the odd spin structure $\R+$ has a zero mode, so it corresponds to the nontrivial element of $\Omega_2^{\spin}$ and hence any odd number of $\R+$ tori cannot bound a spin 3-manifold. The same conclusion applies for an odd number of any set of surfaces with non-trivial mod 2 index, as shown in Figure \ref{fig:zero}.

The pin${}^\pm$ bordism groups of 2-manifolds are given by \cite{Kirby:1991, Kapustin:2014dxa}
\begin{equation}
    \Omega_2^{\pinp}(pt) = \Z_2\,,\quad \Omega_2^{\pinm}(pt) = \Z_8 \,.
\end{equation}
For $\pinm$, the topological invariant on orientable manifolds reduces to the Arf invariant, so we again find that the even spin structure $\R+$ on the torus cannot admit a $\pinm$ bulk filling. For $\pinp$, the relevant topological invariant is the nonchiral mod 2 index, which vanishes on orientable manifolds. In this case all four torus spin structures can bound nontrivial 3-manifolds.

One immediate consequence in the $\pinp$ case is that an odd number of $\R+$ tori has a nonvanishing partition function, and so in particular
\begin{equation}
    Z_{\R+}\neq0\,,\quad Z_{\R+\R-}\neq0\,,\quad Z_{\R+\NS\pm}\neq0 \,.
\end{equation}
Then the density of states in the $\R$ Hilbert space is no longer symmetric between bosons and fermions:
\begin{equation}
    \left\langle\rho^{\text{bos}}_{\R}(E, j)\right\rangle \neq \left\langle\rho^{\text{fer}}_{\R}(E, j)\right\rangle
\end{equation}
and for the two-boundary wormholes, the joint probability distribution on NS/R sectors is no longer symmetric under exchange of bosonic/fermionic R sector Hamiltonians:
\begin{equation}
\begin{split}
    &\left\langle\rho^{\text{bos}}_{\R}(E_1, j_1)\rho^{\text{bos}}_{\R}(E_2, j_2)\right\rangle  \neq  \left\langle\rho^{\text{fer}}_{\R}(E_1, j_1)\rho^{\text{fer}}_{\R}(E_2, j_2)\right\rangle \\
    &\left\langle\rho^{\text{bos}}_{\R}(E_1, j_1)\rho_{\NS}(E_2, j_2)\right\rangle \neq \left\langle\rho^{\text{fer}}_{\R}(E_1, j_1)\rho_{\NS}(E_2, j_2)\right\rangle \,.
\end{split}
\end{equation}
An example of an off-shell topology that contributes to $Z_{\R+}$ is the (Mobius strip)$\times S^1$, whose contribution to the gravitational path integral is discussed in \cite{Yan:2023rjh}. This topology contributes to $Z_{\R+}$ for $\pinp$ but not for $\pinm$ since crosscaps have $\R$ boundary condition in $\pinp$ and $\NS$ boundary condition in $\pinm$.

\subsubsection*{Cobordism groups and bulk TQFT}
We just saw that bordism groups of boundary 2-manifolds determine the existence of bulk fillings. We will now introduce the cobordism groups of bulk 3-manifolds as a tool to classify the 3d invertible topological quantum field theories (TQFTs) that can be consistently coupled to our pure gravity theory. 

An invertible TQFT can be understood as a quantum system with a one-dimensional gapped vacuum, such that its partition function on a compact manifold $X$ is simply given by a complex phase,
\begin{equation} \label{eq:SPT_Z}
    Z[\mathfrak{s}] = e^{\ii\pi\Phi[\mathfrak{s}]}\,,
\end{equation}
where $\mathfrak{s}$ is the spin or pin$^\pm$ structure on $X$.\footnote{Depending on the context, invertible TQFTs are sometimes referred to as symmetry-protected topological (SPT) phases of matter, and in the presence of fermion number symmetry as topological superconductors. An example of the relationship between cobordism invariants and low-energy SPT partition functions is given by the definition of the Arf invariant on 2d spin manifolds in terms of the continuum version of the Kitaev chain partition function: $(-1)^{\zeta[\mathfrak{s}]} = Z_{\text{ferm}}(m\gg0;\mathfrak{s})/Z_{\text{ferm}}(m\ll0;\mathfrak{s})$. See e.g. \cite{Kitaev:2000nmw, Witten:2015aba, Kaidi:2019tyf} and references therein.} In the context of \eqref{eq:couple_to_M}, we could imagine coupling the metric to an arbitrary QFT (rather than a single massive fermion) and studying its low-energy effective field theory, which may contribute nontrivial topological dependence. Since the partition function \eqref{eq:SPT_Z} describes the Hilbert space of a gapped vacuum, we can ignore gappable degrees of freedom such as from massive fermions.\footnote{See \cite{Witten:2015aba} for a regularization argument for why anomalies in QFT only come from massless fermions, which in particular implies that $|Z|$ is well-defined for any relativistic theory of fermions, while the overall phase can be ambiguous in the presence of anomalies.}

Each partition function \eqref{eq:SPT_Z} that is consistent with unitarity and locality corresponds to an equivalence class of the cobordism group ${\rm TP}_3$ (topological phases in 3d). For $\spin$ manifolds, this is determined by
\begin{equation}
    \Hom \left(\Omega_3^{\spin,tors}(pt), U(1)\right) \,,
\end{equation}
the Pontryagin-dual of the torsion subgroup of the $\spin$ bordism group \cite{Kapustin:2014tfa, Kapustin:2014dxa}.\footnote{In the case of local (sometimes called perturbative \cite{Witten:2015aba}) anomalies, the cobordism group ${\rm TP}_3$ contains a free part (e.g. the thermal Hall conductivity on 3-manifolds, which we will discuss in section \ref{subsec:RT}) in addition to $\Hom (\Omega_3^{\spin,tors}(pt), U(1))$. The formal definitions can be found in \cite{Freed:2016rqq}.}

So far, the discussion of the TQFT \eqref{eq:SPT_Z} has only applied to compact manifolds $X$. Trying to make sense of \eqref{eq:SPT_Z} on a manifold $M$ with boundary one discovers an anomaly. The boundary CFT will also present an anomaly which reproduces that arising from defining \eqref{eq:SPT_Z} on $M$. This is similar but fundamentally different than the anomaly inflow mechanism between bulk and boundary theories that live together on the same spacetime \cite{Callan:1984sa, Witten:2019bou}.

\begin{figure}[t!]
    \centering \includegraphics[width=0.7\linewidth]{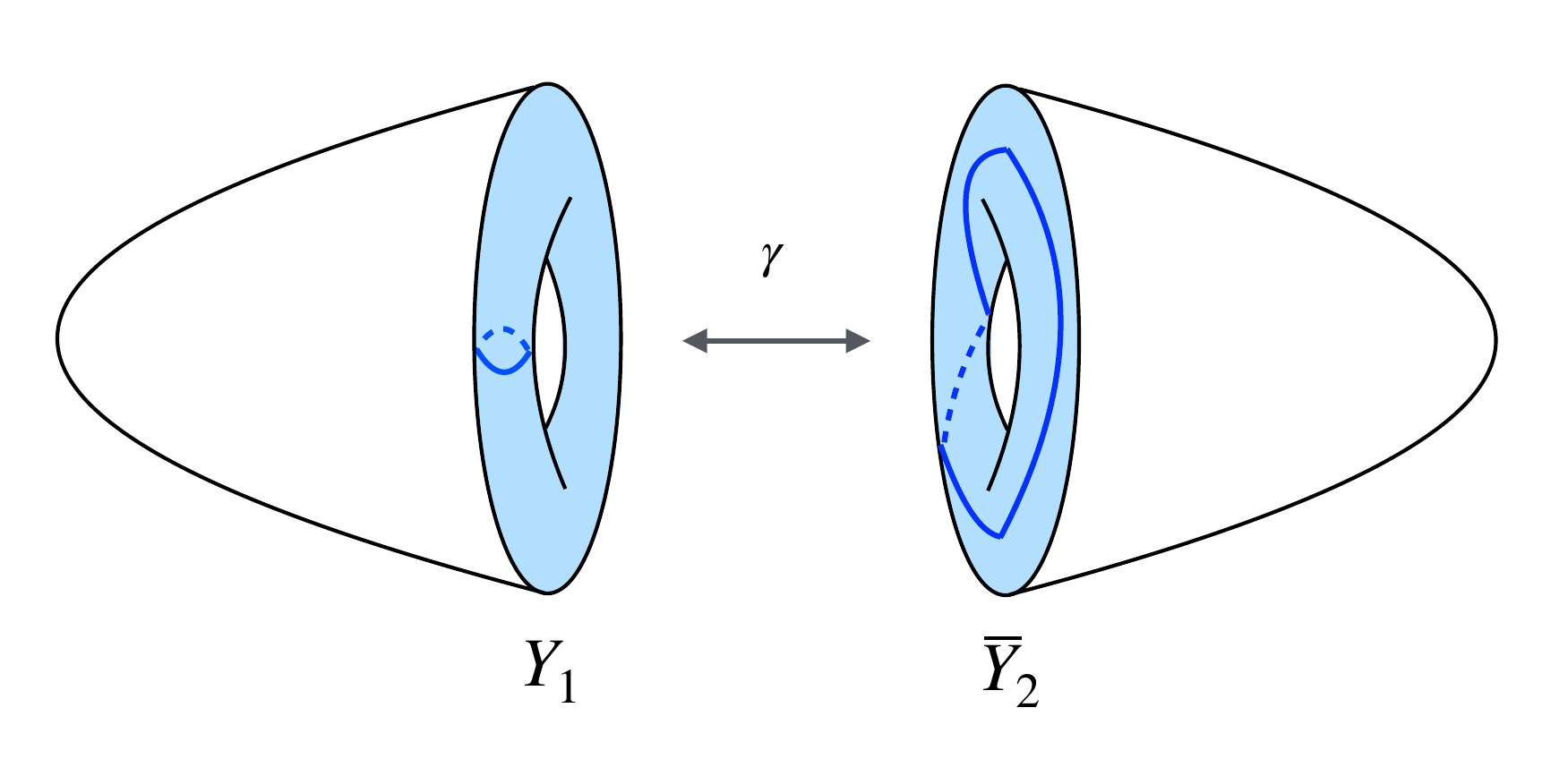}
    \caption{To determine how the cobordism invariant on a solid torus depends on the modular image we do the following. We take two solid tori $Y_1$ and $Y_2$ with different modular parameters and glue them along their boundary after performing a large diffeomorphism $\gamma$. In blue we indicate an example of a cycle on $\partial Y_1$ that could be mapped to a cycle on $\partial Y_2$. This construction results in lens space, with a choice of framing.}
    \label{fig:lens}
\end{figure}

In the context of the gravitational path integral, the implication is that the TQFT \eqref{eq:SPT_Z} contributes a phase to each bulk 3-manifold, which reproduces an anomaly in the boundary field theory.\footnote{This was observed in the context of JT gravity in \cite{Stanford:2019vob}. We review and summarize the method they used to compute cobordism invariants on 2-manifolds with boundary in order to apply it to 3-manifolds with boundary.} In practice, to compute this phase, we recall that on a manifold $M$, any cobordism invariant $\Phi[\mathfrak{s}]$ is local in the sense that if we consider two spin structures $\mathfrak{s}_1,\mathfrak{s}_2$ that differ only in some patch $V\subset M$, then their ratio $e^{\ii\pi\Phi[\mathfrak{s}_1]}/e^{\ii\pi\Phi[\mathfrak{s}_2]}$ depends only on the restriction of $\mathfrak{s}_1,\mathfrak{s}_2$ to $V$.

\smallskip

For our purposes we will need to compute these actions in two situations. The first involves the gravitational path integral on the solid torus and the possibility that $\exp{\i\pi\Phi}$ depends on the modular image. To answer this question we can consider two solid tori $Y_1$ and $Y_2$ with the same boundary spin structures $\mathfrak{s}_1=\mathfrak{s}$ and $\mathfrak{s}_2=\mathfrak{s}$, obtained by a modular transformation $\gamma_1,\gamma_2 \in \text{PSL}(2,\mathbb{Z})$ acting on thermal AdS. For these two images to contribute to the same partition function $\gamma_{1,2}$ should be in certain spin-structure-dependent subgroups of $\text{PSL}(2,\mathbb{Z})$. All solid torus path integrals take this form as explained in section \ref{sec:Path_integral_fermionic_3d_gravity}. We can compute the ratio of TQFT partition functions 
$
Z[Y_1,\mathfrak{s}]/Z[Y_2,\mathfrak{s}]
$
by reversing the orientation of $Y_2$, producing a solid torus we call $\overline{Y}_2$, namely
$$
\frac{Z[Y_1,\mathfrak{s}]}{Z[Y_2,\mathfrak{s}]} = e^{\i \pi \Phi[Y_1]+ \i \pi \Phi[\overline{Y}_2]} = e^{\i \pi \Phi[ Y_1 \cup_{\gamma} \overline{Y}_2]}.
$$
The space $Y_1 \cup_{\gamma} \overline{Y}_2$ arises from taking the two solid tori and gluing them across their boundary after performing a large diffeomorphism characterized by a modular transformation $\gamma = \gamma_1\cdot \gamma_2^{-1} $, as shown in Figure \ref{fig:lens}. This construction leads to lens space and this presentation gives its genus-1 Heegaard split.\footnote{The lens space $L(p,q)$ is defined as $S^3/\mathbb{Z}_p$ where $S^3$ is parametrized by $(z_1,z_2)\in\mathbb{C}^2$ with $|z_1|^2+|z_2|^2=1$ and the $\mathbb{Z}_p$ identification is $(z_1,z_2)\sim (w\,z_1, w^q\,z_2)$ with $w=e^{\frac{2\pi\i}{p}}$ and $p,q$ coprime integers. The surface $|z_1|=|z_2|$ is a torus (parametrized by the two phases) and it divides $S^3$ into two solid tori, one with $|z_1|\leq|z_2|$ and another with $|z_2|\leq|z_1|$. One can check that these spaces remain being solid tori after imposing the $\mathbb{Z}_p$ identification and that $(p,q)$ parametrizes how the contractible circle of one solid torus is identified to that of the other solid torus.} While the geometry only cares of $\gamma$ up to multiplication by $T$, this operation changes the framing. In the presence of gravitational anomalies the path integral will depend on the framing as well.

\smallskip
\begin{figure}[t!]
    \centering
    \includegraphics[width=0.7\linewidth]{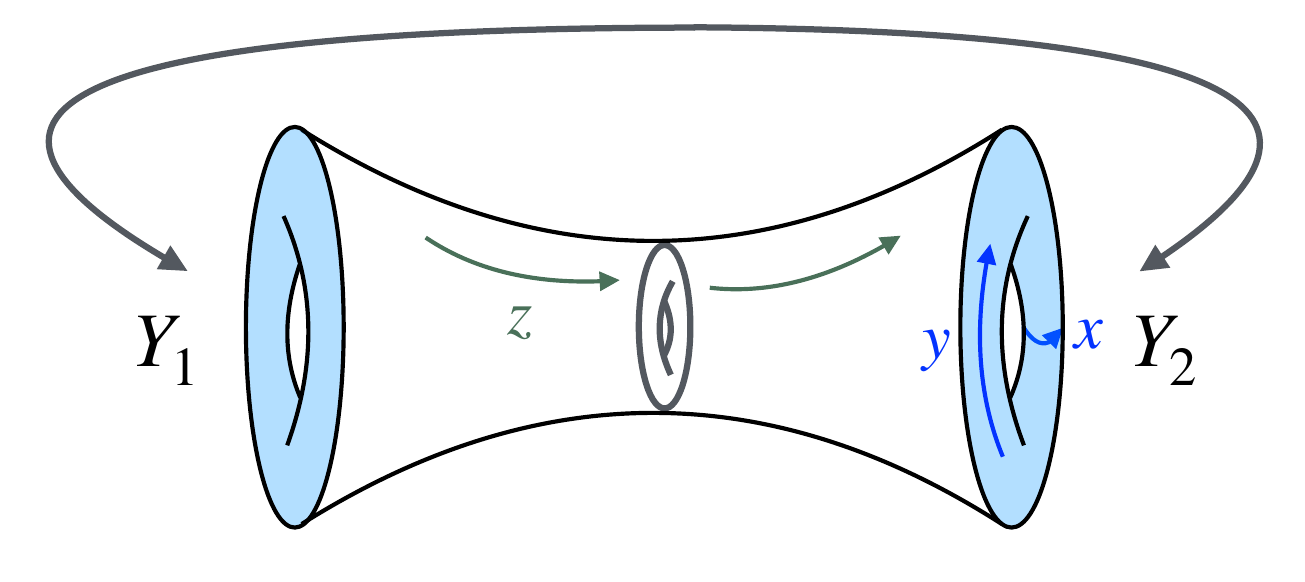}
    \caption{To determine the cobordism invariant on a manifold with boundary, we use locality to instead compute it on a compact manifold. We patch on the manifold connecting two 2-torus boundaries in which the spin structures differ due to the insertion of a codimension one defect along the gray torus. The 3d mapping torus is obtained after gluing the two outer boundaries together.}
    \label{fig:2torus}
\end{figure}

The second situation we will be interested in is the two-boundary torus wormhole, for which gauging the global symmetry $G$ (which in our discussion could be fermion parity, $\r\T$, or time reversal) implies summing over wormhole fillings with or without the insertion of a $G$ codimension one defect at the gluing seam between the two trumpets, depicted as a gray torus in Figure \ref{fig:2torus}.\footnote{See \cite{Harlow:2018tng, Harlow:2023hjb, Gomis:2025gzb} for discussions of gauging global symmetries in quantum gravity and \cite{Petkova:2000ip, Frohlich:2006ch, Kapustin:2014gua, Gaiotto:2014kfa, Cordova:2022ruw} for discussions of representing global symmetries as topological defect operators on codimension one manifolds.} This corresponds to two choices of parallel transport from boundary $Y_1$ to boundary $Y_2$, for example:
\begin{align}
    (-1)^{\sf F}&: \psi_1(t,x) = \psi_2(t,x) \text{ or } \psi_1(t,x) = -\psi_2(t,x) \nonumber\\
    \r\T&: \psi_1(t,x) = \psi_2(t,x) \text{ or } \psi_1(t,x) = \psi_2(-t,-x) \\
    \T&: \psi_1(t,x) = \psi_2(t,x) \text{ or } \psi_1(t,x) = \psi_2(-t,x) \,. \nonumber
\end{align}
We would like to compare the cobordism invariants in the case with and without the insertion of the $G$ defect. Since the two cases differ only in a neighborhood of the gluing seam, shown as a gray torus in Figure \ref{fig:2torus}, we can cut and glue this patch into a 3-torus, shown in Figure \ref{fig:2torus} by gluing the torus boundaries. 

Another operation we can do on the two-boundary torus wormhole is a relative modular transformation $\gamma$, a large diffeomorphism, between the two ends. To determine how the TQFT partition function depends on $\gamma$ we can again glue the two ends together. We can think of the resulting mapping torus as having a codimension one defect implementing the large diffeo. Locality implies that the dependence with $\gamma$ in the wormhole has to be exactly the same as the one computed in lens space.

In the following sections, we will use this cut and glue technique to evaluate the effect of invertible TQFTs on the solid torus and on two-boundary wormholes in 3d gravity.

\subsection{Incorporating $(-1)^{\sf F}$ and ${\sf RT}$} \label{subsec:RT}

We first consider spin 3-manifolds, with both $\r\T$ and $(-1)^{\sf F}$ symmetries gauged. With no insertion in Figure \ref{fig:2torus}, the resulting topology is a three-torus $T^3$, for which a standard description is to start from $\mathbb{R}^3$ and divide by the three symmetries $T_x,T_y,T_z$, where
\begin{equation} \label{eq:T3}
    T_x(x,y,z)=(x+1,y,z)\,,\quad T_y(x,y,z)=(x,y+1,z)\,,\quad T_z(x,y,z)=(x,y,z+1)\,.
\end{equation}
With the insertion of an $\r\T$ defect, the resulting topology involves dividing $\mathbb{R}^3$ by the following symmetries
\begin{equation} \label{eq:T3 RT}
    T_x(x,y,z)=(x+1,y,z)\,,\quad T_y(x,y,z)=(x,y+1,z)\,,\quad T_z(x,y,z)=(-x,-y,z+1)\,.
\end{equation}
This corresponds to the half-turn torus bundle (named dicosm in \cite{Conway:2003tp}), which is an orientable manifold.

The implication of $\r\T$ for the two boundary partition function was observed in \cite{Yan:2023rjh}: the two ways of gluing the torus trumpets lead to a factor of two multiplying the Cotler-Jensen wormhole \cite{Cotler:2020ugk}. Gauging the $(-1)^{{\sf F}}$ symmetry leads to two additional ways of gluing the torus trumpets, corresponding to the insertion of a $(-1)^{\sf F}$ defect in Figure \ref{fig:2torus}. The combination of $\r\T$ and $(-1)^{\sf F}$ leads to an overall factor of 4 relative to the two-boundary wormhole with no symmetries:\footnote{This factor is analogous to that found for JT gravity with both $\T$ and $(-1)^{{\sf F}}$ symmetry in \cite{Stanford:2019vob}.}
\begin{equation} \label{eq: fermionic wormhole RT}
    Z_{\mathfrak{s}_1\mathfrak{s}_2}(\tau_1,\bar\tau_1,\tau_2,\bar\tau_2) = 4\,\,  Z_0(\tau_1)Z_0(\tau_2) \sum_{\gamma\in G_{\mathfrak{s}_1 \mathfrak{s}_2}} \frac{\Im(\tau_1) \Im(\gamma\tau_2)}{2\pi^2|\tau_1+\gamma\tau_2|^2} \,,
\end{equation}
where $G_{\mathfrak{s}_1 \mathfrak{s}_2}$ is the relevant congruence subset for boundaries of spin structure $\mathfrak{s}_1$ and $\mathfrak{s}_2$, for $\mathfrak{s}_1, \mathfrak{s}_2\in\{ \R+, \R-, \NS+, \NS-\}$, and recall $Z_0(\tau)= 1/ \sqrt{\text{Im}(\tau)} |\eta(\tau)|^2$ takes into account descendants in a modular invariant way. This factor of 4 appeared in the two-boundary wormholes mentioned in \eqref{eq:NSMNSM} and in \eqref{eq:mixed wormholes}.

We study the spectral form factor of the primaries by multiplying by $|\eta(\tau_1)|^2|\eta(\tau_2)|^2$ then analytically continuing $y_1=\frac{1}{2\pi}(\beta+\ii t)$ and $y_2=\frac{1}{2\pi}(\beta-\ii t)$:
\begin{equation} \label{eq:SFF}
\begin{split}
    K^{(j_1,j_2)}_{\mathfrak{s}_1 \mathfrak{s}_2}(\beta,t) \equiv \frac{1}{w^2}\int_0^w \d x_1 &\int_0^w \d x_2 e^{-2\pi\ii j_1x_1} e^{-2\pi\ii j_2x_2} |\eta(\tau_1)|^2|\eta(\tau_2)|^2 \\
    & \times Z_{\mathfrak{s}_1\mathfrak{s}_2}(\tau_1,\bar\tau_1,\tau_2,\bar\tau_2) \Big|_{{\rm Im}(\tau_1)=\frac{1}{2\pi}(\beta+\ii t)\,,~{\rm Im}(\tau_2)=\frac{1}{2\pi}(\beta-\ii t)}
\end{split}
\end{equation}
where $w$ is the width of the cusp at infinity: $w=1$ in the $\R$ Hilbert space and $w=2$ in the $\NS$ Hilbert space. Using \eqref{eq:ZNSMNSMDC}, we see that in the $\NS$ Hilbert space, the basis in which the Hamiltonian is block-diagonal involves blocks labeled by half-integer spin $j\in\frac12\Z$, each of which obeys the late time behavior
\begin{equation} \label{eq: NS late time}
\begin{split}
    &K^{(j,j)}_{\mathfrak{s}\mathfrak{s}}(\beta,t) \sim \frac{t}{2\pi\beta} e^{-2\beta|j|} \quad ({\rm GOE}) \\
    &K^{(j_1,j_2)}_{\mathfrak{s}\mathfrak{s}}(\beta,t) \sim 0\,,\quad j_1\neq j_2 \,.
\end{split}
\end{equation}
for $\mathfrak{s}\in\{\NS+,\NS-\}$. On the other hand, in the $\R$ Hilbert space, we see from \eqref{eq:R poisson} that we label the blocks by integer spin $j\in\Z$, with an extra factor of 2 in the late time behavior relative to the $\NS$ result:
\begin{equation} \label{eq: R late time}
\begin{split}
    &K^{(j,j)}_{\mathfrak{s}\mathfrak{s}}(\beta,t) \sim \frac{t}{\pi\beta}e^{-2\beta|j|} \quad (2\,\text{GOE}) \\
    &K^{(j_1,j_2)}_{\mathfrak{s}\mathfrak{s}}(\beta,t) \sim 0 \,,\quad j_1\neq j_2
\end{split}
\end{equation}
for $\mathfrak{s}\in\{\R+,\R-\}$. In the $\R$ Hilbert space, the RMT interpretation of this additional factor of 2 arises from the further decomposition of each fixed spin block into eigenvalues of $(-1)^{{\sf F}}=\left(\begin{smallmatrix}I&0\\0&-I\end{smallmatrix}\right)$. For the two-boundary wormhole, pairs of states with eigenvalue $1$ or $-1$ of $(-1)^{{\sf F}}$ can contribute, leading to an additional factor of 2. This does not apply to the $\NS$ Hilbert space, for which $(-1)^{{\sf F}}=e^{2\pi\ii J}$ is already embedded in the spin group.

The linear in time ramp is a universal feature of the spectral form factor in RMT that results from a Fourier transform of the $1/(E_1-E_2)^2$ contribution to the connected density-density correlator $\langle\rho(E_1)\rho(E_2)\rangle$ \cite{Dyson:1962es}. This contribution encodes long-range repulsion between energy eigenvalues, and its coefficient provides a hint about the RMT symmetry class. For example, in this case the extra factor of 2 in each fixed eigenvalues block is a signature of GOE/GSE statistics. Since in both Hilbert spaces, each block is invariant under $\r\T$ and $(\r\T)^2=1$, we conclude that the spectral correlations obey GOE statistics and that blocks of different spin $j_1\neq j_2$ have statistically independent eigenvalues.

\subsubsection*{Gravitational Chern-Simons and chiral anomaly}
The classification of topological phases on 3D ${\rm spin}$ manifolds is given by the cobordism group \cite{Kapustin:2014dxa}
\begin{equation} \label{eq:spin cob}
    {\rm TP}^{\rm Spin}_3 = \Z\,,
\end{equation}
which corresponds to adding a gravitational Chern-Simons (CS) term in the bulk \cite{Alvarez-Gaume:1983ihn, Witten:1988hf, Kitaev:2005hzj, Kraus:2005zm}:
\begin{equation}
    I = - \frac{1}{16 \pi G_N} \int_M \d^3 x \, \sqrt{g} (R-2\Lambda) + \ii\pi\Delta c\left[\frac{\eta_{\rm grav}}{2} + \frac{1}{96\pi^2} \int_M \Tr\left(\omega \d\omega + \frac23\omega^3\right) \right]\,,
\end{equation}
for $\Delta c\equiv c_L-c_R$ the difference between left- and right-moving central charge in the boundary dual CFT, for spin connection $\omega$, and eta invariant $\eta_{\rm grav}$ \cite{Witten:1988hf}. If one tries to define such a theory on a three-manifold with boundary, then one runs into a boundary anomaly, which is precisely the anomaly of a 2d field theory with different central charge for left- and right-moving fermionic fields. The gravitational CS term is referred to as the anomaly inflow action of the 2d CFT, and its coupling obeys the quantization condition \cite{Kitaev:2005hzj, Kapustin:2014dxa}
\begin{equation} \label{eq: fermionic Delta c}
    \Delta c = \frac k2\,,\quad k\in\Z \,.
\end{equation}

\subsubsection*{Warm-up: gravitational anomaly in bosonic theories}

We first analyze the bosonic case, which has the same cobordism group as \eqref{eq:spin cob}: ${\rm TP}^{\rm SO}_3 = \Z$, but where the quantum of conductivity $k$ is 16 times larger than in the fermionic case \cite{Kapustin:2014dxa}:
\begin{equation} \label{eq:boson Delta c}
    \Delta c \equiv c_L - c_R = 8k\,,\quad k\in\Z \,.
\end{equation}
The gravitational CS term affects both the solid torus partition function of \cite{Maloney:2007ud} and the two-boundary wormhole of \cite{Cotler:2020ugk}. For the solid torus, the same derivation that the gravitational path integral is the modular completion of the vacuum character still holds in presence of the gravitational CS term \cite{Maloney:2007ud}. For example consider
$$
Z_{\rm bos}(\tau) \overset{?}{=} \sum_{\gamma \in \mathbb{Z} \backslash PSL(2,\Z)} \chi^L_{\text{vac}}(\gamma \tau)\bar{\chi}^R_{\text{vac}}(\gamma \bar{\tau}),
$$
where $\chi^L_{\text{var}}$ and $\chi^R_{\text{var}}$ correspond to the Virasoro vacuum characters with central charges $c_L$ and $c_R$. Taking this formula at face value leads immediately to issues. The modular sum $\gamma \in \mathbb{Z}\backslash PSL(2,\Z)$ mods out by left multiplication by $T$, but the path integral over a given solid torus is \textit{not} invariant under this operation
$$
\chi^L_{\text{vac}}(T\gamma \tau)\bar{\chi}^R_{\text{vac}}(T\gamma \bar{\tau}) =e^{-\frac{2\pi\i k}{3}}\, \, \chi^L_{\text{vac}}(\gamma \tau)\bar{\chi}^R_{\text{vac}}(\gamma \bar{\tau}).
$$
The interpretation of this equation is well-known \cite{Witten:1988hf}, in the context of 2d CFTs see \cite{Castro:2014tta}. Acting with $T$ on the left does not change the solid torus but it does change its framing. We propose one should replace the partition function by
\beq\label{eq:framingbos}
Z_{\rm bos}(\tau) = \sum_{\gamma \in \mathbb{Z} \backslash PSL(2,\Z)} e^{\frac{2\pi\i k}{3} \Phi(\gamma)}\, \chi^L_{\text{vac}}(\gamma \tau)\bar{\chi}^R_{\text{vac}}(\gamma \bar{\tau}),
\eeq
where $\Phi(\gamma)$ is the Rademacher function.\footnote{This function is defined by $\Phi(\gamma) = \frac{a+d}{s} - 12 \, \text{sgn}(s)\, {\sf s}(d,|s|)$ for $s\neq 0$ or $b/d$ for $s=0$, where ${\sf s}(d,s)=\frac{1}{4s}\sum_{k=1}^{s-1} \cot \frac{\pi k}{s} \cot \frac{\pi d k}{s}$ is the Dedekind sum. A useful identity is $\Phi(\gamma_1 \gamma_2) = \Phi(\gamma_1) + \Phi(\gamma_2) - 3 \, \text{sgn}(s_1 s_2 s_{12})$ where $s_{1,2}$ is the bottom left entry in $\gamma_{1,2}$ and $s_{12}$ that of $\gamma_1\gamma_2$, see \cite{jeffrey2010etainvariantsanomaliesu1chernsimonstheory} for example.
} The phase arises from the gravitational CS term, and the dependence on the modular image can be diagnosed by evaluating its contribution in lens space. The result is given in Proposition 32 of \cite{jeffrey2010etainvariantsanomaliesu1chernsimonstheory}. One can show that $\Phi(T \gamma) = \Phi(\gamma)+1$ and therefore the quotient in the modular sum is well-defined. 

\smallskip

Relatedly, the gravitational CS term leads to a shift in the angular momentum quantization. Start with equation \eqref{eq:framingbos} for $Z_{\text{bos}}(T \tau)$ which involves the vacuum characters evaluated at $\gamma T \tau$. Then shift $\gamma \to T \gamma T^{-1}$ and use $\Phi(\gamma) = \Phi ( T \gamma T^{-1})$. Using the transformation of the vacuum character we get 
\beq \label{eq:bos spin quant}
Z_{\text{bos}}(\tau+1) = e^{- \frac{2\pi\i k}{3}}\, Z_{\text{bos}}(\tau),~~~~\Rightarrow~~~~j \in \Z - \frac{k}{3}.
\eeq
We can verify the transformation law under S transformation as well. For that we can compute $Z_{\text{bos}}(S\tau)$ and shift $\gamma \to \gamma S^{-1}$. Then we can use the fact that $\Phi(\gamma S ) = \Phi(\gamma) ~\text{mod}~3$ (which can be proven using the composition formula of the Rademacher function) to conclude that $Z_{\text{bos}}(-1/\tau)=Z_{\text{bos}}(\tau)$. This is consistent with the analysis of how modular invariance is affected by gravitational anomalies in \cite{Seiberg:2018ntt, Chang:2020aww}.

\smallskip

Next, we analyze the effect of the gravitational CS term on the two-boundary torus wormhole. Recall that in bosonic 3d gravity, the two-boundary wormhole in the presence of $\r\T$ symmetry is given by \cite{Cotler:2020ugk}
\begin{equation} \label{eq: bosonic wormhole RT}
    Z_{\rm bos}(\tau_1,\bar\tau_1,\tau_2,\bar\tau_2) =  2\,Z_0(\tau_1) Z_0(\tau_2) \sum_{\gamma\in PSL(2,\Z)} \frac{\Im(\tau_1)\Im(\gamma\tau_2)}{2\pi^2|\tau_1+\gamma\tau_2|^2} \,.
\end{equation}
where the factor of two comes from gauging $\r\T$ \cite{Yan:2023rjh}. One can generalize the calculation done in \cite{Cotler:2020ugk} in a straightforward fashion when the left- and right-moving central charges are different. We find the final result, for a given modular image, to be identical. But in the presence of the TQFT we also need to incorporate the fact that its partition function depends on $\gamma$ in the same way controlled by lens space. This implies when summing over modular images we need to add the same phase factor as in the solid torus \eqref{eq:framingbos}:
\begin{equation}
    Z_{\rm bos}(\tau_1,\bar\tau_1,\tau_2,\bar\tau_2, k) =  Z_0(\tau_1) Z_0(\tau_2) \sum_{\gamma\in PSL(2,\Z)} \frac{\Im(\tau_1) \Im(\gamma\tau_2)}{\pi^2|\tau_1+\gamma\tau_2|^2} \,  e^{\frac{2\pi\i k}{3} \Phi(\gamma)} \,.
\end{equation}
We can check that this expression is compatible with the anomaly in angular momentum quantization using the fact that $\Phi(T^n \gamma T^m) = \Phi(\gamma) + n + m$. This results in a shift by $k/3$ in the eigenvalues of spin in the late-time limit after analytic continuation to Lorentzian signature.

It would be interesting to extend RMT$_2$ to incorporate anomalies in modular transformations, perhaps by adding the Rademacher phases in the definition of the Eisenstein series, but we leave it for future work.

\subsubsection*{Gravitational anomaly in fermionic theories}

Similar arguments hold for the fermionic solid torus partition functions, with $k$ related to central charge via \eqref{eq: fermionic Delta c}. For example,
\beq\label{eq:fer framing}
Z_{\NS + }(\tau) = \sum_{\gamma \in \mathbb{Z} \backslash \Gamma^0(2)} e^{\frac{\i \pi k}{24} \Phi_\spin(\gamma)}\, \chi^L_{\text{vac}}(\gamma \tau)\bar{\chi}^R_{\text{vac}}(\gamma \bar{\tau}).
\eeq
The path integral of the TQFT on lens space is denoted by $\Phi_\spin(\gamma)$. We will not attempt to compute it explicitly, but there are some general properties one can immediately infer. For the sum over saddles to be compatible with the coset structure the partition function must satisfy $\Phi_\spin(T\gamma) = \Phi_\spin(\gamma)+1$. The transformation of the vacuum character results in
\beq
Z_{\NS+}(\tau+2) = e^{- \frac{\i \pi k}{12}}\, Z_{\NS+}(\tau),~~~~\Rightarrow~~~~j_{\NS} \in \frac{\mathbb{Z}}{2}- \frac{k}{48}.
\eeq
We can also apply a similar procedure to see how the relation between $\NS+$ and $\NS-$ is modified. Using that $T^{-1} \Gamma^0(2) T = \Gamma_\theta$ we find $Z_{\NS+}(\tau+1) = e^{-\i \pi k/24}\, Z_{\NS-}(\tau)$. The spin-statistics relation in the NS Hilbert space becomes anomalous
\beq
\text{NS:}~~~~~~(-1)^{\sf F} = e^{2\pi \i J} e^{\frac{\i \pi k}{24}}.
\eeq 
We can repeat this analysis in the R sector. In this case the sum is over $\gamma\in \Gamma_0(2)$ acting on $\tau$ and can be written as a sum over the $S \gamma$ images of thermal AdS. To compute $Z_{\R-}(\tau+1)$ we can shift 
$
    \gamma \to S^{-1} T^2 S \gamma T^{-1}
$
and use that $\Phi_\spin(T^2 S \gamma T^{-1}) = \Phi_\spin(S\gamma) +4$. We get
\beq
Z_{\R-}(\tau+1) = e^{\frac{\i \pi k}{12}} Z_{\R-}(\tau),~~~~\Rightarrow~~~~j_{\R} \in \mathbb{Z}+ \frac{k}{24}.
\eeq

\smallskip
The gravitational CS term modifies the two-boundary torus wormhole in a similar manner,\footnote{The overall factor of four relative to \cite{Cotler:2020ugk} is the same factor as found in \eqref{eq: fermionic wormhole RT} because the topological term given by gravitational CS corresponded to a perturbative anomaly, which assigned the same weight to each global symmetry insertion in Figure \ref{fig:2torus}.}
\beq \label{eq:CJGCS}
    Z_{\mathfrak{s}_1\mathfrak{s}_2}(\tau_1,\tau_2) =  Z_0(\tau_1) Z_0(\tau_2) \sum_{\gamma \in G_{\mathfrak{s}_1\mathfrak{s}_2}}\frac{2}{\pi^2}\,\frac{\text{Im}(\tau_1)\text{Im}(\gamma \tau_2)}{|\tau_1 + \gamma \tau_2|^2}\,e^{\frac{\i \pi k}{24} \Phi_\spin(\gamma)}\,. 
\eeq
After analytic continuation $y_1=\frac{1}{2\pi}(\beta+\ii t)$ and $y_2=\frac{1}{2\pi}(\beta-\ii t)$, the late-time limit is dominated by $T$ or $T^2$ images and therefore the result is fixed just using the properties of $\Phi_\spin(\gamma)$ under $T$ multiplication:
\begin{equation}
\begin{split}
    |\eta(\tau_1)|^2|\eta(\tau_2)|^2 Z_{\mathfrak{s} \mathfrak{s}}(\tau_1,\tau_2, k) &\supset \frac{2}{\pi^2} \sum_{n\in\Z} \frac{\sqrt{y_1y_2}}{|\tau_1+\tau_2+wn|^2} e^{\frac{2\pi\i k}{48}n} \\
    &= \frac{2}{w\pi} \sum_{j\in\Lambda_{\mathfrak{s}}} \frac{\sqrt{y_1y_2}}{y_1+y_2} e^{2\pi\ii j(x_1+x_2) - 2\pi|j|(y_1+y_2)} \,,
\end{split}
\end{equation}
where $w=2$ and $\Lambda_{\mathfrak{s}}=\frac12\Z-\frac{k}{48}$ in the $\NS$ Hilbert space, while $w=1$ and $\Lambda_{\mathfrak{s}}=\Z+\frac{k}{24}$ in the $\R$ Hilbert space. This leads to the same late-time limit as in \eqref{eq: NS late time} and \eqref{eq: R late time}, but with the anomalous spin quantization condition, which reduces to integer and half-integer spin when $\Delta c\in 24\Z$. 

\subsection{Incorporating $(-1)^{\sf F}$, ${\sf RT}$ and parity} \label{subsec: T}
We now consider theories with time-reversal symmetry $\T$ and parity $\r$ in addition to $\r\T$. On the gravity side, this corresponds to allowing non-orientable fillings of the two-boundary wormhole. There are eight possible symmetry defects we could insert in Figure \ref{fig:2torus}, consisting of products of $\{\mathds{1}, (-1)^{\sf F}, \r\T, \T\}$. With the insertion of a $\T$ defect, we divide $\mathbb{R}^3$ by the symmetries
\begin{equation} \label{eq: KBS1 T}
    T_x(x,y,z)=(x+1,y,z)\,,\quad T_y(x,y,z)=(x,y+1,z)\,,\quad T_z(x,y,z)=(-x,y,z+1)\,,
\end{equation}
which corresponds to $\KB\times S^1$, a non-orientable 3-manifold involving the Klein bottle $\KB$. With the insertion of an $\r\T\circ\T\equiv\r$  defect, we divide $\mathbb{R}^3$ by the symmetries
\begin{equation} \label{eq: KBS1 R}
    T_x(x,y,z)=(x+1,y,z)\,,\quad T_y(x,y,z)=(x,y+1,z)\,,\quad T_z(x,y,z)=(x,-y,z+1)\,,
\end{equation}
which also corresponds to KB$\times S^1$. 

When gauging $\r$ and $\T$, we have to account for the way they act on the torus modular parameter,\footnote{Acting with $\T$ on the geometry also complex conjugates the partition function, but since the two-boundary partition function \eqref{eq: fermionic wormhole RT} is already real, we will not worry about this here.}
\begin{equation}
    (\tau,\bar\tau) \mapsto (-\bar\tau,-\tau) \,.
\end{equation}
In this subsection we write the full two-boundary partition functions in which $\r\T$, $\r$, and $\T$ symmetry are gauged for the cases of (1) bosonic theory with $\T^2=1$, (2) fermionic theory with $\T^2=1$, (3) fermionic theory with $\T^2=(-1)^{\sf F}$.

\subsubsection*{Bosonic case with ${\sf R}/{\sf T}$}
There are no nontrivial topological phases for bosons with $\T$ symmetry since the cobordism group for 3D manifolds with spacetime symmetry $O(3)$ is given by \cite{Thom:1954}
\begin{equation}
    {\rm TP}^{\rm O}_3 = 0\,.
\end{equation}
Then the wormhole with both $\r\T$ and $\T$ gauged weighs all contributions equally:
\begin{equation} \label{eq:Z bos T}
    Z_{\rm bos, \T}(\tau_1,\bar\tau_1,\tau_2,\bar\tau_2) = Z_{\rm bos}(\tau_1,\bar\tau_1,\tau_2,\bar\tau_2) + Z_{\rm bos}(\tau_1,\bar\tau_1,-\bar\tau_2,-\tau_2) \,.
\end{equation}
where the first term includes the two contributions corresponding to no defect insertion \eqref{eq:T3} and $\r\T$ insertion \eqref{eq:T3 RT}, while the second term includes the two contributions corresponding to $\T$ defect insertion \eqref{eq: KBS1 T} and combined $\T$ and $\r\T$ insertion \eqref{eq: KBS1 R}.

After analytic continuation $y_1=\frac{1}{2\pi}(\beta+\ii t)$ and $y_2=\frac{1}{2\pi}(\beta-\ii t)$, we would like to write down the RMT statistics for each fixed spin $j$, where the algebra involving $J$ and the discrete spacetime symmetries is given by (see section \ref{subsec:anomalies})
\begin{equation}
    \{J,\T\}=0\,,\quad \{J,\r\}=0\,,\quad [\r,\T] = 0\,,\quad [J,\r\T]=0 \,.
\end{equation}
Only the spin $j=0$ block can be further decomposed into even and odd parity blocks corresponding to eigenvalues of $\r$. The structure of the Hilbert space is 
$$
\mathcal{H}=\mathcal{H}_{0}^{{\sf R}=1}\oplus \mathcal{H}_{0}^{{\sf R}=-1}   \bigoplus_{j\neq 0} \mathcal{H}_{j}
$$ The anti-unitary symmetry $\T$ is not preserved in any block of the Hamiltonian: $\T$ swaps the parity even and odd blocks for $j=0$, and it swaps the spin $\pm j$ blocks for $j\neq0$. 
For nonzero spin, $\r$ and $\T$ identify the Hamiltonian in a sector of spin $j$ with another of spin $-j$.

For the two-boundary wormhole, in the spin $j=0$ sector, pairs of states with opposite parity make no contribution, while pairs of states with the same eigenvalue of $\r$ make equal contributions, which leads to an extra factor of $2$.  This correlation arises directly from the second term in \eqref{eq:Z bos T}:
\begin{equation}
    \sum_{n \in \mathbb{Z}} \frac{\text{Im}(\tau_1)\text{Im}(-\bar\tau_2+n)}{|\tau_1 - \bar\tau_2+n|^2} =\pi \sum_{j \in  \mathbb{Z}} \frac{y_1 y_2}{y_1 + y_2} e^{-2\pi |j| (y_1+y_2)} e^{2\pi \i j (x_1 - x_2)} \,.
\end{equation}
In summary, at late times, the spectral form factor simplifies to
\begin{equation}
\begin{split}
    &K^{(0,0)}_{\rm bos, \T}(\beta,t) \sim \frac{t}{\pi\beta} e^{-2\beta|j|} \quad (2\,\text{GOE}) \\
    &K^{(j,\pm j)}_{\rm bos, \T}(\beta,t) \sim \frac{t}{2\pi\beta} e^{-2\beta|j|} \,,\quad j\neq 0 \quad ({\rm GOE}) \\
    &K^{(j_1,j_2)}_{\rm bos, \T}(\beta,t) \sim 0\,,\quad |j_1|\neq|j_2|
\end{split}
\end{equation}
for $j\in\Z$.

\subsubsection*{Fermionic case with $\T^2=1$ at the classical level}
In the presence of fermion number symmetry $(-1)^{\sf F}$ and time-reversal symmetry $\T$ satisfying $\T^2=1$, the relevant classification of topological phases is on 3d ${\rm pin^-}$ manifolds, given by the trivial cobordism group \cite{Kapustin:2014dxa}
\begin{equation} \label{eq: cobordism fermions T^2=1}
    {\rm TP}^{\rm pin^-}_3 = 0 \,,
\end{equation}
so there are no nontrivial topological phases. Therefore both the solid torus and the torus wormhole solutions weigh all fillings equally. The $\pinm$ torus wormhole is simply given by
\begin{equation} \label{eq:Z pinm}
    Z^{\pinm}(\tau_1,\bar\tau_1,\tau_2,\bar\tau_2) = Z^{\rm spin}(\tau_1,\bar\tau_1,\tau_2,\bar\tau_2) + Z^{\rm spin}(\tau_1,\bar\tau_1,-\bar\tau_2,-\tau_2) \,,
\end{equation}
where the first term arises from insertions of $\r\T$ and $(-1)^{\sf F}$ defects while the second term arises from the additional insertion of a $\T$ defect. Next we take the double cone limit and compare with the relevant RMT ensemble.

After analytic continuation, the late-time limit in the $\NS$ Hilbert space is identical to the bosonic case, only with $j\in\frac12\Z$,
\begin{equation} \label{eq: late time fermions T NS}
\begin{split}
    &K^{(0,0)}_{\mathfrak{s}\mathfrak{s}}(\beta,t) \sim \frac{t}{\pi\beta}e^{-2\beta|j|} \quad (2\,\text{GOE}) \\
    &K^{(j,\pm j)}_{\mathfrak{s}\mathfrak{s}}(\beta,t) \sim \frac{t}{2\pi\beta}e^{-2\beta|j|}\,,\quad j\neq 0 \quad ({\rm GOE}) \\
    &K^{(j_1,j_2)}_{\mathfrak{s}\mathfrak{s}}(\beta,t) \sim 0\,,\quad |j_1|\neq|j_2|
\end{split}
\end{equation}
for $\mathfrak{s}\in\{\NS+,\NS-\}$. On the other hand, the R Hilbert space has a slightly different decomposition $$
\mathcal{H}_R=\mathcal{H}_{R,0}^{\text{bos,{\sf R}=1}}\oplus \mathcal{H}_{R,0}^{\text{bos,{\sf R}=-1}} \oplus \mathcal{H}_{R,0}^{\text{fer,{\sf R}=1}}\oplus \mathcal{H}_{R,0}^{\text{fer,{\sf R}=-1}}  \bigoplus_{j\neq 0} \mathcal{H}_{R,j}^{\text{fer}}\oplus\mathcal{H}_{R,j}^{\text{bos}}.
$$
The late-time limit in the $\R$ Hilbert space has an extra factor of two from gauging $(-1)^{\sf F}$, with now $j\in\Z$,
\begin{equation} \label{eq: late time fermions T R}
\begin{split}
    &K^{(0,0)}_{\mathfrak{s}\mathfrak{s}}(\beta,t) \sim \frac{2t}{\pi\beta}e^{-2\beta|j|} \quad (4\,\text{GOE}) \\
    &K^{(j,\pm j)}_{\mathfrak{s}\mathfrak{s}}(\beta,t) \sim \frac{t}{\pi\beta}e^{-2\beta|j|}\,,\quad j\neq 0 \quad (2\,\text{GOE}) \\
    &K^{(j_1,j_2)}_{\mathfrak{s}\mathfrak{s}}(\beta,t) \sim 0\,,\quad |j_1|\neq|j_2|
\end{split}
\end{equation}
for $\mathfrak{s}\in\{\R+,\R-\}$. Since $[\r\T,(-1)^{\sf F}]=[\r\T,J]=0$, $\r\T$ is an anti-unitary symmetry that squares to 1 in each block of the Hamiltonian, and so all cases with $|j_1|=|j_2|$ again obey GOE statistics. From the RMT perspective, the extra factor of 2 in the $\R$ Hilbert space SFF arises from the sum of the bosonic and fermionic sectors. For both Hilbert spaces, the first line gets contributions from the two even/odd parity blocks at spin 0 resulting in twice the GOE answer. In the double-cone limit all sectors are statistically independent. 

\subsubsection*{$\T^2=(-1)^{\sf F}$ at the classical level}

The cases with time-reversal symmetry so far involved summing over orientable and non-orientable bulk manifolds with the same weight assigned to each, in other words with a trivial topological field theory in the bulk. The discussion for bulk ${\rm pin^+}$ is more subtle due to the existence of a finite-order nontrivial cobordism group \cite{Kapustin:2014dxa}:
\begin{equation} \label{eq: cobordism fermions T}
    {\rm TP}^{\rm pin^+}_3 = \Z_2\,.
\end{equation}
This results in two different ways to sum over ${\rm pin^+}$ structures in the bulk: one in which we assign the same weight to each 3-manifold, and one in which we include the factor $(-1)^\zeta$, where $\zeta$ is the mod 2 index of the Dirac operator on the 3-manifold. The mod 2 index is a topological invariant that counts the number of zero-modes mod 2 of the Dirac equation $\bar D\lambda=0$.

\smallskip

We briefly analyze the possible effect this $\mathbb{Z}_2$ invariant can have on the solid torus. As we explained earlier, its contribution to the sum over images can be diagnosed by studying lens space. To evaluate the invariant we recall the following isomorphism \cite{Kirby:1991}
\beq \label{eq:smith pinp}
\text{TP}_3^{\pinp} \to \text{TP}_2^{\spin}.
\eeq
Indeed the 3d $\pinp$ topological invariant can be written, schematically, as \cite{Kapustin:2014dxa, Wan:2019soo}
\begin{equation} \label{eq:PD w1}
    \zeta[M^3] \equiv \int_{M^3} w_1 \cup \text{Arf} = \int_{\PD(w_1)} \text{Arf} \equiv \text{Arf}[\PD(w_1)]
\end{equation}
where $w_1$ is the first Stiefel-Whitney class, measuring the obstruction to orientability and $\PD(w_1)$ is the Poincaré dual to $w_1$. This expression makes manifest that to evaluate $\zeta$ we should compute the mod 2 (Arf) invariant of the spin structure on the 2-manifold $\PD(w_1)$. If we try to apply this to lens space we immediately realize that it is orientable and therefore $w_1=0$, implying the Poincaré dual is empty and therefore the invariant is trivial.   

\smallskip

The next step is to study the path integral on the torus wormhole first without and then with the addition of the $\mathbb{Z}_2$ TQFT. The results are the following. 

\bigskip

\noindent \textbf{Trivial topological field theory}: Similarly to the case of $\pinm$, we have
\begin{equation}
    Z^{\pinp}(\tau_1,\bar\tau_1,\tau_2,\bar\tau_2) = Z^{\rm spin}(\tau_1,\bar\tau_1,\tau_2,\bar\tau_2) + Z^{\rm spin}(\tau_1,\bar\tau_1,-\bar\tau_2,-\tau_2) \,,
\end{equation}
where the discrete symmetries satisfy the non-anomalous algebra \cite{Seiberg:2023cdc}
\begin{equation}
    [J,(-1)^{\sf F}] = [\T,(-1)^{\sf F}] = [\r,(-1)^{\sf F}] = \{J,\T\} = \{J,\r\} = [J,\r\T]=0 \,.
\end{equation}
After analytic continuation $y_1=\frac{1}{2\pi}(\beta+\ii t)$ and $y_2=\frac{1}{2\pi}(\beta-\ii t)$, the late-time limit is identical to \eqref{eq: late time fermions T R} and \eqref{eq: late time fermions T NS}. The surviving anti-unitary symmetry in each block of the Hamiltonian is $\r\T$, so RMT statistics are still in the GOE class for $|j_1|=|j_2|$.

\bigskip

\noindent \textbf{Non trivial topological field theory}: We begin by briefly reviewing the mod 2 index $\zeta$ for 3-manifolds, which was computed in \cite{Witten:2015aba}. Since $\zeta$ vanishes on orientable 3-manifolds, the distinction from the trivial topological field theory only appears for non-orientable 3-manifolds. In particular, for the two-boundary torus wormhole with insertion of $G\in\{\mathds{1}, (-1)^{\sf F}, \r\T\}$ defects, the 3-manifold is orientable so $\zeta=0~\text{mod}~2$. It only remains to probe the mod 2 index for the $\KB\times S^1$ topology, corresponding to either $\r$ or $\T$ defects, both with and without the $(-1)^{\sf F}$ defect.

Without loss of generality, consider the case of $\T$ reflection, where the 3-manifold formed by gluing the outer torus boundaries can be described by dividing $\mathbb{R}^3$ by the symmetries
\begin{equation}
    T_x(x,y,z)=(x+1,y,z)\,,\quad T_y(x,y,z)=(x,y+1,z)\,,\quad T_z(x,y,z)=(-x,y,z+1)\,.
\end{equation}
resulting in the $\KB\times S^1$ topology. A fermionic field would obey the periodicities,
\begin{equation} \label{eq: 3D kb periodicities}
\begin{split}
    \lambda(x+1,y,z)&=(-1)^\alpha \lambda(x,y,z)\,,\quad \lambda(x,y+1,z)=(-1)^\beta \lambda(x,y,z)\\
    &\lambda(-x,y,z+1) = (-1)^\mu \gamma_x\lambda(x,y,z)\,,
\end{split}
\end{equation}
where $(\alpha,\beta)$ control the periodicities of the external boundaries, $\mu$ controls whether fermions pick up an extra sign when parallel-transported across the wormhole, and $\gamma_x^2=1$ on pin$^+$ manifolds. In the case of the $(\alpha,\beta)=(0,0)$ spin structure, i.e. $\R+$ on both boundaries, there is a zero-mode for $\mu=0,1$ whose restriction to $\KB$ is the zero-mode identified in \cite{Witten:2015aba, Stanford:2019vob}.\footnote{A single zero-mode is present both with and without insertion of $(-1)^{\sf F}$: for $\mu=0$, the zero-mode exists for the $+1$ eigenvalue of $\gamma_x$, while for $\mu=1$, the zero-mode exists for the $-1$ eigenvalue of $\gamma_x$. On a pin$^-$ manifold, $\gamma_x^2=-1$, so there would be no zero-mode.} Then $\zeta=1~\text{mod}~2$ for the non-orientable wormhole with two $\R+$ boundaries, and $\zeta=0~\text{mod}~2$ for all other cases. This result can be reproduced using the isomorphism \eqref{eq:smith pinp} of $\text{TP}_3^{\pinp}$ to $\text{TP}_2^{\spin}$, in which case $\alpha$ and $\beta$ determine the periodicities of the Poincaré dual 2-torus and the computation reduces to studying the Arf invariant on this 2-torus.

The two-boundary path integral in the presence of $(-1)^\zeta$ is given by
\begin{equation} \label{eq: Z pin+ zeta}
    Z^{\pinp,\,\zeta} = \left\{\begin{matrix}
        Z^{\spin}(\tau_1,\bar\tau_1,\tau_2,\bar\tau_2) - Z^{\spin}(\tau_1,\bar\tau_1,-\bar\tau_2,-\tau_2) &\quad \R+,\R+ \\
        Z^{\spin}(\tau_1,\bar\tau_1,\tau_2,\bar\tau_2) + Z^{\spin}(\tau_1,\bar\tau_1,-\bar\tau_2,-\tau_2) &\quad \text{else}\,.
    \end{matrix} \right. 
\end{equation}
The algebra of discrete symmetries is anomalous in the $\R$ Hilbert space \cite{Seiberg:2023cdc}:
\begin{equation} \label{eq: R sector RMT algebra}
    [J,(-1)^{\sf F}] = \{\T,(-1)^{\sf F}\} = \{\r,(-1)^{\sf F}\} = \{J,\T\} = \{J,\r\} = [J,\r\T]=0
\end{equation}
where the relationship between $(-1)^{\sf F}$ and $\T,\r$ is anomalous. Since $J$ commutes with $(-1)^{\sf F}$ we can still decompose the Hilbert space according to $j$ and fermion parity. Since $\r\T$ still commutes with $J$ and $(-1)^{\sf F}$, the fixed spin random matrix statistics are given by GOE blocks with fixed overall fermion number. Between two boundaries with the same $j\neq0$, we see from \eqref{eq: R sector RMT algebra} that bosons and fermions are not correlated. This happens because $\r$ anticommutes with $(-1)^{\sf F}$ and $J$ and a bosonic block with spin $j$ gets mapped to a fermionic state with spin $-j$. The correlator between same $j$ is given by
\begin{equation}
    Z_{\R\pm\R\pm}^{(j,j)} = \int \d E_1 \d E_2 \, \left\langle \rho_{\R}^{\text{bos}}(E_1,j) \rho_{\R}^{\text{bos}}(E_2,j) + \rho_{\R}^{\text{fer}}(E_1,j) \rho_{\R}^{\text{fer}}(E_2,j)\right\rangle \chi_{E_1,j}(\tau_1) \chi_{E_2,j}(\tau_2)
\end{equation}
while the correlation between boundaries with spin $\pm j$ is only between bosons and fermions:
\begin{equation}
\begin{split}
    Z_{\R\pm\R\pm}^{(j,-j)} = \mp \int \d E_1 \d E_2 \, &\left\langle \rho_{\R}^{\text{bos}}(E_1,j) \rho_{\R}^{\text{fer}}(E_2,-j) + \rho_{\R}^{\text{fer}}(E_1,j) \rho_{\R}^{\text{bos}}(E_2,-j) \right\rangle \\
    &\qquad\times \chi_{E_1,j}(\tau_1) \chi_{E_2,-j}(\tau_2) \,.
\end{split}
\end{equation}
When $j=0$ the situation is slightly different. The two terms in \eqref{eq: Z pin+ zeta} cancel each other, leading to a vanishing spectral form factor. $J$ acts trivially but we still have ${\sf R}$ anticommuting with $(-1)^{\sf F}$. The Hilbert space decomposes as $\mathcal{H}_{R,0} = \mathcal{H}_{R,0}^{\text{bos}}\oplus \mathcal{H}_{R,0}^{\text{fer}}$ (instead of the four sectors that appear in the theory without the TQFT). Parity, instead of introducing a separate grading, identifies the two blocks $H_{R,0} = \left( \begin{smallmatrix}
    \text{GOE} & 0\\
    0 & \text{GOE}
\end{smallmatrix} \right)$. At late times, for two $\R+$ torus boundaries, we find\footnote{For $j_1=j_2=0$, the two-boundary wormhole vanishes exactly, not just in the late time limit.}
\begin{equation}
\begin{split}
    &K^{(j_1,j_2)}_{\R+\R+}(\beta,t) \sim 0\,,\quad j_1=j_2=0 \text{ or } |j_1|\neq|j_2| \\
    &K^{(j,j)}_{\R+\R+}(\beta,t) \sim \frac{t}{\pi\beta}e^{-2\beta|j|}\,,\quad j\neq 0 \quad (2\,\text{GOE}) \\
    &K^{(j,-j)}_{\R+\R+}(\beta,t) \sim -\frac{t}{\pi\beta}e^{-2\beta|j|}\,,\quad j\neq 0 \quad (2\,\text{GOE}) \,,
\end{split}
\end{equation}
while the $\R-,\NS\pm$ late-time limits are still given by \eqref{eq: late time fermions T R} and \eqref{eq: late time fermions T NS}. The first line is consistent with the two GOE copies which cancel each other if $(-1)^{\sf F}$ is inserted. The last line is negative since the contribution comes from terms with product of bosonic and fermionic spectrum in the $\R$ Hilbert space. 

\subsection{Anomalies in fermionic 2d CFT} \label{subsec:anomalies}
Unlike in 2d gravity, we do not have an explicit holographic dual theory for which we can study the effects of the anomaly. However, since the cobordism groups \eqref{eq: cobordism fermions T^2=1}, \eqref{eq: cobordism fermions T}, and \eqref{eq: cobordism fermions (-1)^FL} are realized by the free fermion SPT phases, we show an example 2d field theory realization of the anomalies using free fermions. The discussion here is mostly a review of \cite{Seiberg:2023cdc}, with a modified focus to illustrate the cases of time-reversal symmetry more explicitly. We stress that although the free fermion reproduces the classification of anomalies, its spectrum will not display level repulsion since it is certainly not chaotic.

We will consider a system of $N$ non-chiral massless free Majorana fermions with left- and right- moving components $\psi_L^a$ and $\psi_R^a$ for $a=1,\ldots,N$, with Lagrangian
\begin{equation} \label{eq: massless L}
    L = \ii \int dx \sum_{a=1}^N \left[\psi_L^a(\p_t-\p_x)\psi_L^a + \psi_R^a(\p_t+\p_x)\psi_R^a\right] \,.
\end{equation}
We will study anomalies that arise in the symmetry algebra in the case of time-reversal symmetry on $\pinp$ and $\pinm$. The case with a unitary $\mathbb{Z}_2$ will be the topic of the next section.

\subsubsection*{$\T^2=1$: no anomaly}
The relevant symmetry algebra for $\pinm$ is generated by
\begin{align}\label{eq: T^2=1 sym}
    &(-1)^{\sf F}: \quad \psi_L^a(t,x)\to -\psi_L^a(t,x)\,,\quad \psi_R^a(t,x)\to -\psi_R^a(t,x) \nonumber\\
    &\T: \quad \psi_L^a(t,x)\to -\psi_R^a(-t,x)\,,\quad \psi_R^a(t,x)\to -\psi_L^a(-t,x)\,,\quad \ii\to -\ii \\
    &\r: \quad \psi_L^a(t,x)\to \psi_R^a(t,-x)\,,\quad \psi_R^a(t,x)\to -\psi_L^a(t,-x)\nonumber
\end{align}
and a spatial momentum operator $J=h_L-h_R$.\footnote{The Lagrangian \eqref{eq: massless L} is also invariant under separate left and right fermion number, but for now we consider the restricted group of symmetries in \eqref{eq: T^2=1 sym}.}  These symmetries satisfy the algebra
\begin{equation} \label{eq:T^2=1 algebra}
\begin{split}
    &\left((-1)^{\sf F}\right)^2=1\,,\quad \T^2=1\,,\quad \r^2 = (-1)^{\sf F}\,,\quad (\r\T)^2 = 1 \\
    &[(-1)^{\sf F},\T] = [(-1)^{\sf F},\r] = [(-1)^{\sf F},\r\T] = 0 \\
    &[J,(-1)^{\sf F}] = \{J,\r\} = \{J,\T\} = 0
\end{split}
\end{equation}
for any $N$. These relations hold for both Hilbert spaces, the only difference being that $e^{2\pi\ii J}=(-1)^{\sf F}$ in the $\NS$ Hilbert space, while $e^{2\pi\ii J}=1$ in the $\R$ Hilbert space. The realization of the algebra \eqref{eq:T^2=1 algebra} without obstruction for any $N$ is consistent with the trivial cobordism group \eqref{eq: cobordism fermions T^2=1}.

Since we could add a mass term $\ii m \psi_L^a\psi_R^a$ for even a single fermion, the fermions with symmetries \eqref{eq: T^2=1 sym} are gappable and hence cannot contribute anomalies \cite{Witten:2015aba}.

\subsubsection*{$\T^2=(-1)^{\sf F}$: $\Z_2$ anomaly}
The relevant symmetry algebra for $\pinp$ is generated by
\begin{align} \label{eq: R^2=1 sym}
    &(-1)^{\sf F}: \quad \psi_L^a(t,x)\to -\psi_L^a(t,x)\,,\quad \psi_R^a(t,x)\to -\psi_R^a(t,x) \nonumber\\
    &\T: \quad \psi_L^a(t,x)\to \psi_R^a(-t,x)\,,\quad \psi_R^a(t,x)\to -\psi_L^a(-t,x)\,,\quad \ii\to -\ii  \\
    &\r: \quad \psi_L^a(t,x)\to \psi_R^a(t,-x)\,,\quad \psi_R^a(t,x)\to \psi_L^a(t,-x) \nonumber
\end{align}
and a spatial momentum operator $J=h_L-h_R$.

If $N$ is even, these symmetries satisfy the algebra
\begin{equation} \label{eq: NS algebra}
\begin{split}
    &\left((-1)^{\sf F}\right)^2=1\,,\quad \T^2=(-1)^{\sf F}\,,\quad \r^2 = 1\,,\quad (\r\T)^2 = 1 \\
    &[(-1)^{\sf F},\T] = [(-1)^{\sf F},\r] = [(-1)^{\sf F},\r\T] = 0\,,\quad \r\T = (-1)^{\sf F}\T\r \\
    &[J,(-1)^{\sf F}] = \{J,\r\} = \{J,\T\} = 0
\end{split}
\end{equation}
for both sectors, with again $e^{2\pi\ii J}=(-1)^{\sf F}$ in the $\NS$ Hilbert space, while $e^{2\pi\ii J}=1$ in the $\R$ Hilbert space.

If $N$ is odd, the symmetries \eqref{eq: R^2=1 sym} satisfy the algebra \eqref{eq: NS algebra} in the $\NS$ Hilbert space, while in the $\R$ Hilbert space, they satisfy the algebra
\begin{equation} \label{eq: R algebra}
\begin{split}
    &\left((-1)^{\sf F}\right)^2=1\,,\quad \T^2=\ii(-1)^{\sf F}\,,\quad \r^2 = 1\,,\quad (\r\T)^2 = 1 \\
    &\{(-1)^{\sf F},\T\} = \{(-1)^{\sf F},\r\} = [(-1)^{\sf F},\r\T] = 0\,,\quad \r\T = -\ii(-1)^{\sf F}\T\r \\
    &e^{2\pi\ii J}=1\,,\quad \{J,(-1)^{\sf F}\} = \{J,\r\} = \{J,\T\} = 0\,,
\end{split}
\end{equation}
The $\Z_2$-valued anomaly of \eqref{eq: cobordism fermions T} is detected by the $\R$ Hilbert space algebra \eqref{eq: R algebra}, which only projectively realizes the algebra \eqref{eq: NS algebra}. This anomaly is realized when we have an odd number of fermions, and cannot be more subtle by an argument reproduced in \cite{Witten:2015aba, Stanford:2019vob}: for any pair of fermions we could add the mass term $\ii m(\psi_L^0\psi_R^1 - \psi_L^1\psi_R^0)$ and consider $m\to\infty$ such that this pair of fermions are removed from the low energy effective field theory without breaking any symmetry.

\section{Fermionic 2d CFT with unitary $\Z_2$ symmetry} \label{sec:FL}

In this section, we incorporate a boundary global unitary $\mathbb{Z}_2$ symmetry. The discussion below will apply generally but it is useful to think of the $\mathbb{Z}_2$ symmetry as corresponding to a left-moving fermion parity $(-1)^{\sf F_L}$. This corresponds to adding a $\mathbb{Z}_2$ gauge field into our theory of fermionic 3d gravity.

\subsection{Solid torus}
We consider 3d gravity with fermion parity $(-1)^{\sf F}$ and an additional $\mathbb{Z}_2$ symmetry generated by 
$$
\mathbb{Z}_2:~~~(-1)^{{\sf F_L}},
$$
identified with left-moving fermion number with right-moving fermion number $(-1)^{\sf F_R} \equiv (-1)^{\sf F} (-1)^{\sf F_L}$, at least classically. The bulk theory can be thought of as fermion gravity as in section \ref{sec:Path_integral_fermionic_3d_gravity} now coupled to an extra bulk $\mathbb{Z}_2$ gauge theory \cite{Dijkgraaf:1989pz}.

\smallskip

Consider a single torus boundary with moduli $\tau,\bar{\tau}$. The full specification of this boundary involves a choice of spin structure which we can label as $\mathfrak{s} \in \{ (\mu,\nu) \in (\mathbb{Z}/2\mathbb{Z})^2\}$. $\mu$ describes the sign a spectator fermion gets around the time circle and $\nu$ the sign it gets around the spatial circle. A background gauge field for the global $\mathbb{Z}_2$ symmetry is equivalent to a connection on the principal $\mathbb{Z}_2$-bundle. We can characterize this by the pair 
$$
\mathfrak{c} = \{ (\alpha,\beta) \in (\mathbb{Z}/2\mathbb{Z})^2\},
$$
such that the holonomy around the spatial (temporal) circle is $e^{\pi \i \beta}$ ($e^{\pi \i \alpha}$) in analogy to the conventions we use for the spin structure. A non-trivial holonomy around a cycle means that the periodicity condition for left-moving fermions is flipped.

\smallskip

In its 2d CFT interpretation, the torus has now 4 Hilbert spaces depending on the holonomies and spin structure along the spatial circle. The NSNS Hilbert space corresponds to $\nu=1$ and $\beta=0$, the NSR Hilbert space to $\nu=0$ and $\beta=1$, the RNS Hilbert space to $\nu=\beta=1$, and the RR Hilbert space to $\nu=\beta=0$. Simultaneously we can have either no insertion of fermion parity corresponding to $\mu=1$ and $\alpha=0$, an insertion of $(-1)^{\sf F}$ corresponding to $\mu=\alpha=0$, an insertion of $(-1)^{\sf F_L}$ corresponding to $\mu=\alpha=1$, and an insertion of $(-1)^{\sf F_R}$ corresponding to $\mu=0$ and $\alpha=1$. This results in a total of 16 spin structures and $\mathbb{Z}_2$ connections that can be labeled as $(\mu,\nu;\alpha,\beta)$. We can also label them by the sector and periodicity for left- and right-moving fermion, e.g. $(1,1;0,1)$ corresponds to the partition function $\NS\NS+-$ in the NSNS Hilbert space with an $(-1)^{\sf F_L}$.

\smallskip

We saw there are 16 different torus partition functions. What are their corresponding modular groups? One can answer this question by analyzing the change in both spin structure and $\mathbb{Z}_2$ connection under a modular transformation
\beq
\text{Spin:}~~\left(\begin{array}{c}
     \mu  \\
     \nu
\end{array} \right) \to \left(\begin{array}{cc}
     a & b \\
     \cc & d
\end{array} \right) \left(\begin{array}{c}
     \mu  \\
     \nu
\end{array} \right),~~~~\mathbb{Z}_2:~~\left(\begin{array}{c}
     \alpha  \\
     \beta
\end{array} \right) \to \left(\begin{array}{cc}
     a & b \\
     \cc & d
\end{array} \right) \left(\begin{array}{c}
     \alpha  \\
     \beta
\end{array} \right),
\eeq
with $a,b,s,d\in\mathbb{Z}$ and $ad-sb=1$. In each sector, one should select a subset of all modular transformations that preserve the spin structure and $\mathbb{Z}_2$ connection $(\mu,\nu;\alpha,\beta)$. We analyze two cases below and present the results for the rest in Table \ref{tab:mgflfr}. Equivalently one can look at the right- and left-moving spin structures and take the intersection of their corresponding modular groups.

\smallskip
\begin{table}
    \centering
   \hspace{-0cm} \begin{tabular}{c|cccc}\toprule
         & ${I}$ & $(-1)^{\sf F}$ & $(-1)^{\sf F_L}$ & $(-1)^{\sf F_R}$  \\\midrule\midrule
        NSNS  &  \textcolor{blue}{$\Gamma_\theta$} & \textcolor{blue}{$ \Gamma^0(2)$} & \textcolor{blue}{$\Gamma(2)$} & \textcolor{blue}{$\Gamma(2)$} \\
        NSR  & \textcolor{blue}{$\Gamma(2)$} & \textcolor{red}{$\Gamma^0(2)$} & \textcolor{blue}{$\Gamma(2)$} & \textcolor{red}{$\Gamma_\theta$}\\
        RNS  & \textcolor{blue}{$\Gamma(2)$} & \textcolor{red}{$\Gamma^0(2)$} & \textcolor{red}{$\Gamma_\theta$} & \textcolor{blue}{$\Gamma(2)$} \\
        RR  & \textcolor{blue}{$\Gamma_0(2)$} & \textcolor{red}{$\text{PSL}(2,\mathbb{Z})$} & \textcolor{red}{$\Gamma_0(2)$} & \textcolor{red}{$\Gamma_0(2)$} \\ \bottomrule
    \end{tabular}
    \caption{Table of modular groups leaving each left- and right-moving spin structures invariant. The modular groups colored blue (red) corresponds to boundary torus in the trivial (non-trivial) component of $\Omega_2^{\text{spin}}(B\mathbb{Z}_2)=\mathbb{Z}_2\times \mathbb{Z}_2$. Therefore the partition functions with the modular groups in red vanish non-perturbatively in pure 3d gravity. }
    \label{tab:mgflfr}
\end{table}

Consider the partition function in the NSNS Hilbert space with no fermion parity insertion. In this case $(\mu,\nu;\alpha,\beta)=(1,1;0,0)$. Since $\alpha=\beta=0$, the group preserving the $\mathbb{Z}_2$ connection is the whole modular group $\text{PSL}(2,\mathbb{Z})$. The group preserving the spin structure is $\Gamma_\theta$ since $\mu=\nu=1$. Therefore the modular group is the intersection 
$$
\text{PSL}(2,\mathbb{Z}) \, \cap \, \Gamma_\theta = \Gamma_\theta,
$$
the same that appears for the partition function in the NS Hilbert space of gravity with only the diagonal fermion parity. 

\smallskip

The second case we analyze involves a new ingredient. Consider the partition function in the NSR Hilbert space with no fermion parity insertion, namely $(0,1;1,0)$. The subgroup preserving the spin structure is $\Gamma^0(2)$ while the group preserving the $\mathbb{Z}_2$ connection is $\Gamma_0(2)$. In this case the intersection generates a new modular subgroup
\beq
\Gamma^0(2) \cap \Gamma_0(2) = \Gamma(2) = \left\{\left(\begin{matrix}
    a & b \\
    \cc & d
\end{matrix}\right), b=\cc=0~\text{mod}~2~\&~ a=d=1~\text{mod}~2 \right\}.
\eeq
This subgroup has not appeared in the fermionic theory and only becomes relevant when a $\mathbb{Z}_2$ symmetry is incorporated. Since
\beq
\Gamma^0(2) \cap \Gamma_0(2)=\Gamma^0(2) \cap \Gamma_\theta =\Gamma_0(2) \cap \Gamma_\theta = \Gamma(2),
\eeq
$\Gamma(2)$ is the only new ingredient we need. The other options reduce to the groups that already appeared in the case without any global symmetry.

\smallskip

The calculation of the gravitational path integral on the solid torus is now straightforward. In the NSNS Hilbert space we have thermal AdS and its modular transformations that preserve spin and $\mathbb{Z}_2$ structure
\begin{eqnarray}
  Z_{\NS\NS--} (\tau) &=& \sum_{\gamma \in \mathbb{Z} \backslash \Gamma_\theta} |\chi_{\text{vac}}(\gamma\,\tau)|^2  =Z_{\NS-}(\tau),\\
  Z_{\NS\NS++} (\tau) &=& \sum_{\gamma \in \mathbb{Z} \backslash \Gamma^0(2)} |\chi_{\text{vac}}(\gamma\,\tau)|^2 =Z_{\NS+}(\tau),\\
  Z_{\NS\NS+-} (\tau) &=& \sum_{\gamma \in \mathbb{Z} \backslash \Gamma(2)} |\chi_{\text{vac}}(\gamma\,\tau)|^2 ,\\
    Z_{\NS\NS-+} (\tau) &=& \sum_{\gamma \in \mathbb{Z} \backslash \Gamma(2)} |\chi_{\text{vac}}(T\,\gamma\,\tau)|^2=Z_{\NS\NS+-} (\tau+1).
\end{eqnarray}
The path integral of the bulk $\mathbb{Z}_2$ gauge theory is trivial since, in the presence of boundaries, a global $\mathbb{Z}_2$ transformation is not treated as a gauge symmetry. As we discussed in section \ref{sec:Path_integral_fermionic_3d_gravity} the mod $\mathbb{Z}$ in these expressions will always mean the subset of left $T$ action which preserve spin structure (and in this case $\mathbb{Z}_2$ connection as well). The new restriction $\mathbb{Z}\backslash\Gamma(2)$ identifies $\gamma \sim T^{2\mathbb{Z}} \, \gamma$ for $\gamma\in\Gamma(2)$. We also indicated partition functions that happen to coincide with that of fermionic gravity without any $\mathbb{Z}_2$ symmetries. 

One can prove that not only $Z_{\NS\NS+-}(\tau+1)=Z_{\NS\NS-+}(\tau)$, which is always true, but also the nontrivial relation $Z_{\NS\NS+-}(\tau+1)=Z_{\NS\NS+-}(\tau)$, which is valid for pure gravity. To show this one can evaluate $Z_{\NS\NS+-}(T\tau)$ and move $T$ past $\gamma$ using the identity $T^{-1} \Gamma(2) T = \Gamma(2)$ so that $\gamma T = T \gamma'$ with $\gamma, \gamma'\in\Gamma(2)$. Moreover, since $|\chi_{\text{vac}}(T \tau)|^2 = |\chi_{\text{vac}}(\tau)|^2$ the partition function $Z_{\NS\NS+-}(\tau)$ has period $\tau \to \tau+1$ even though the modular group would suggest $\tau \to \tau+2$. This implies 
$$
Z_{\NS\NS-+}(\tau) = Z_{\NS\NS+-}(\tau),
$$
which will be important later when we discuss the spectrum. This derivation would fail in the presence of 3d matter fields that are fermionic.

\smallskip

Next we consider the partition functions in the NSR Hilbert space. If no fermion parity is inserted then the gravitational path integral gets contributions from the BTZ black hole and its modular images that preserve the spin and $\mathbb{Z}_2$ structure, analogous to the $\R-$ structure in fermionic gravity. The same observation holds in the RNS case. The path integral without insertions is
\begin{eqnarray}
Z_{\NS\R--}(\tau) &=& \sum_{\gamma\in \mathbb{Z}\backslash\Gamma(2)} |\chi_{\text{vac}}(S\,\gamma\,\tau)|^2  = Z_{\R\NS--}(\tau).
\end{eqnarray}The restriction $\mathbb{Z}\backslash\Gamma(2)$ in this equations identifies $\gamma \sim S^{-1}T^{2\mathbb{Z}}S \gamma$. The case of $\NS\R+-$ or $\R\NS-+$ is interesting since neither the BTZ nor the thermal AdS saddles contribute. Instead we need a black hole where the sum over the spatial and temporal circle contracts. This can be achieved by acting on thermal AdS with the transformation
$$
ST = \pm \begin{pmatrix}
    0 & -1\\
    1 & 1
\end{pmatrix},
$$
and the full gravitational path integral is a sum over  $\Gamma(2)$ images of this saddle $Z(\tau) = |\chi_{\text{vac}}(ST\tau)|^2$, namely
\beq
Z_{\NS\R+-}(\tau) =Z_{\R\NS-+}(\tau)= \sum_{\gamma\in \mathbb{Z}\backslash  \Gamma(2)} |\chi_{\text{vac}}(S\,T\,\gamma\,\tau)|^2.
\eeq
The restriction $\mathbb{Z}\backslash\Gamma(2)$ in this equation identifies $\gamma \sim T^{-1}S^{-1} T^{2\mathbb{Z}} S T \gamma$. All the remaining partition functions vanish. We will discuss the implications of this on the spectrum below. One can verify that e.g. the $\NS\R--$ partition function and $\NS\NS-+$ one are related by an S move since $S^{-1} \Gamma(2) S = \Gamma(2)$. One can also check that $Z_{\NS\R--}(\tau+1) =Z_{\NS\R+-}(\tau)$ and similarly $Z_{\R\NS--}(\tau+1) = Z_{\R\NS-+}(\tau)$, using that $T^{-1} \Gamma(2) T = \Gamma(2)$. 

\smallskip

Finally we consider the RR structure. The gravitational path integral gets a contribution from the relevant modular images of BTZ when no fermion parity is inserted
\begin{eqnarray}
Z_{\R\R--}(\tau) &=&\sum_{\gamma\in \mathbb{Z}\backslash\Gamma_0(2)} |\chi_{\text{vac}}(S\,\gamma\,\tau)|^2= Z_{\R-}(\tau), \\
Z_{\R\R-+}(\tau)&=&Z_{\R\R+\pm}(\tau)=0.
\end{eqnarray}
Since this partition function equals the one of gravity with only a spin structure, computed in section \ref{sec:Path_integral_fermionic_3d_gravity}, this case does not require any new calculation. Path integrals in the RR Hilbert space with any fermion parity insertion vanish non-perturbatively since it is a non-trivial element of the bordism group.

\smallskip

We are now ready to analyze the implications on the spectrum of the dual 2d CFT ensemble:

\smallskip

\noindent\textbf{RR:} The Hilbert space decomposes as
\beq \label{eq:H RR}
\mathcal{H}_{\R\R} = \bigoplus_{j\in\mathbb{Z}} \mathcal{H}^{\text{bos/bos}}_j\oplus \mathcal{H}^{\text{bos/fer}}_j\oplus \mathcal{H}^{\text{fer/bos}}_j\oplus \mathcal{H}^{\text{fer/fer}}_j,
\eeq
where $\text{bos/bos}$ means the state is even under both $(-1)^{\sf F_L}$ and $(-1)^{\sf F_R}$, $\text{fer/bos}$ means the state is odd under $(-1)^{\sf F_L}$ and even under $(-1)^{\sf F_R}$, and so on. 
The observation in the previous paragraph automatically implies that the density of states for $j\in\mathbb{Z}$ is given by
\beq \label{eq: (-1)^FL rho avg}
\rho_{\R\R}^{\text{bos/bos}}(E,j) = \rho_{\R\R}^{\text{bos/fer}}(E,j)=\rho_{\R\R}^{\text{fer/bos}}(E,j)=\rho_{\R\R}^{\text{fer/fer}}(E,j) = \frac{1}{2}\, \rho_{\R}(E,j),~~~j\in\mathbb{Z}.
\eeq
 On the right hand side we have the same density computed in section \ref{sec:Path_integral_fermionic_3d_gravity}. There are no states with half-integer spin in the RR Hilbert space and there is no spin-statistics connection. The $\R\R$ spectrum has the same features and pathologies as the $\R$ spectrum of the fermionic theory.

\smallskip

\noindent\textbf{NSR and RNS:} These sectors are twisted under the $\mathbb{Z}_2$ symmetry. Since their partition functions are identical we consider NSR for concreteness. The spectrum now includes both integer and half-integer spin $j$, due to the restriction $\Gamma_0(2) \to \Gamma(2)$ which makes the partition function $\tau\to\tau+2$ periodic, or equivalently $\Omega \to \Omega + 4\pi \i /\beta$, periodic. $Z_{\NS\R--}(\tau+1)=Z_{\NS\R+-}(\tau)$ implies
\beq
\mathcal{H}_{\NS\R}:~~~~e^{2\pi \i J} =(-1)^{\sf F_L},~~~~\mathcal{H}_{\R\NS}:~~~~e^{2\pi \i J} =(-1)^{\sf F_R}.
\eeq
Any state with energy $E$ and spin $j$ therefore comes in 2 copies that we label with `bos' or `fer'. In the $\NS\R$ Hilbert space bos (fer) indicates even (odd) parity with respect to $(-1)^{\sf F_R}$. In the $\R\NS$ Hilbert space bos (fer) indicates even (odd) parity with respect to $(-1)^{\sf F_L}$, namely
\beq\label{eq:NSRHH}
\mathcal{H}_{\NS\R} = \bigoplus_{j\in \mathbb{Z}/2} \mathcal{H}^{\text{bos}}_j\oplus \mathcal{H}^{\text{fer}}_j,~~~~\mathcal{H}_{\R\NS} = \bigoplus_{j\in \mathbb{Z}/2} \mathcal{H}^{\text{bos}}_j\oplus \mathcal{H}^{\text{fer}}_j
\eeq
(If left- or right-moving fermion parity is anomalous, which is a possibility that will arise in the presence of bulk TQFTs, these labels might not apply. For now we focus on the non-anomalous theory.) From the fact that two of the four partition functions in the $\NS\R$ Hilbert space vanish we get\footnote{One can think of a pair of free left-moving $\psi$ and right-moving $\tilde{\psi}$ 2d fermion for simplicity. The mode expansion in the NSR Hilbert space is $\psi_n$ and $\tilde{\psi}_m$ with $n\in\mathbb{Z}+1/2$ and $m\in\mathbb{Z}$. For any state $|E,j\rangle$ we can change the statistics without changing the spin by acting with $\tilde{\psi}_0|E,j\rangle$ for example. Similarly, we can change the spin by a half without changing the statistics, e.g. $\psi_{-1/2} \tilde{\psi}_0|E,j\rangle$ has the same statistics as $|E,j\rangle$ but it has spin $j+1/2$. But every time we act with the left-moving fermion, left-moving fermion parity changes and the spin shifts by a half. }
\begin{equation} \label{eq: (-1)^FL NSR rho avg}
\begin{split}
\rho_{\NS\R}^{\text{bos}}(E,j) &= \rho_{\NS\R}^{\text{fer}}(E,j) = \text{Spectrum of}\left(\frac{Z_{\NS\R--}(\tau)}{2}\right),\\
\rho_{\R\NS}^{\text{bos}}(E,j) &= \rho_{\R\NS}^{\text{fer}}(E,j) = \text{Spectrum of}\left(\frac{Z_{\R\NS--}(\tau)}{2}\right).
\end{split}
\end{equation}
The right hand side indicates to take the spectrum of the partition function by performing an inverse Laplace transform similar to the one in Appendix \ref{app:spectrum}. Since $Z_{\NS\R--}(\tau)=Z_{\R\NS--}(\tau)$ all four densities of states are equal. The derivation is very similar to that of the appendix so we simply point out the differences. We need to input the restriction on $(s,d)$ in going from $\Gamma_0(2) \to \Gamma(2)$. Before incorporating the new $\Gamma(2)$ restrictions, the answer is given by equation \eqref{eq:rhoFG} with $a=\R$. For $\gamma\in \Gamma(2)$ we have $\gamma = \left(\begin{smallmatrix}
    \text{odd} & \text{even}\\
   \text{even} & \text{odd}    
\end{smallmatrix} \right)$ and therefore $S\gamma =\left(\begin{smallmatrix}
    \text{even} & \text{odd}\\
   \text{odd} & \text{even}    
\end{smallmatrix} \right) $. The sum is therefore restricted to odd $s$ but now also even $d$. This has two effects. First, when we decompose $d=d'+n s$ we need $n\in 2\mathbb{Z}$ to preserve the parity of $d$. This implies we now get half-integer spins in the spectrum as previously anticipated. Second, we restrict the sum over $d$ in the definition of $S_{\R}(j,j';s)$ to only even $d$. The final answer is given by \eqref{eq:rhoFG} replacing the R-Klostermaan sum by
\beq
S_{\NS\R}(j,j';s) = \begin{cases}
    \sum_{\substack{0\leq d<2s\\ \gcd(s,d)=1\\ 
    \text{$d$ even}}} \exp\left(2\pi\i \frac{dj+a_{\R}(d,s)j'}{s}\right) ~~~~\text{for $s$ odd}\\
    0 ~~~\text{for $s$ even}
\end{cases}
\eeq
as well as allowing $j$ to be half integer. This gives the leading density of states in the RNS/NSR Hilbert spaces.

\smallskip

\noindent\textbf{NSNS:} One can easily verify that a $T$ transform maps $Z_{\NS\NS--}(\tau+1) = Z_{\NS\NS++}(\tau)$. Therefore, and only for NSNS states, spin-statistics holds in its naive form 
\beq
\mathcal{H}_{\NS\NS}:~~~~(-1)^{\sf F} =e^{2\pi \i J}.
\eeq
There is no a priori relation between $(-1)^{\sf F_L}$ or $(-1)^{\sf F_R}$ and $J$ other than $(-1)^{\sf F_L} = (-1)^{\sf F_R} e^{2\pi \i J}$ which essentially follows from the definition of $(-1)^{\sf F_R}$. This implies that states with $j\in\mathbb{Z}+1/2$ have opposite left/right statistics, while those with $j\in\mathbb{Z}$ have diagonal left/right statistics. We therefore have 
\beq\label{eq:NSNSHH}
\mathcal{H}_{\NS\NS}=\left(\bigoplus_{j\in\mathbb{Z}} \mathcal{H}_j^{\text{bos/bos}}\oplus \mathcal{H}_j^{\text{fer/fer}}\right)\oplus \left( \bigoplus_{j\in\mathbb{Z}+1/2} \mathcal{H}_j^{\text{bos/fer}}\oplus \mathcal{H}_j^{\text{fer/bos}}\right)
\eeq
and the following a priori independent density of states
\begin{eqnarray}
j\in\mathbb{Z}:&&~~~~\rho_{\NS\NS}^{\text{bos/bos}}(E,j),~\rho_{\NS\NS}^{\text{fer/fer}}(E,j),\\
j\in\mathbb{Z}+1/2:&&~~~~\rho_{\NS\NS}^{\text{bos/fer}}(E,j),~~\rho_{\NS\NS}^{\text{fer/bos}}(E,j).
\end{eqnarray}
Using the partition function without any fermion parity insertion we get 
\bea
\rho_{\NS\NS}^{\text{bos/bos}}(E,j)+\rho_{\NS\NS}^{\text{fer/fer}}(E,j)&=&  \rho_{\NS}(E,j),~~~j\in\mathbb{Z}.\\
\rho_{\NS\NS}^{\text{bos/fer}}(E,j)+\rho_{\NS\NS}^{\text{fer/bos}}(E,j)&=&  \rho_{\NS}(E,j),~~~j\in\mathbb{Z}+1/2.
\ea
The difference between these sectors can be extracted from the other non-trivial partition function
\bea
\rho_{\NS\NS}^{\text{bos/bos}}(E,j)-\rho_{\NS\NS}^{\text{fer/fer}}(E,j)&=& \text{Spectrum of}\left(Z_{\NS\NS+-}(\tau)\right),~~~j\in\mathbb{Z}.\\
\rho_{\NS\NS}^{\text{bos/fer}}(E,j)-\rho_{\NS\NS}^{\text{fer/bos}}(E,j)&=& \text{Spectrum of}\left(Z_{\NS\NS+-}(\tau)\right),~~~j\in\mathbb{Z}+1/2.
\ea
The distribution between these two components can only be extracted from a careful calculation of $Z_{\NS\NS+-}(\tau)$. Since we are working with pure gravity, from the fact that $Z_{\NS\NS+-}(\tau)=Z_{\NS\NS+-}(\tau+1)$, we can conclude that the spectrum of $Z_{\NS\NS+-}$ has only contributions with $j\in \mathbb{Z}$. This immediately implies that
\beq
\rho_{\NS\NS}^{\text{bos/fer}}(E,j)=\rho_{\NS\NS}^{\text{fer/bos}}(E,j)= \frac{1}{2} \rho_{\NS}(E,j),~~~j\in\mathbb{Z}+1/2,
\eeq
We can now focus on $j\in\mathbb{Z}$ and compute the spectrum from Laplace transforming $Z_{\NS\NS+-}$. Following the steps outlined in Appendix \ref{app:spectrum} we get that for either $j$ integer or half integer, the difference in spectral densities is given by equation \eqref{eq:rhoFG} with $S_{\NS}(j,j';s)$, and the main restriction from $\Gamma(2)$ is that now the sum over $s$ includes only even $s$  contributions. Having computed the difference between bos/bos and fer/fer we can separately evaluate $\rho_{\NS\NS}^{\text{bos/bos}}(E,j)$ and $\rho_{\NS\NS}^{\text{fer/fer}}(E,j)$ completing the characterization of the NSNS Hilbert space. Finally, only in this sector is thermal AdS with $E=-c/12$ and $j=0$ part of the spectrum.

\smallskip

To complete this discussion, we can consider what happens if we gauge the $(-1)^{\sf F_L}$ symmetry. One can check explicitly that this leads to the fermionic theory of section \ref{sec:Path_integral_fermionic_3d_gravity} directly by using the following identities $\Gamma(2) \cup S \Gamma(2) = \Gamma_\theta$, $\Gamma(2) \cup T^{-1} S T \Gamma(2) = \Gamma^0(2)$ and $S\Gamma(2) \cup ST \Gamma(2) = S \Gamma_0(2)$. A further gauging of fermion parity automatically leads to bosonic 3d gravity.

\smallskip

It is interesting to analyze the solid torus partition functions that vanish from the point of view of the bordism groups. The relevant bordism group for spin surfaces with $\mathbb{Z}_2$ gauge field is 
\beq
\Omega_2^{\text{spin}}(B\mathbb{Z}_2) = \mathbb{Z}_2 \times \mathbb{Z}_2.
\eeq
Here $B\mathbb{Z}_2$ refers to the space of isomorphism classes of principal $\Z_2$ bundles \cite{Chen:2011pg, Kapustin:2014tfa}. There is an invariant distinguishing each of the four components for an arbitrary $\mathbb{Z}_2$ symmetry. This invariant is equivalent to the sum of invariants of the left- and right-moving mod 2 indices. If any the right- or left-moving structure has periodic conditions on both circles then the surface is not the boundary of any 3d surface with a spin structure and a $\mathbb{Z}_2$ gauge field. This shows that the partition functions that we find to vanish for the solid torus actually vanish non-perturbatively. Moreover, any wormhole with an odd number of such surfaces also vanish. This implies that in the 2d CFT ensemble the spectrum in the $\NS\R$ or $\R\NS$ Hilbert spaces is 2-fold exactly degenerate  on each member of the ensemble and not only on average. Similarly the spectrum in the $\R\R$ Hilbert space is exactly 4-fold degenerate.

\subsection{CFT anomalies from gravity}
Having analyzed the leading spectrum of the ensemble we can move on to discussing the bulk TQFTs dual to anomalies in the $(-1)^{\sf F_L}$ symmetry. The classification of topological phases on 3D Spin manifolds in the presence of a unitary $\Z_2$ symmetry is given by the cobordism group \cite{Kapustin:2014dxa, Delmastro:2021xox}
\begin{equation} \label{eq: cobordism fermions (-1)^FL}
    {\rm TP}^{\spin}_3(B\Z_2) = \Z\times\Z_8\,,\quad {\rm TP}^{\spin,\,tors}_3(B\Z_2) = \Z_8
\end{equation}
where we focus on the torsion subgroup, which ignores the free part considered in section \ref{subsec:RT}. We label the eight topological field theories of \eqref{eq: cobordism fermions (-1)^FL} by $N$ mod $8$, with cobordism invariant $\eta$.

We compute $\eta$ using the Smith isomorphism, which relates the cobordism invariant on a $D$ dimensional $\spin$ manifold with a unitary $\Z_2$ symmetry to the cobordism invariant on a $D-1$ dimensional $\pinm$ manifold with a time reversal symmetry $\T^2=1$ \cite{Kapustin:2014dxa, Hason:2020yqf, Cordova:2019wpi}. In particular, between a 3d $\spin$ manifold and a 2d $\pinm$ manifold, we have
\begin{equation}
    \Hom \left(\Omega_3^{\spin,\,tors}(B\Z_2),U(1)\right) \cong \Hom\left(\Omega_2^\pinm(pt),U(1)\right) \cong \Z_8 \,.
\end{equation}
Here the Smith isomorphism relates the value of $\eta$ on a closed 3-manifold $M^3$ to the 2d Arf-Brown-Kervaire invariant ABK computed on the 2-manifold $\PD(A)\subset M^3$ that is Poincaré dual to the $(-1)^{\sf F_L}$ gauge field $A\in H^1(M^3,\Z_2)$ \cite{Kapustin:2014dxa}. Schematically, this can be written as \cite{Putrov:2016qdo}
\begin{equation}
    \eta[M^3] \equiv \int_{M^3} A \cup \ABK = \int_{\PD(A)} \ABK \equiv \ABK[\PD(A)] \,,
\end{equation}
analogous to the relationship \eqref{eq:PD w1} between the 3d $\pinp$ invariant and the Arf invariant in 2d. This explain also our choice of label, we call the invariant $\eta$ since it is the Atiyah-Patodi-Singer invariant \cite{Atiyah:1975jf} on a certain 2d surface inside the geometry.

\subsubsection*{Solid torus}

When performing the gravity path integral on the solid torus, we can use the Smith isomorphism to analyze how the $\mathbb{Z}_8$ TQFT depends on the modular images of the boundary. As we explained in Section \ref{sec:discretesymm}, this is diagnosed by evaluating its path integral on lens space and leads to possible shifts in the angular momentum quantization. Although lens space is orientable, there can be unorientable 2d manifolds inside of it leading to non-trivial $\pinm$ structures.\footnote{As the simplest example take the orientable lens space $L(2,1)=\mathbb{RP}^3$ defined as a 3-sphere in $\mathbb{C}^2$, $|z_1|^2 + |z_2|^2=1$ with the identification $(z_1,z_2) \sim (-z_1,-z_2)$. Consider the 2d surface $\text{Im}(z_2)=0$ and call $x=\text{Re}(z_1)$, $y=\text{Im}(z_1)$ and $z=\text{Re}(z_2)$. This surface is a sphere $S^2$ inside $S^3$ since $x^2 + y^2 + z^2=1$, but the identification $(z_1,z_2) \sim (-z_1,-z_2)$ becomes $(x,y,z) \to (-x,-y,-z)$ leading to $\mathbb{RP}^2$ which is unorientable.} The $\gamma$-dependent phase introduced by the TQFT is given by its path integral on Lens space obtained gluing up to a modular transformation $\gamma$
$$
Z_\text{TQFT}(\gamma) = e^{-\i\pi N \eta/2}
$$
where the invariant $\eta$ on the relevant lens space has an implicit dependence on the spin structure $\mathfrak{s}$, the $(-1)^{\sf F_L}$ gauge field $A$, and the modular transform $\gamma$ that is used to construct the space. 

We will not attempt to provide a full account of how the averaged density of states of the 2d CFT ensemble is modified, but we can at least consider how angular momentum quantization is affected by it. This can be read off from the anomalous change in the TQFT partition function under the action of $T^{2n}$. The path integral on lens space with a relative $T^{2n}$ gluing was computed in \cite{Grigoletto:2021zyv} and it is given by
\beq
Z_{\text{TQFT}}(T^{2n})=  \begin{cases}
    \exp{-\frac{\i\pi n N}{4}} ~~~~\text{for $\NS\R+-$, $\NS\R--$}\\
    \exp{\frac{\i\pi n N}{4}} ~~~~\hspace{0.3cm}\text{for $\R\NS-+$, $\R\NS--$}\\
    ~~~~1 ~~\hspace{2.1cm}~\text{else}
\end{cases}
\eeq
In the R sector we need to consider gluing by odd powers of $T$, and it was found in \cite{Grigoletto:2021zyv} that $Z_{\text{TQFT}}(T^{n})=1$. These results imply the following
\beq
j_{\R\NS} \in \frac{\mathbb{Z}}{2} + \frac{N}{16},~~~j_{\NS\R} \in \frac{\mathbb{Z}}{2} - \frac{N}{16},~~~j_{\NS\NS}\in \frac{\mathbb{Z}}{2},~~~j_{\R\R} \in \mathbb{Z},
\eeq
so angular momentum quantization becomes anomalous in the presence of the $\mathbb{Z}_8$ anomaly. This also implies an anomaly under modular transformations.

\subsubsection*{Two-boundary torus wormhole}

\begin{figure}
    \centering
    \includegraphics[width=0.6 \linewidth]{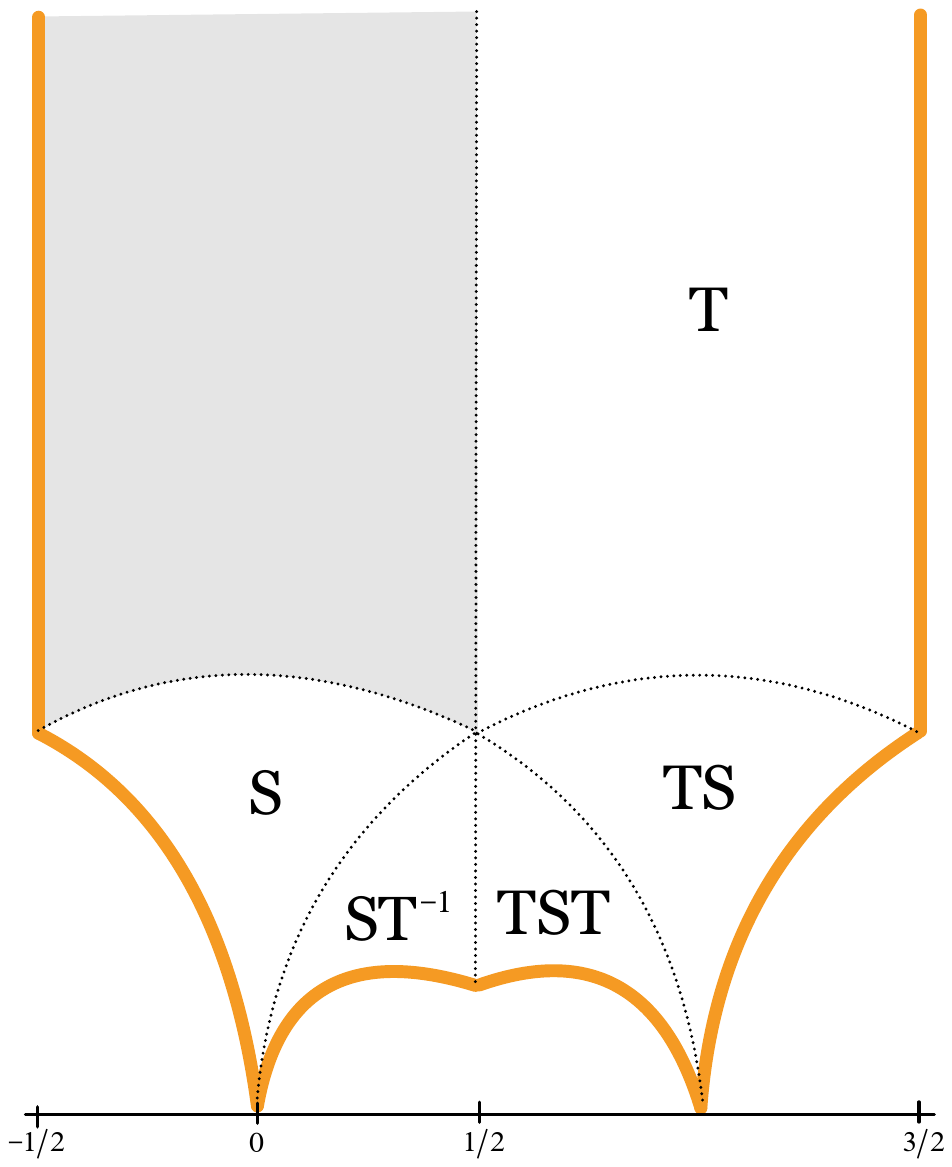}
    \caption{The fundamental domain of $\Gamma(2)$ has cusps at $\tau=0,1,\infty$. The gray region is the fundamental domain of $SL(2,\Z)$, for comparison.}
    \label{fig:Gamma2}
\end{figure}

For $N=0$ mod $8$ (trivial TQFT), the only new case for the torus wormhole is when the congruence subgroup is $\Gamma(2)$, for which we can apply a similar analysis as in section \ref{sec:rmt2}, with the fundamental domain shown in Figure \ref{fig:Gamma2}. Even though the fundamental domain of $\Gamma(2)$ has three cusps, the statements of \ref{subsec:RMT2_two_boundary_wormhole} and in particular the application of the Kuznetsov trace formula still apply \cite{Iwaniec2002SpectralMO}, resulting in the wormhole
\begin{equation} \label{eq:CJ (-1)^FL}
    Z_{\mathfrak{s}_L \mathfrak{s}_R}(\tau_1,\tau_2,N=0) = 8\, Z_0(\tau_1) Z_0(\tau_2) \sum_{\gamma \in G_{\mathfrak{s}_L\mathfrak{s}_R}}\frac{\text{Im}(\tau_1)\text{Im}(\gamma \tau_2)}{2\pi^2|\tau_1 + \gamma \tau_2|^2},
\end{equation}
where for simplicity we consider wormholes with the same spin structure on each boundary, for which $\mathfrak{s}_L$ and $\mathfrak{s}_R$ label the left- and right-moving spin structures, respectively. The groups $G_{\mathfrak{s}_L\mathfrak{s}_R}$ are listed in Table \ref{tab:mgflfr}. The factor of eight in \eqref{eq:CJ (-1)^FL} relative to the Cotler-Jensen two-boundary wormhole arises from gauging $\r\T$, $(-1)^{\sf F}$, and $(-1)^{\sf F_L}$. The twist by $(-1)^{\sf F_L}$ can be thought of a arising from a non-trivial $\mathbb{Z}_2$ connection in the bulk.

We now analyze how the two-boundary torus wormhole changes in the presence of the $\Z_8$ TQFTs \eqref{eq: cobordism fermions (-1)^FL}. Just as in the case of the gravitational anomaly \eqref{eq:CJGCS}, the partition function of each modular image depends on $\gamma$ in a manner controlled by lens spaces:
\begin{equation} \label{eq:bare}
    Z_{\mathfrak{s}_L \mathfrak{s}_R}^{\text{bare}}(\tau_1,\tau_2,N) =  Z_0(\tau_1)Z_0(\tau_2) \sum_{\gamma\in G_{\mathfrak{s}_L \mathfrak{s}_R}} \frac{\Im(\tau_1) \Im(\gamma\tau_2)}{2\pi^2|\tau_1+\gamma\tau_2|^2} e^{-\i\pi N \eta/2} \,,
\end{equation}
where $\eta$ has an implicit dependence on the spin structure, modular transformation and $\mathbb{Z}_2$ connection. In \eqref{eq:bare}, we use ``bare'' to denote the wormhole before we gauge the $\r\T$, $(-1)^{\sf F}$, and $(-1)^{\sf F_L}$ symmetries.

 In order to gauge these symmetries, we will need the value of $\eta$ for the closed 3-manifolds constructed as in Figure \ref{fig:2torus} with appropriate insertions of symmetry defects along the gray torus. This will correspond to computing $\ABK[\PD(A)]$ for any choice of spin structure $\mathfrak{s}$ and gauge field $A$, within two classes of 3-manifolds: the 3-torus \eqref{eq:T3} and the dicosm \eqref{eq:T3 RT}. The path integral will then weigh each two-boundary wormhole with a phase of
\begin{equation}
    e^{-\ii\pi\eta N/2} \,.
\end{equation}

\smallskip

\noindent\textbf{3-torus:} Consider a 3-torus parametrized by $(x,y,z)$ such that
\begin{equation}
    (x,y,z)\sim (x+1,y,z) \sim (x,y+1,z)\sim(x,y,z+1)\,.
\end{equation}
The Poincaré dual, when it exists, is given by a 2-torus. This is orientable, so the $\Z_8$-valued ABK invariant $\eta$ reduces to the $\Z_2$-valued Arf invariant $\zeta$. Let $\mathfrak{s}_x$ denote the spin structure of right-moving fermions along the $x$ cycle (given by $\NS$ or $\R$ periodicity). There are four classes of possibilities for the Poincaré dual of $A=(A_x,A_y,A_z)$, where $A_i$ indicates whether $(-1)^{\sf F_L}$ has been inserted along the $i$th cycle:
\begin{enumerate}
    \item $A=(0,0,0)$: There is no Poincaré dual, so $\eta=0$.
    \item $A=(0,0,1)$: The Poincaré dual is given by a torus at fixed $z$, for which 
    \begin{equation*}
        \eta = 2\zeta(\mathfrak{s}_x, \mathfrak{s}_y)\,.
    \end{equation*}
    \item $A=(0,1,1)$: The Poincaré dual is given by a torus at fixed $y+z$, for which
    \begin{equation*}
        \eta = 2\zeta(\mathfrak{s}_x, \mathfrak{s}_y+\mathfrak{s}_z)\,.
    \end{equation*}
    Here we take $\mathfrak{s}_y+\mathfrak{s}_z$ to mean the periodicity of a spectator fermion that goes along both $y$ and $z$ cycles. For example $\R+\NS=\NS$, where the fermion picks up a single $(-1)^{\sf F}$, while $\NS+\NS=\R$, where the fermion picks up $(-1)^{\sf F}(-1)^{\sf F}=1$.
    \item $A=(1,1,1)$: The Poincaré dual is given by a torus at fixed $x+y+z$, for which one example is given by
    \begin{equation*}
        \eta = 2\zeta(\mathfrak{s}_x+\mathfrak{s}_y, \mathfrak{s}_y+\mathfrak{s}_z)\,.
    \end{equation*}
\end{enumerate}

We would like to apply this analysis to the two-boundary wormhole. As shown in Figure \ref{fig:2torus}, the boundary conditions on the wormhole fix the periodicities on cycles $x$ and $y$, while there are four choices of periodicity on cycle $z$:
\begin{equation} \label{eq:4defects}
\begin{split}
    \text{no defect insertion: }& \NS\NS \\
    (-1)^{\sf F}\text{ defect insertion: }& \R\R \\
    (-1)^{\sf F_L}\text{ defect insertion: }& \R\NS \\
    (-1)^{\sf F_L}(-1)^{\sf F} \text{ defect insertion: }& \NS\R \,.
\end{split}
\end{equation}
For example:
\begin{enumerate}
    \item $\NS\NS$ on $x$, any on $y$: either $A$ has no Poincaré dual, or $\eta=2\zeta(\NS,\cdot)=0$.
    \item $\R\R$ on $x$ and $y$: for either of the first two cases in \eqref{eq:4defects}, $A$ has no Poincaré dual and $\eta=0$. For either of the latter two cases in \eqref{eq:4defects}, the Poincaré dual is a 2-torus with $x$ and $y$ cycles, so $\eta=2\zeta(\R,\R)=2$.
    \item $\R\NS$ on $x$ and $\NS\R$ on $y$: all four cases in \eqref{eq:4defects} result in $\eta=2\zeta(\NS,\cdot)=0$.
    \item $\R\NS$ on $x$ and $\R\NS$ on $y$: the second and third cases in \eqref{eq:4defects} result in $\eta=2\zeta(\R,\R)=2$, while the first and fourth cases result in $\eta=2\zeta(\NS,\cdot)=0$.
\end{enumerate}
This is summarized by the following
\begin{equation}
    Z_{\mathfrak{s}_L \mathfrak{s}_R}(\tau_1,\tau_2,N) \supset \left\{\begin{matrix}
        2\left[1+(-1)^N\right] Z_{\mathfrak{s}_L\mathfrak{s}_R}^{\text{bare}}(\tau_1,\tau_2,N) &\quad \zeta(\mathfrak{s}_L)=1 \text{ or } \zeta(\mathfrak{s}_R)=1 \\
        2\left[1+(-1)^N\right] Z_{\mathfrak{s}_L \mathfrak{s}_R}^{\text{bare}}(\tau_1,\tau_2,N) &\quad \zeta(\mathfrak{s}_L)=\zeta(\mathfrak{s}_R)=1 \\
        4Z_{\mathfrak{s}_L \mathfrak{s}_R}^{\text{bare}}(\tau_1,\tau_2,N) &\quad \text{else}
    \end{matrix} \right. 
\end{equation}
where we refer to the spin structure of left- and right-moving spectator fermions as $\mathfrak{s}_L$ and $\mathfrak{s}_R$, respectively, and we label the TQFT by the parameter $N$ mod $8$.

\bigskip

\noindent\textbf{Dicosm:} Consider a dicosm parametrized by $(x,y,z)$ such that
\begin{equation}
    (x,y,z)\sim (x+1,y,z) \sim (x,y+1,z)\sim(-x,-y,z+1)\,.
\end{equation}
While the dicosm is an orientable topology, the Poincaré dual of $A$, when it exists, can result in either a 2-torus (orientable), or a Klein bottle (non-orientable). On the Klein bottle, it was shown in \cite{Stanford:2019vob} that when the orientable cycle has $\NS$ periodicity, $\eta=0$, and when the orientable cycle has $\R$ periodicity, the two choices of periodicity on the non-orientable cycle contribute $\eta=1$ and $\eta=-1$.

For the dicosm, $A=(0,0,0)$ and $A=(0,0,1)$ are the same as in the 3-torus analysis. There are two new classes of possibilities for the Poincaré dual:
\begin{enumerate}
    \item $A=(0,1,0)$ or $A=(0,1,1)$: The Poincaré dual is given by a Klein bottle at fixed $y$ or $y+z$, respectively. In both cases, $\eta = 0$ if $\mathfrak{s}_x=\NS$ and $\eta = \pm1$ if $\mathfrak{s}_x=\R$.
    \item $A=(1,1,0)$ or $A=(1,1,1)$: The Poincaré dual is given by a Klein bottle at fixed $x+y$ or $x+y+z$, respectively. In both cases, $\eta = 0$ if $\mathfrak{s}_x+\mathfrak{s}_y=\NS$ and $\eta = \pm1$ if $\mathfrak{s}_x+\mathfrak{s}_y=\R$.
\end{enumerate}
For example:
\begin{enumerate}
    \item $\R\R$ on $x$ and $\R\NS$ on $y$: two cases in \eqref{eq:4defects} result in $\eta=1$ and the other two result in $\eta=-1$.
    \item $\R\NS$ on $x$ and $\NS\R$ on $y$: all four cases in \eqref{eq:4defects} result in $\eta=0$.
    \item $\R\NS$ on $x$ and $\R\NS$ on $y$: two cases in \eqref{eq:4defects} result in $\eta=1$ and the other two result in $\eta=-1$.
\end{enumerate}

\smallskip

The two-boundary wormhole after gauging $\r\T$, $(-1)^{\sf F}$, and $(-1)^{\sf F_L}$ combines the saddles from the 3-tori and dicosms, resulting in
\begin{equation} \label{eq:PD wormhole}
    Z_{\mathfrak{s}_L \mathfrak{s}_R}(\tau_1,\tau_2,N) = \left\{\begin{matrix}
        2\left[1+(-1)^N+2\cos\frac{2\pi N}{4}\right] Z_{\mathfrak{s}_L\mathfrak{s}_R}^{\text{bare}}(\tau_1,\tau_2,N) &\quad \zeta(\mathfrak{s}_L)=1 \text{ or } \zeta(\mathfrak{s}_R)=1 \\
        4\left[1+(-1)^N\right] Z_{\mathfrak{s}_L \mathfrak{s}_R}^{\text{bare}}(\tau_1,\tau_2,N) &\quad \zeta(\mathfrak{s}_L)=\zeta(\mathfrak{s}_R)=1 \\
        8Z_{\mathfrak{s}_L \mathfrak{s}_R}^{\text{bare}}(\tau_1,\tau_2,N) &\quad \text{else}
    \end{matrix} \right. 
\end{equation}
where the first line results in $8Z_{\mathfrak{s}_L\mathfrak{s}_R}^{\text{bare}}(\tau_1,\tau_2,N)$ when $N\equiv0$ mod $4$ and vanishes otherwise, while the second line results in $8Z_{\mathfrak{s}_L\mathfrak{s}_R}^{\text{bare}}(\tau_1,\tau_2,N)$ when $N$ is even and vanishes otherwise.

\smallskip

\subsection{RMT statistics at late times}
We compute the late-time limit of \eqref{eq:PD wormhole} in the untwisted and twisted Hilbert spaces.

\subsubsection*{NSNS Hilbert space}
For all values of $N$ mod $8$, the two-boundary wormhole is given by
\begin{eqnarray}
    Z_{\NS\pm}(\tau_1,\tau_2) = 8Z^{\text{bare}}_{\NS\pm}(\tau_1,\tau_2)
\end{eqnarray}
and the analytic continuation  $y_1=\frac{1}{2\pi}(\beta+\ii t)$ and $y_2=\frac{1}{2\pi}(\beta-\ii t)$ of the spectral form factor \eqref{eq:SFF} at late times results in
\begin{equation} \label{eq:NSNS GOE}
    K^{(j,j)}_{\mathfrak{s}_L \mathfrak{s}_R}(\beta,t) \sim \frac{t}{\pi\beta}e^{-2\beta|j|} \quad (2\,\text{GOE})
\end{equation}
for $\mathfrak{s}_L,\mathfrak{s}_R\in\{\NS\pm\}$. The two contributions from the GOE ensemble come from the two sectors for fixed spin in \eqref{eq:NSNSHH}.

\subsubsection*{RR Hilbert space}
We label the TQFTs by $N$ mod $8$.

\smallskip

\noindent\textbf{$N\equiv0,4\text{ mod }8$:} There are four different nonvanishing wormholes in this case, which at fixed $j$ are given by 
\begin{equation} \label{eq: general (-1)^FL 2bdy}
    Z_{\mathfrak{s}_L \mathfrak{s}_R}^{(j,j)}(\tau_1,\tau_2,N) = \int \d E_1 \d E_2 \, \left\langle \rho_{\mathfrak{s}_L \mathfrak{s}_R}(E_1,j) \rho_{\mathfrak{s}_L \mathfrak{s}_R}(E_2,j) \right\rangle \chi_{E_1,j}(\tau_1) \chi_{E_2,j}(\tau_2) 
\end{equation}
where
\begin{equation} \nonumber
\begin{split}
    \rho_{\R\R--} &= \rho_{\R\R}^{\text{bos}/\text{bos}} + \rho_{\R\R}^{\text{bos}/\text{fer}} + \rho_{\R\R}^{\text{fer}/\text{bos}} + \rho_{\R\R}^{\text{fer}/\text{fer}} \\
    \rho_{\R\R++}&= \rho_{\R\R}^{\text{bos}/\text{bos}} - \rho_{\R\R}^{\text{bos}/\text{fer}} - \rho_{\R\R}^{\text{fer}/\text{bos}} + \rho_{\R\R}^{\text{fer}/\text{fer}} \\
    \rho_{\R\R+-} &= \rho_{\R\R}^{\text{bos}/\text{bos}} + \rho_{\R\R}^{\text{bos}/\text{fer}} - \rho_{\R\R}^{\text{fer}/\text{bos}} - \rho_{\R\R}^{\text{fer}/\text{fer}} \\
    \rho_{\R\R-+} &= \rho_{\R\R}^{\text{bos}/\text{bos}} - \rho_{\R\R}^{\text{bos}/\text{fer}} + \rho_{\R\R}^{\text{fer}/\text{bos}} - \rho_{\R\R}^{\text{fer}/\text{fer}}\,.
\end{split}
\end{equation}
The second corresponds to the partition function with $(-1)^{\sf F}$ insertion, the third includes $(-1)^{\sf F_L}$, and the fourth includes $(-1)^{\sf F_L}(-1)^{\sf F}$.

\smallskip

\noindent\textbf{$N\equiv1,3,5,7\text{ mod }8$:}  The wormholes with any insertion vanish: $Z_{\R\R++} = Z_{\R\R+-} = Z_{\R\R-+} = 0$. In this case, we will see in the next subsection that $\{(-1)^{\sf F},(-1)^{\sf F_L}\}=0$, so without loss of generality we only label the Hilbert space by the eigenvalue under $(-1)^{\sf F}$. Then
\begin{equation} \label{eq:RR rho variances}
    \rho_{\R\R}^{\text{bos}}(E,j) = \rho_{\R\R}^{\text{fer}}(E,j)
\end{equation}
holds at the level of variances. We anticipate that this equality, which was discussed in \eqref{eq: (-1)^FL rho avg} at level of averages, becomes exact in the RMT approximation for each member of the ensemble.

\smallskip

\noindent\textbf{$N\equiv2,6\text{ mod }8$:} The wormholes with $(-1)^{\sf F_L}$ insertion vanish: $Z_{\R\R+-} = Z_{\R\R-+} = 0$. In this case, the relations
\begin{equation}\label{eq:RR bos bos variances}
    \rho_{\R\R}^{\text{bos/bos}}(E,j) = \rho_{\R\R}^{\text{fer/fer}}(E,j)\,,\quad \rho_{\R\R}^{\text{bos/fer}}(E,j)=\rho_{\R\R}^{\text{fer/bos}}(E,j)
\end{equation}
hold at the level of variances. 

\bigskip

After analytic continuation, the nonvanishing wormholes have the late-time limit 
\begin{equation}
    \label{eq:RRZ2NZ}
K^{(j,j)}_{\mathfrak{s}_L \mathfrak{s}_R}(\beta,t,N) \sim \frac{2t}{\pi\beta}e^{-2\beta|j|} \quad (8\,\text{GUE}/4\,\text{GOE})
\end{equation}
for $\mathfrak{s}_L,\mathfrak{s}_R\in\{\R\pm\}$. We will explain in subsection \ref{subsec:(-1)^FL algebra} why the RMT statistics are either GUE or GOE, depending on $N$, as well as the origin of the prefactors.

\subsubsection*{NSR Hilbert space}
There are again four wormholes of the form \eqref{eq: general (-1)^FL 2bdy}, with now
\begin{equation} \label{eq: (-1)^FL rho 2bdy}
\begin{split}
    &\rho_{\NS\R--} = (-1)^{2j} \rho_{\NS\R+-} = \rho_{\NS\R}^{\text{bos}} + \rho_{\NS\R}^{\text{fer}} \\
    &\rho_{\NS\R-+} = (-1)^{2j}\rho_{\NS\R++} = \rho_{\NS\R}^{\text{bos}} - \rho_{\NS\R}^{\text{fer}}
\end{split}
\end{equation}
as functions of $E$ and $j$.

\smallskip

\noindent\textbf{$N\equiv0,4\text{ mod }8$:} all four wormholes are nonvanishing with a late-time limit given below.

\smallskip

\noindent\textbf{$N\equiv2,6\text{ mod }8$:} The wormholes with $(-1)^{\sf F}$ or $(-1)^{\sf F_L}(-1)^{\sf F}$ insertion, corresponding to the second line of \eqref{eq: (-1)^FL rho 2bdy}, vanish: $Z_{\NS\R-+} = Z_{\NS\R++} = 0$. Then the relations \eqref{eq: (-1)^FL NSR rho avg} hold at the level of variances:
\begin{equation} \label{eq:NSR rho bos/fer}
    \rho_{\NS\R}^{\text{bos}}(E,j) = \rho_{\NS\R}^{\text{fer}}(E,j) \,.
\end{equation}
and we will see this is exact in the RMT ensemble.

\smallskip

\noindent\textbf{$N\equiv1,3,5,7\text{ mod }8$:} The wormholes with $(-1)^{\sf F}$ or $(-1)^{\sf F_L}(-1)^{\sf F}$ insertion vanish again, but in this case $(-1)^{\sf F}$ is not a symmetry, so the $\rho_{\NS\R}$ density of states has no additional bos/fer label. We will discuss this in more detail in the next subsection.

\bigskip

After analytic continuation, the nonvanishing wormholes have the late-time limit 
\begin{equation} \label{eq:(-1)^FL SFF NSR}
    K^{(j,j)}_{\mathfrak{s}_L \mathfrak{s}_R}(\beta,t,N) \sim \frac{t}{\pi\beta}e^{-2\beta|j|} \quad (4\,\text{GUE}/2\,\text{GOE}/2\,\text{GSE})
\end{equation}
for $\mathfrak{s}_L\in\{\NS\pm\}$ and $\mathfrak{s}_R\in\{\R\pm\}$. Here we will see in the next subsection that all three Dyson ensembles GUE, GOE, and GSE appear, depending on the topological field theory labeled by $N$.

\subsection{2d CFT symmetry algebra} \label{subsec:(-1)^FL algebra}
In this section, we check the consistency of \eqref{eq:PD wormhole} with the free fermion symmetry algebra, which depends on the number of fermions $N$ mod $8$. We consider the Lagrangian \eqref{eq: massless L} and the group of symmetries generated by
\begin{equation} \label{eq:(-1)^FL sym alg}
\begin{split}
    (-1)^{\sf F_L}:&\quad \psi_L^a(t,x)\to -\psi_L^a(t,x)\,,\quad \psi_R^a(t,x)\to \psi_R^a(t,x) \\
     (-1)^{\sf F}:& \quad \psi_L^a(t,x)\to -\psi_L^a(t,x)\,,\quad \psi_R^a(t,x)\to -\psi_R^a(t,x) \\
    \r\T:& \quad \psi_L^a(t,x)\to \psi_L^a(-t,-x)\,,\quad \psi_R^a(t,x)\to -\psi_R^a(-t,-x)\,,\quad \ii\to -\ii \,.
\end{split}
\end{equation}
The symmetry algebras in the different Hilbert spaces for different values of $N$ were computed in \cite{Seiberg:2023cdc, Seiberg:2025zqx}.\footnote{Our $(-1)^{\sf F_L}$ corresponds to their $\C$, our $\T$ corresponds to their $\C\T$, and our $\r\T$ corresponds to their $\Theta=\C\r\T$.}

\subsubsection*{NSNS Hilbert space}
For any value of $N$, the symmetries \eqref{eq:(-1)^FL sym alg} satisfy the algebra
\begin{equation} \label{eq:NSNS (-1)^FL alg}
\begin{split}
    &\left((-1)^{\sf F}\right)^2=1\,,\quad \left((-1)^{\sf F_L}\right)^2=1\,,\quad (\r\T)^2 = 1 \\
    &[\r\T, (-1)^{\sf F}] = [\r\T, (-1)^{\sf F_L}] = [(-1)^{\sf F},(-1)^{\sf F_L}] = 0 \\
    &[J,(-1)^{\sf F}] = [J,(-1)^{\sf F_L}] = [J,\r\T] = 0 \,.
\end{split}
\end{equation}
Recall we can decompose the Hilbert space into the basis \eqref{eq:NSNSHH}, such that
\begin{equation} 
\begin{split}
    &H_j=\left(\begin{array} {c|c}  H_{bb} & 0\cr \hline 0 & H_{ff}\end{array}\right),~~~~(-1)^{\sf F_L}=\left(\begin{array} {c|c}  1 & 0\cr \hline 0 & -1\end{array}\right),~~~~(-1)^{\sf F_R}=\left(\begin{array} {c|c}  1 & 0\cr \hline 0 & -1\end{array}\right),~~~~ j\in\Z \\
    &H_j=\left(\begin{array} {c|c}  H_{bf} & 0\cr \hline 0 & H_{fb}\end{array}\right),~~~~(-1)^{\sf F_L}=\left(\begin{array} {c|c}  1 & 0\cr \hline 0 & -1\end{array}\right),~~~~(-1)^{\sf F_R}=\left(\begin{array} {c|c}  -1 & 0\cr \hline 0 & 1\end{array}\right),~~~~ j\in\Z+\frac12 \,.
\end{split}
\end{equation}
Since the anti-unitary symmetry $\r\T$ squares to 1 and all the discrete symmetries commute with each other, this is consistent with our findings in \eqref{eq:NSNS GOE} that all choices of boundary conditions lead to the same late-time limit with two statistically independent GOE blocks.

\subsubsection*{RR Hilbert space}
The symmetries \eqref{eq:(-1)^FL sym alg} satisfy the algebra
\begin{equation} \label{eq:RR (-1)^FL alg}
\begin{split}
    &\left((-1)^{\sf F}\right)^2=1\,,\quad \left((-1)^{\sf F_L}\right)^2=1\,,\quad (\r\T)^2 = 1 \\
    &\r\T(-1)^{\sf F_L} = (-1)^{N(N-1)/2}(-1)^{\sf F_L}\r\T \,,\quad (-1)^{\sf F}(-1)^{\sf F_L} = (-1)^N(-1)^{\sf F}(-1)^{\sf F_L} \\
    &[J,(-1)^{\sf F}] = [J,(-1)^{\sf F_L}] = [J,\r\T] = [(-1)^{\sf F},\r\T] = 0
\end{split}
\end{equation}

\bigskip

\noindent\textbf{For $N\equiv0,4\text{ mod }8$,} the algebra \eqref{eq:RR (-1)^FL alg} reduces to the anomaly-free algebra \eqref{eq:NSNS (-1)^FL alg}. The symmetry $\r\T$ squares to 1 and commutes with both $(-1)^{\sf F}$ and $(-1)^{\sf F_L}$. Recall that one can choose the basis \eqref{eq:H RR} for the Hilbert space, such that at spin $j$ we have (in the absence of anomalies)
\begin{equation} \label{eq:(-1)^FL 4 GOE}
    H_j=\left(\begin{array} {c|c|c|c}  H_{bb} & 0 & 0& 0 \cr \hline 0 & H_{bf} & 0 & 0 \cr \hline 0 & 0 & H_{fb} & 0 \cr \hline 0 & 0 & 0 & H_{ff} \end{array}\right)\,,~ (-1)^{\sf F}=\left(\begin{array} {c|c|c|c}  1 & 0 & 0& 0 \cr \hline 0 & -1 & 0 & 0 \cr \hline 0 & 0 & -1 & 0 \cr \hline 0 & 0 & 0 & 1 \end{array}\right)\,,~ (-1)^{\sf F_L}=\left(\begin{array} {c|c|c|c}  1 & 0 & 0& 0 \cr \hline 0 & 1 & 0 & 0 \cr \hline 0 & 0 & -1 & 0 \cr \hline 0 & 0 & 0 & -1 \end{array}\right).
\end{equation}
Then each block is invariant under $\r\T$, and a maximally random matrix consistent with the symmetries consists of four GOE blocks that become statistically independent at late times.

\smallskip

\noindent\textbf{For $N\equiv1,3,5,7\text{ mod }8$,} the relationship between $(-1)^{\sf F}$ and $(-1)^{\sf F_L}$ is anomalous:\footnote{For $N\equiv3,7\text{ mod }8$, there is an additional subtlety due to $\{\r\T, (-1)^{\sf F_L}\} = 0$, but we can redefine $\r\T$ by $\r\T'=\r\T\cdot (-1)^{\sf F}$ to recover an anti-unitary symmetry that commutes with $(-1)^{\sf F_L}$.}
\begin{equation}
    \{(-1)^{\sf F},(-1)^{\sf F_L}\} = 0\,,\quad [\r\T, (-1)^{\sf F}] = 0 \,,
\end{equation}
so we can only diagonalize the Hamiltonian under either $(-1)^{\sf F}$ or $(-1)^{\sf F_L}$. The Hilbert space therefore decomposes as $\mathcal{H}_{\R\R} = \bigoplus_{j\in\mathbb{Z}} \mathcal{H}_j^{\text{bos}} \oplus \mathcal{H}_j^{\text{fer}}$. Without loss of generality, under $(-1)^{\sf F}$ we have
\begin{equation} \label{eq:(-1)^FL 2 GOE}
    H=\left(\begin{array} {c|c}  H_b & 0\cr \hline 0 & H_f\end{array}\right),~~~~~~(-1)^{\sf F}=\left(\begin{array} {c|c}  1 & 0\cr \hline 0 & -1\end{array}\right).
\end{equation}
Since $\r\T$ commutes with $(-1)^{\sf F}$, the two blocks have GOE statistics and are not statistically independent, as we explain next. The symmetry $(-1)^{\sf F_L}$ acts as a fermionic generator that anticommutes with $(-1)^{\sf F}$ and commutes with the Hamiltonian and angular momentum, and its behavior can be understood as follows. Let $\ket{B}\in\mathcal{H}^{\text{bos}}_j$ denote a bosonic state. Then since $(-1)^{\sf F_L}$ is a fermionic operator: for every $\ket{B}$ state, there exists a state 
\begin{equation}
    \ket{F} = (-1)^{\sf F_L}\ket{B} ,~~~~\ket{F}\in\mathcal{H}^{\text{fer}}_j,
\end{equation}
with the same energy and spin, but different statistics. Note that a state cannot be annihilated by $(-1)^{\sf F_L}$ since the operator squares to the identity. A similar analysis applies to $(-1)^{\sf F_R}$ with the same conclusion and therefore $H_b = H_f$. This would explain why all partition functions vanish with the insertion of any fermion parity operator. Moreover, the ramp for $Z_{\R\R--}$ is four times the GOE answer since there are two equal GOE blocks in the Hamiltonian that add up, reproducing \eqref{eq:RRZ2NZ}.

\smallskip

\noindent\textbf{For $N\equiv2,6\text{ mod }8$,} the relationship between $\r\T$ and $(-1)^{\sf F_L}$ is anomalous:
\begin{equation}
    [(-1)^{\sf F}, (-1)^{\sf F_L}] = 0\,,\quad \{\r\T, (-1)^{\sf F_L}\} = 0 \,.
\end{equation}
Since $(-1)^{\sf F_L}$ commutes with $(-1)^{\sf F}$ we can still simultaneously diagonalize them and expand the Hilbert space according to \eqref{eq:(-1)^FL 4 GOE}. Since $\r\T$ commutes with $(-1)^{\sf F}$, it cannot relate bosons and fermions, but it can relate $\ket{BB}\in \mathcal{H}_j^{\text{bos/bos}}$ to $\ket{FF}\in \mathcal{H}_j^{\text{fer/fer}}$, and it can also relate $\ket{BF}\in \mathcal{H}_j^{\text{bos/fer}}$ to $\ket{FB}\in \mathcal{H}_j^{\text{fer/bos}}$. This is consistent with \eqref{eq:RR bos bos variances}, and also implies that the relationships between densities of states, found at the level of the wormhole, hold exactly. Since $\r\T$ now anticommutes with $(-1)^{\sf F_L}$, the maximally random matrix consistent with the symmetries now consists of the decomposition \eqref{eq:(-1)^FL 4 GOE} with four GUE blocks that are not statistically independent, instead $H_{bb} = H_{ff}$ and $H_{bf} = H_{fb}$. This explains the factor of $8$ times GUE in the ramp \eqref{eq:RRZ2NZ}.

\subsubsection*{NSR Hilbert space}

The $\NS\R$ Hilbert space is twisted by $(-1)^{\sf F_L}$, whose anomalies affect the behavior of $\r\T$. When $N$ is even,
\begin{equation} \label{eq:NSR (-1)^FL even}
\begin{split}
    &\left((-1)^{\sf F}\right)^2=1 \,,\quad \left((-1)^{\sf F_L}\right)^2=1 \,,\quad (\r\T)^2 = (-1)^{N(N-2)/8} \\
    &\r\T(-1)^{\sf F} = (-1)^{N/2}(-1)^{\sf F}\r\T\,,\quad [\r\T,(-1)^{\sf F_L}]=0\,,\quad [(-1)^{\sf F},(-1)^{\sf F_L}]=0\\
    &e^{2\pi\ii J} = (-1)^{\sf F_L} e^{-\frac{\pi\ii}{8}N} 
\end{split}
\end{equation}
while for $N$ odd, $(-1)^{\sf F}$ is not a symmetry:
\begin{equation}
    \left((-1)^{\sf F_L}\right)^2=1 \,,\quad (\r\T)^2 = \left\{\begin{matrix}
        (-1)^{(N-1)/4} & \text{for $N=4k+1$} \\
        (-1)^{(N+1)/4} & \text{for $N=4k+3$}
    \end{matrix}\right.
\end{equation}
where what we call $\r\T$ for $N=4k+3$ is related to the usual $\r\T$ by $(-1)^{\sf F}$ \cite{Seiberg:2023cdc, Seiberg:2025zqx}. The algebra involving $\r\T$ and $(-1)^{\sf F}$ reproduces precisely the RMT statistics in the eight cases shown in Table 1 of \cite{Stanford:2019vob}, corresponding to the $\Z_8$ anomaly of 2d $\pinm$ manifolds. In particular, the results for the TQFTs labeled by $N$ and $8-N$ are identical.

\bigskip

\noindent\textbf{For $N\equiv0\text{ mod }8$:} the symmetry $\r\T$ squares to 1 and commutes with both $(-1)^{\sf F}$ and $(-1)^{\sf F_L}$. We have two statistically independent GOE blocks, corresponding to the basis \eqref{eq:NSRHH}:
\begin{equation} \label{eq:(-1)^FL 2 GOE}
\begin{split}
    &H_j=\left(\begin{array} {c|c}  H_b & 0\cr \hline 0 & H_f\end{array}\right),~~~~~~(-1)^{\sf F_R}=\left(\begin{array} {c|c}  1 & 0\cr \hline 0 & -1\end{array}\right),~~~~~~ j\in\Z \\
    &H_j=\left(\begin{array} {c|c}  H_b & 0\cr \hline 0 & H_f\end{array}\right),~~~~~~(-1)^{\sf F_R}=\left(\begin{array} {c|c}  -1 & 0\cr \hline 0 & 1\end{array}\right),~~~~~~ j\in\Z+\frac12\,.
\end{split}
\end{equation}
\smallskip

\noindent\textbf{For $N\equiv1,7\text{ mod }8$:} there is no $(-1)^{\sf F}$ so the Hilbert space decomposes as $\mathcal{H}_{\NS\R}=\bigoplus_j \mathcal{H}_j$ and only the left-moving fermion number graded by $j$ survives. Besides this one has $(\r\T)^2=1$, so the maximally random matrix consistent with the symmetries has a single GOE block. 

A similar argument as presented in \cite{Stanford:2019vob} in the context of SYK applies here: when $(-1)^{\sf F_R}$ is inserted in the path integral, each right-moving free fermion $\psi^a_R$ has a zero-mode $\psi_{R,0}$. For odd $N$, in either sector of the Hilbert space, there is a an odd number of fermionic zero modes $\psi_{R,0}^a$. Hence in the free fermion 2d CFT, the partition functions in the $\NS\R$ Hilbert space with $(-1)^{\sf F_R}$ insertion vanish exactly.

The slope of the ramp in \eqref{eq:(-1)^FL SFF NSR} is a factor of two greater than what we should expect from a random matrix with one GOE block. This factor of two has a similar origin as in the SYK odd $N$ anomaly explained in \cite{Stanford:2019vob}. In this case, the path integral is related to a trace in the Hilbert space via \cite{Seiberg:2023cdc}:
\begin{equation} \label{eq: NSR anomalous Z}
    Z_{\NS\R\pm\pm}(\tau) = \sqrt2\, \Tr_{\H_{\NS\R}}\left[(\mp1)^{\sf F_L}(\mp1)^{\sf F_R} q^{L_0 - \frac{c}{24}} \bar{q}^{\bar{L}_0 - \frac{c}{24}} \right] \,.
\end{equation}
For a two-boundary partition function, this $\sqrt2$ explains the missing factor of two in \eqref{eq:(-1)^FL SFF NSR}.

\smallskip

\noindent\textbf{For $N\equiv2,6\text{ mod }8$:} the relationship between $\r\T$ and $(-1)^{\sf F}$ is anomalous, with
\begin{equation}
    (\r\T)^2 = 1\,,\quad \{\r\T,(-1)^{\sf F}\} = 0\,,\quad [\r\T,(-1)^{\sf F_L}] = 0\,.
\end{equation}
Then $\r\T$ relates right-moving bosons and fermions, which would predict that \eqref{eq:NSR rho bos/fer} holds exactly. Since $\r\T$ now anticommutes with $(-1)^{\sf F}$, the maximally random matrix consistent with the symmetries consists of the decomposition \eqref{eq:(-1)^FL 2 GOE} with two GUE blocks that are not statistically independent, instead $H_{b} = H_{f}$. This explains the factor of $4$ times GUE in the ramp \eqref{eq:(-1)^FL SFF NSR}.

\smallskip

\noindent\textbf{For $N\equiv3,5\text{ mod }8$:} there is no $(-1)^{\sf F}$, so the Hilbert space again decomposes as $\mathcal{H}_{\NS\R}=\bigoplus_j \mathcal{H}_j$, and since $(\r\T)^2=-1$, the maximally random matrix consistent with the symmetries has a single GSE block. The partition functions in the $\NS\R$ Hilbert space with $(-1)^{\sf F_R}$ insertion vanish for the same reasons as $N\equiv1,7\text{ mod }8$, and the missing factor of two in comparing to the ramp coefficient is similarly explained by \eqref{eq: NSR anomalous Z}.

\smallskip

\noindent\textbf{For $N\equiv4\text{ mod }8$:} Since
\begin{equation}
    (\r\T)^2 = -1\,,\quad [\r\T,(-1)^{\sf F}] = 0
\end{equation}
we have a similar decomposition as in \eqref{eq:(-1)^FL 2 GOE}, but with two independent GSE blocks.

\section{Discussion \label{sec:discussion}}

We introduced a fermionic version of pure 3d gravity and analyzed its implications to models of quantum chaos in fermionic 2d CFTs as well as random matrix universality. This is a theory of pure gravity without the presence of low-energy fermionic degrees of freedom which nonetheless has fermionic black holes in its spectrum. It is therefore ideal to explore notions of quantum chaos in fermionic CFTs that are not supersymmetric. We studied the gravitational path integral on the solid torus and the torus two-boundary wormhole. The restriction to spin manifolds leads to results which are consistent with characteristic signatures of quantum chaos but with important refinements tied to fermion parity and other discrete symmetries such as spacetime parity ${\sf R}$. Our results are consistent with a recently-introduced model of 2d quantum chaos referred to as RMT${}_2$ \cite{Boruch:2025ilr}. We also studied the implications of adding non-trivial topological field theories in the gravitational path integral weighting different spin (or pin) structures and show that the results for the torus two-boundary wormhole is always compatible with the classification of 2d CFT anomalies. 

\smallskip

Our analysis also highlights several directions for generalization. One natural extension is toward anyonic or more general spin-like structures that can arise only in 3d since $\pi_1 (\text{SO}(2,1)) = \mathbb{Z}$. This would lead to a theory of pure gravity with a gap and anyonic black hole microstates. For example, the results for the gravitational path integral on the solid torus would be similar to section \ref{sec:Path_integral_fermionic_3d_gravity} but with the modular groups replaced by congruent subgroups of $\text{PSL}(2,\mathbb{Z})$ of higher order. As emphasized in section \ref{sec:rmt2}, the mathematical theorems behind the connection between 3d gravity and RMT${}_2$ still hold for more general congruent (or any) subgroups of $\text{PSL}(2,\mathbb{Z})$. It would also be interesting to extend this analysis to the gravitational path integral in $\mathcal{N}\geq 1$ supergravity in asymptotic AdS$_3$ spacetimes \cite{Jensen:wip} (see also \cite{Bhattacharyya:2024vnw}).

\smallskip

It would be interesting to extend the analysis here to more general on-shell geometries in fermionic gravity following the approach of either \cite{Collier:2023fwi} or \cite{Hartman:2025ula}. Similar to what we saw for the solid torus or the wormhole, the effect of the sum over spin structures becomes more subtle since one needs to preserve the boundary spin structure restricting the sum over modular images. Nevertheless, no current approach can reliably deal with off-shell geometries and this would be the most interesting direction since fermionic 3d gravity presents the same issues as its bosonic counterpart near threshold. 

\smallskip

The techniques used in section \ref{sec:discretesymm}, borrowed from the theory of SPT phases of matter, can potentially be powerful in higher-dimensional gravity as well. This could happen in two ways. First, a proper understanding of the boundary bordism groups in the presence of spin/pin structures and other global symmetries can allow us to derive non-trivial vanishing theorems for the gravitational path integral with certain boundary conditions. (This was used in this paper to derive non-trivial symmetry properties of the ensemble of fermionic 2d CFTs.) Second, the bulk cobordism groups can classify possible invertible TQFTs that should be included in the proper definition of the gravity theory. Incorporating them was crucial in section \ref{sec:discretesymm} when comparing gravity with the set of possible anomalies that fermionic 2d CFTs could display. 

\smallskip
\begin{figure}
    \centering
\includegraphics[width=0.9\linewidth]{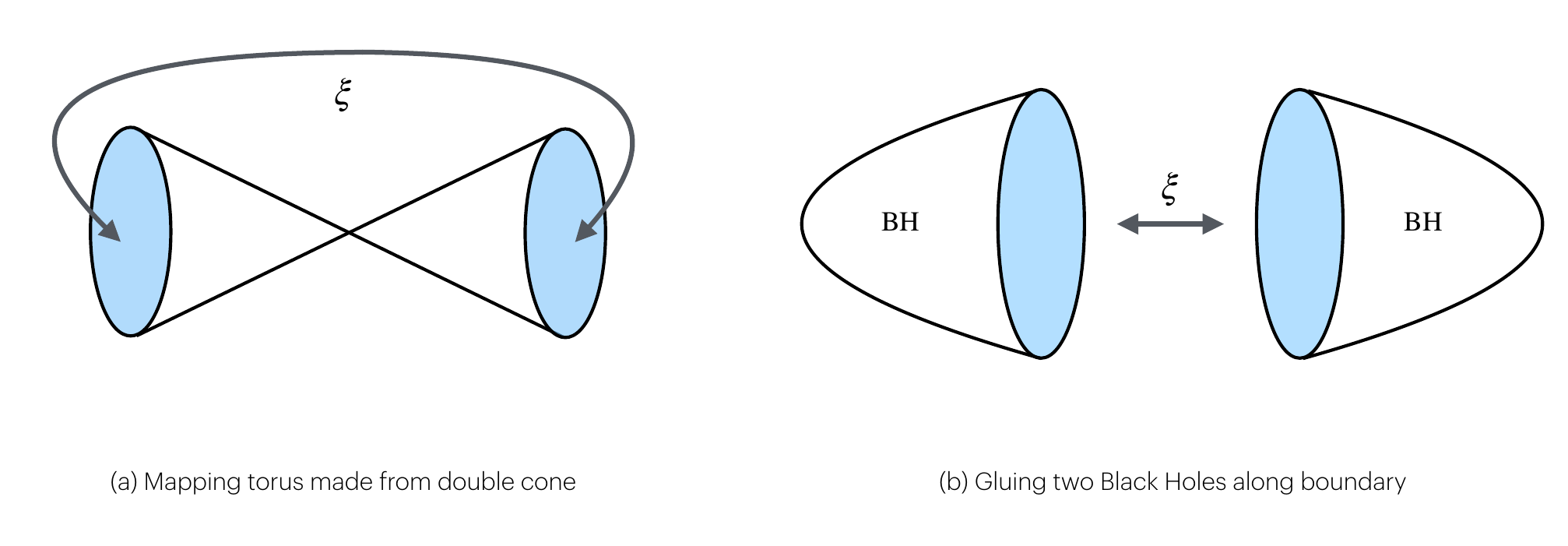}
    \caption{Bulk TQFT effects in higher dimensions can be probed by twisted gluings. \textbf{Left:} ramp/level repulsion from gluing the boundaries of the black hole double-cone after acting with $\xi$ (which could be a large diffeomorphism, fermion parity, ${\sf RT}$, etc.). \textbf{Right:} the induced modification of the black hole spectrum from gluing two black holes along their boundaries with the same $\xi$-twist.}
    \label{fig:glue_dc}
\end{figure}

Although we do not have a higher-dimensional exact evaluation of any wormhole path integral, we do have approximate solutions such that the double-cone and its universal ramp contribution to the spectral form factor \cite{Saad:2018bqo,Chen:2023hra}. Consider for example a theory of fermionic asymptotically AdS${}_4$ gravity. One can show that the cobordism group in this case is trivial $\text{TP}_4^{\text{Spin}} = 0$ and therefore no corresponding TQFT exist. But suppose that we assume the boundary theory is parity invariant such that the relevant bulk structure is $\text{Pin}^+$. In that case we know that $\text{TP}_4^{\text{Pin}^+} = \mathbb{Z}_{16}$ leading to non-trivial TQFTs to include in the gravitational path integral. The presence of this TQFT actually in principle affects the result for the black hole spectrum as well as the double-cone path integral. To see this we need to first realize the presence of twisted double cones where we act with either fermion parity, ${\sf RT}$ or ${\sf R}$ on one boundary relative to the other. Each member of the $\mathbb{Z}_{16}$ family leads to different phases for each of these contributions, and the result should match with a reasonable ensemble of 3d CFTs with the corresponding anomalies in fermion parity and time-reversal. We leave a more thorough study of the implications of this for future work. For now we can say that to evaluate the relative phases between different twisted double cone geometries we can do a similar procedure as we did for the torus and use the locality of the SPT phases to reduce the calculation to that of a mapping torus constructed out of identifying the two ends of a double-cone. This arises from gluing together the two boundaries of the double cone as we show in figure \ref{fig:glue_dc}. We can similarly study the effect on the black hole spectrum by gluing two black hole geometries along their boundary. Some recent work relevant to this direction are \cite{Gomis:2025gzb} or \cite{Cummings:2026giw}.

\smallskip

The inclusion of these bulk TQFTs might also lead to interesting effects in the evaluation of protected quantities in the context of supergravity and string theory. This is an interesting setup since one can compare rigorously the gravitational result with an exact boundary calculation. A closely related example of this feature that we are aware of has been given in \cite{Heydeman:2024fgk}. In that reference it was observed that $\mathcal{N}=2$ SCFT arising by twisted compactification of wrapped M5-branes can present mixed 't Hooft anomalies that affect their superconformal index and cause them to vanish. This was understood from the bulk as a consequence of a bulk theta term leading to a Witten effect for the dual AdS${}_4$ black holes.

\section*{Acknowledgements}

It is a pleasure to thank Yiming Chen, Gabriele Di Ubaldo, Jaume Gomis, Felix Haehl, Daniel Harlow, Luca Iliesiu, Kristan Jensen, Mukund Rangamani, Moshe Rozali, Shu-Heng Shao, Douglas Stanford, Cynthia Yan, Matt Yu, and especially Ho Tat Lam
for interesting and useful discussions. JB was supported by the Department of Energy QuantISED award DE-SC0019380. GJT would like to thank Jonah Librande for initial collaboration and assistance with some of the calculations in section 3. GJT was supported by the DOE Early Career Award DE-SC0026287 and funds from the University of Washington.

\bigskip

\appendix
\section{Fourier transform of solid torus path integral}\label{app:spectrum}

Here we provide some details on the integral transform necessary to extract the density of states for pure fermionic 3d gravity. The calculation is parallel to the one in \cite{Benjamin:2020mfz} but the fermionic case has some details worth emphasizing, in particular regarding to the quantization conditions for angular momentum.

Let us begin with the path integral on the solid torus with boundary spin structure $(\mu,\nu)$. (It will be more convenient in this Appendix to use this label than $\NS\pm$ or $\R\pm$.) We want to expand
$$
Z_{(\mu,\nu)}(\tau,\bar{\tau}) = \sum_{j \in  \mathbb{Z}+ \nu} \int_{0}^\infty \d \Delta ~ \rho^{(\mu,\nu)}\left(\frac{\Delta+j}{2},\frac{ \Delta-j}{2}\right) \, \chi_{\frac{\Delta+j}{2}}(q) \chi_{\frac{\Delta-j}{2}}(\bar{q}),
$$
where $\Delta= h + \bar{h}$ and $j = h- \bar{h}$.  All partition functions can be written as sums of $|\chi_{\text{vac}}(\gamma \cdot \tau)|^2$ over some sector-dependent set of modular transformations. We use the identity \cite{MaxfieldUnp, Benjamin:2020mfz}
\begin{equation}
    \chi_h(\gamma\tau) = \int_{\frac{c-1}{24}}^\infty \d h' \, \K_{h'h}^{(\gamma)} \, \chi_{h'}(\tau)
\end{equation}
where the modular crossing kernel is
\begin{equation}
    \K_{h'h}^{(\gamma)} = \epsilon(\gamma) e^{\frac{2\pi\ii}{s}\left[a\left(h-\frac{c-1}{24}\right)+d\left(h'-\frac{c-1}{24}\right)\right]} \underbrace{\frac{\cos\left(\frac{4\pi}{s}\sqrt{\left(h-\frac{c-1}{24}\right)\left(h'-\frac{c-1}{24}\right)}\right)}{\sqrt{s/2}\sqrt{h'-\frac{c-1}{24}}}}_{=d_h(h',s)} \,,
\end{equation}
and we also defined the function $d_h(h',s)$ to simplify some expressions below. $\epsilon(\gamma)$ is an unimportant phase that arises from the modular transformation properties of the Dedekind eta function. It will cancel below when combining right- and left-movers. This expression cannot be used when $\gamma=\text{id}$ is in the set of relevant modular images. For spin structures that include $\gamma=\text{id}$ its contribution has to be separated by hand.

In the solid torus path integral, we can find its modular properties rewriting the vacuum character as $\chi_{\text{vac}}=\chi_{h=0} - \chi_{h=1}$ and applying the modular kernel to the two terms separately. The partition function becomes
\beq
\begin{split}\label{eq:ZintofK}
    Z^{(\mu,\nu)} &=  \chi_{\text{vac}}(q) \chi_{\text{vac}}(\bar{q}) \delta_{\nu,1/2} \\
    &+ \int \d h \, \d \bar{h} \sum_{\gamma\neq \text{id}}\left[\K_{h0}^{(\gamma)} \overline\K_{\bar h0}^{(\gamma)} - \K_{h0}^{(\gamma)} \overline\K_{\bar h1}^{(\gamma)} - \K_{h1}^{(\gamma)} \overline\K_{\bar h0}^{(\gamma)} + \K_{h1}^{(\gamma)} \overline\K_{\bar h1}^{(\gamma)}\right] \chi_h(q) \chi_{\bar{h}}(\bar{q})
    \end{split}
\eeq
The trivial modular action is only included for the NS sector. The R sector does not include this state. We focus on non-vacuum contributions from now on.

\smallskip

We can insert the explicit expression for the modular crossing kernel and assemble the terms in the following way
\begin{equation} \label{eq:densitySectors}
\begin{split}
    \rho^{(\mu,\nu)}(h,\bar h) &= \sum_{s=1}^\infty S^{(\mu,\nu)}(j,0;s) d_0(h,s) d_0(\bar h,s) - S^{(\mu,\nu)}(j,-1;s) d_0(h,s) d_1(\bar h,s) \\
    &\qquad - S^{(\mu,\nu)}(j,1;s) d_1(h,s) d_0(\bar h,s) + S^{(\mu,\nu)}(j,0;s) d_1(h,s) d_1(\bar h,s)
\end{split}
\end{equation}
We now outline the derivation and define $S^{(\mu,\nu)}(j,j',s)$ appearing on the RHS in each sector:
\begin{itemize}
    \item \textbf{$\R-$ sector:} For $\gamma = \begin{pmatrix}
    \tilde{a} & \tilde{b} \\
    \tilde{s} & \tilde{d}
\end{pmatrix}$ one gets in our conventions $S\cdot \gamma = \begin{pmatrix}
    -\tilde{s} & -\tilde{d} \\
    \tilde{a} & \tilde{b}
\end{pmatrix} $ acting on $\tau$. Since $\gamma \in \Gamma_0(2)$ we restrict to odd $\tilde{a}$, even $\tilde{s}$ and no constrain on $\tilde{b}$. This implies that we should sum over coprime $s$ and $d$, with odd $s$ with unconstrained $d$. Following \cite{Maloney:2007ud}, we expand $d= d' + n s$ with $d'$ and $s$ coprime and $n \in \mathbb{Z}$, but only for odd $s$. When evaluating $a$ given $s$ and $d$ we need to keep in mind that it should be even. The $n$ dependence in the Fourier transform arises only from the term $e^{\frac{2\pi \i}{s} d (h-\bar{h})}$ and is
$$
\sum_{n\in\mathbb{Z}} e^{2\pi \i n (h-\bar{h})} = \sum_{j\in\mathbb{Z}} \delta(h-\bar{h}-j)
$$
where we used Poisson resummation. We can therefore see precisely which subset of the modular images impose momentum quantization. We also see that the momentum is integral in the R Hilbert space. We can change variables from $\d h \, \d \bar{h} \to \frac{1}{2} \d \Delta \, \d j$ and the Delta function above restricts $j$ to be integer. A sum over $s$ and $d'$ remains which we relabel as $d$. The factor in \eqref{eq:densitySectors} in this sector is
$$
S^{(1,0)}(j,j';s) = \begin{cases}
    \sum_{\substack{0\leq d<s\\ \gcd(s,d)=1}} \exp\left(2\pi\ii \frac{dj+a_{\R}(d,s)j'}{s}\right) ~~~~\text{for $s$ odd}\\
    0 ~~~\text{else}
\end{cases}
$$
where 
\begin{align}
    a_{\R}(d,s) &= \left\{\begin{matrix}
        (d^{-1})_{s} & \text{ if $(d^{-1})_{s}$ is even} \\
        (d^{-1})_{s} + s &\text{ else}
    \end{matrix}\right. 
    \end{align}
Since $j'\in\{-1,0,1\}$ the shift in $s$ is irrelevant, but our expression also holds when $J$ is half-integer which is relevant in supergravity.

\item \textbf{$\NS+$ sector:} We sum over coprime $(s,d)$ with the only constraint that $d$ is odd. We can rewrite now $d= d' + n s$ with $n$ even and $0< d' < 2s$ as in \cite{Maloney:2007ud}.  The $n$ dependence in \eqref{eq:ZintofK} is now 
$$
\sum_{n\in2\mathbb{Z}} e^{2\pi \i n (h-\bar{h})} = \frac{1}{2} \sum_{j\in\mathbb{Z}/2} \delta(h-\bar{h}-j)
$$
So besides the spin being now both integer (bosonic black holes) or half-integer (fermionic black holes), there is an extra factor of a half in the density of states. Finally since $b$ is even in this sector the relevant value of $a$ is $(d^{-1})_{2s}$ and we find
$$
S^{(0,1)}(j,j';s) = \sum_{\substack{0\leq d<2s\\ \gcd(s,d)=1\\ d\text{ odd}}} \exp\left(2\pi\i \frac{dj+(d^{-1})_{2s}j'}{s}\right)
$$

\item \textbf{$\NS-$ sector} The relevant sum with this boundary condition is
$$
S^{(1,1)}(j,j';s) = \sum_{\substack{0\leq d<2s\\ \gcd(s,d)=1\\ d+s\text{ odd}}} \exp\left(2\pi\ii \frac{dj+a_{\NS}(d,s)j'}{s}\right)
$$
where
\begin{align}
    a_{\NS}(d,s) &= \left\{\begin{matrix}
        (d^{-1})_{2s} & \text{ if $s$ is even} \\
        (d^{-1})_{s} &\text{ if $s$ is odd and $(d^{-1})_{s}$ is even} \\
        (d^{-1})_{s} + s &\text{ else}
    \end{matrix}\right.
\end{align}
\end{itemize}

One can prove that the Kloosterman sums appearing in the $\NS+$ and $\NS-$ boundary conditions are not independent
$$
S^{(1,1)}(j,j';s) = (-1)^{2j+2j'} S^{(0,1)}(j,j';s).
$$
This guarantees the NS spin-statistics $(-1)^{\sf F}=e^{2\pi \i J}$. This is true since in the partition function only $j'=-1,0,1$ appear and therefore the only change in the density of states between $\NS+$ and $\NS-$ is a factor of $(-1)^{2j}$. Finally, $S^{(1,1)}(j,j';s)$ is what we refer to as $S_{\NS}(j,j';s)$ in the main text and $S^{(1,0)}(j,j';s)$ is what we refer to as $S_{\R}(j,j';s)$. 

\smallskip

Finally we can determine the discrete set of states that appear at $E=j=0$ when regularizing the sum over $s$. Following \cite{Benjamin:2020mfz}, we regulate the sum by modifying the weight $w>1/2$ of the Dedekind eta function by hand and reevaluating the kernel. To determine the leading effect of the divergence we can Taylor expand the cosines at $s=\infty$ and keep the leading term, higher order terms will decay faster with $s$ and converge without regulator. In the Ramond sector
\begin{eqnarray}
\rho_{\R}(E,0) &\supset& \lim_{w\to1/2}\sum_{s=1}^\infty s^{-2w}E^{2w-2}  \left( 2S_{\R}(0,0;s) - S_{\R}(0,1;s) -S_{\R}(0,-1;s)\right),\nonumber\\
&\supset& \lim_{w\to1/2} \frac{\pi^{2w}}{\Gamma(w)^2E^{2w-2}}\frac{2( \zeta(2w-1)(1-2^{-(2w-1)})-1)}{\zeta(2w)(1-2^{-2w})}\nonumber\\
&=& - 4 \delta(E) + (\text{continuous function of $E$}).
\end{eqnarray}
In the first line we Taylor expanded the cosines to leading order. In the second line we wrote, for odd $s$ and integer $j'$, $S_{\R}(0,j';s)  = \sum_{\text{gcd}(a,d)=1} e^{2\pi \i j' a/s} = c_s(j')$, the Ramanujan sum. The sum of these factors can be done using standard theorems in number theory, see \cite{Maloney:2007ud} for an example. A similar calculation in the NS sector gives
\beq
\rho_{\NS}(E,0) = -2 \delta(E) + (\text{continuous function of $E$})
\eeq
Notice these results are consistent with the $-6\delta(E)$ obtained for bosonic gravity. If one gauges fermion parity, the sum over all spin structures gives $-12 \delta(E)$ which upon dividing by $2$ reproduces bosonic gravity. It would be interesting to repeat the analysis of \cite{Stanford:2025llj} for fermionic gravity to deal with these singularities.

\bibliography{Biblio}
\bibliographystyle{JHEP}

\end{document}